\documentclass[%
 aip,
 amsmath,amssymb,
preprint,%
]{revtex4-1}

\usepackage{graphicx}
\usepackage{dcolumn}
\usepackage{bm}

\usepackage[utf8]{inputenc}
\usepackage[T1]{fontenc}
\usepackage{mathptmx}
\usepackage{makecell}

\begin{document}


\title[]{Nusselt Number for Steady Periodically Developed Heat Transfer in Micro- and Mini-Channels with Arrays of Offset Strip Fins Subject to a Uniform Heat Flux}

\author{A. Vangeffelen}
\email{arthur.vangeffelen@kuleuven.be.}
\author{G. Buckinx}%
\affiliation{ 
Department of Mechanical Engineering, KU Leuven, Celestijnenlaan 300A, 3001 Leuven, Belgium
}%
\affiliation{%
VITO, Boeretang 200, 2400 Mol, Belgium
}%
\affiliation{%
EnergyVille, Thor Park, 3600 Genk, Belgium
}%

\author{C. De Servi}
\affiliation{%
VITO, Boeretang 200, 2400 Mol, Belgium
}%
\affiliation{%
EnergyVille, Thor Park, 3600 Genk, Belgium
}%

\author{M. R. Vetrano}
\affiliation{ 
Department of Mechanical Engineering, KU Leuven, Celestijnenlaan 300A, 3001 Leuven, Belgium
}%
\affiliation{%
EnergyVille, Thor Park, 3600 Genk, Belgium
}%

\author{M. Baelmans}
\affiliation{ 
Department of Mechanical Engineering, KU Leuven, Celestijnenlaan 300A, 3001 Leuven, Belgium
}%
\affiliation{%
EnergyVille, Thor Park, 3600 Genk, Belgium
}%

\date{\today}

\begin{abstract}
In this work, the Nusselt number is examined for periodically developed heat transfer in micro- and mini-channels with arrays of offset strip fins, subject to a constant heat flux.
The Nusselt number is defined on the basis of a heat transfer coefficient which represents the spatially constant macro-scale temperature difference between the fluid and solid during conjugate heat transfer.
Its values are determined numerically on a single unit cell of the array for Reynolds numbers between 1 and 600 and fin height-to-length ratios below 1. 
Two combinations of the Prandtl number and the thermal conductivity ratio are selected, corresponding to air and water.
It is shown that the Nusselt number correlations from the literature mainly apply to air in the transitional flow regime in larger conventional channels if the wall temperature remains uniform. 
As a result, they do not correctly capture the observed trends for the Nusselt number in micro- and mini-channels subject to a constant heat flux. 
Therefore, new Nusselt number correlations, obtained through a least-squares fitting of 2282 numerical simulations, are presented for air and water. 
The suitability of these correlations is assessed via the Bayesian approach for parameter estimation and model validation. 
The correlations respect the observed asymptotic trends and limits of the Nusselt number for all the geometrical parameters of the offset strip fins. 
In addition, they predict a linear dependence of the Nusselt number on the Reynolds number, in good agreement with the data from this work.  
Nevertheless, a detailed analysis reveals a more complex scaling of the Nusselt number with the Reynolds number, closely related to the underlying flow regimes, particularly the weak and strong inertia regimes. 
Finally, through 62 additional simulations, the influence of the material properties on the Nusselt number is illustrated and compared to the available literature. \\

\end{abstract}

\maketitle

\section{\label{sec:intro}Introduction}

Heat transfer in periodic fin arrays in micro- and mini-channels has been investigated over the past twenty years due to the increasing demand for high-power-density heat transfer devices in numerous applications (Refs.~\onlinecite{kandlikar2005heat,khan2006role,izci2015effect,yang2017heatpin}). 
The channels of heat transfer devices are commonly categorized based on their smallest dimension. 
Notably, it is the practice to distinguish between micro- and mini-channels if the smallest dimension ranges from 10 $\mu$m to 200 $\mu$m or 200 $\mu$m to 3 mm, respectively (Refs.~\onlinecite{kandlikar2005heat}). 
Particular attention in the literature has been devoted to micro- and mini-channels with arrays of periodic offset strip fins due to their favorable thermo-hydraulic characteristics (Refs.~\onlinecite{bapat2006thermohydraulic,yang2007advanced, hong2009three,do2016experimental,nagasaki2003conceptual,yang2017heat,jiang2019thermal,yang2014design,pottler1999optimized}). 
For example, micro-channels with an array of periodic offset strip fins have been exploited for the cooling of microelectronic systems to cope with the increased thermal management needs resulting from the continuous miniaturization of such devices (Refs.~\onlinecite{tuckerman1981high, bapat2006thermohydraulic, yang2007advanced, hong2009three, bartolini2012thermal}). 
Offset strip fins in mini-channels are also adopted in various energy conversion applications where a compact design of the heat transfer device or a small temperature difference between two flows is desired. 
This is the case for heat recuperators of small-capacity gas turbines (Refs.~\onlinecite{do2016experimental,nagasaki2003conceptual}), heat exchangers for refrigeration and liquefaction in cryogenic systems (Refs.~\onlinecite{yang2017heat,jiang2019thermal}), and solar air heating collectors (Refs.~\onlinecite{yang2014design,pottler1999optimized}). 


As highlighted in our previous work (Refs.~\onlinecite{vangeffelen2021friction}), the flow regime inside micro- and mini-channels with offset strip fins is typically steady and laminar due to the small dimensions. 
Moreover, as the fin height-to-length ratio commonly remains below 1, the flow regime is characterized by low to moderate Reynolds numbers ranging from 10 to 500 (Refs.~\onlinecite{tuckerman1981high, bapat2006thermohydraulic, yang2007advanced, hong2009three,do2016experimental,nagasaki2003conceptual,yang2017heat,jiang2019thermal,yang2014design,pottler1999optimized}). 

The heat transfer regime in most micro- and mini-channel applications is typically characterized by a fluid Prandtl number $Pr_f$  which equals 0.7 or 7, for air and water, respectively (Refs.~\onlinecite{tuckerman1981high,bapat2006thermohydraulic,yang2007advanced,do2016experimental,nagasaki2003conceptual,hong2009three,bartolini2012thermal,yang2014design,pottler1999optimized}). 
For cryogenic fluids such as helium, hydrogen, and nitrogen, both in the gaseous and liquid state, the Prandtl number is of the same order of magnitude as for air (Refs.~\onlinecite{yang2017heat,jiang2019thermal}). 
Furthermore, the heat transfer process in typical micro- and mini-channel applications can be accurately simulated by assuming that a steady and uniform heat flux is established on the channel wall. 
This so-called \textit{H-type} boundary condition is often assumed to be representative of the heat transfer process occurring in cooling systems for electronics, as well as balanced counter-flow heat exchangers, such as gas turbine recuperators (Refs.~\onlinecite{shah1978laminar,renfer2013microvortex,xia2017micro,zhang2011convective,gong2020heat}). 
Similarly, a steady and uniform heat flux is commonly considered to model the heat transfer process in micro- and mini-channels of solar air heaters and heat exchangers of cryogenic systems (Refs.~\onlinecite{yang2014design,pottler1999optimized,priyam2016thermal,jiang2019thermal}). 

In the literature, both experimental and numerical studies have been conducted to investigate the heat transfer regime occurring in channels with an offset strip fin array (Refs.~\onlinecite{manson1950correlations,wieting1975empirical,kays1984compact,joshi1987heat,manglik1995heat,dong2007air,kim2011correlations}). 
Nonetheless, these studies are primarily limited to conventional channels with fin dimensions in the centimeter range and air as working fluid. 
Then, the investigated conditions are representative of applications such as automotive radiators and air-conditioning condensers and evaporators, which operate at high air flow rates. 
Therefore, the experimental and numerical data in the literature are mainly applicable to transitional and turbulent flows of working fluids with Prandtl numbers around 0.7, and heat transfer processes occurring under the condition of a uniform channel wall temperature. 
In our previous work (Refs.~\onlinecite{vangeffelen2021friction}), we have characterized the friction factor for steady laminar periodically developed flow through offset strip fins arrays in micro- and mini-channels.  
It was found that the friction factor correlations available in the literature, which were constructed based on the data for the transitional flow regime in larger conventional channels, result in discrepancies of 20\% to 80\% for micro- and mini-channels. 
In this work, similar discrepancies are found for the Nusselt number correlations from the literature, in particular when the channel walls are subject to a constant heat flux instead of a constant temperature.
Therefore, the present work aims to study the Nusselt number for offset strip fin arrays in micro- and mini-channels, subject to a constant heat flux. 
In contrast to previous studies, the conjugate heat transfer between the fluid and the solid fins is taken into account, as we aim to include the influence of the material properties of the solid on the Nusselt number.
Besides, it is known that conjugate heat transfer can have a significant effect on the Nusselt number (Refs.~\onlinecite{cukurel2013local,li2016effect}), which may not be recognizable from the available data in the literature.

The experimental Nusselt number data reported in the literature are typically derived from temperature measurements upstream and downstream of an air-side offset strip fin array for a range of flow rates through the array. 
For such measurements, the channel wall temperature is mainly maintained at steady and uniform conditions using a control flow, which is a condensing steam or an evaporating water flow along the outer surface of the channel. 
The temperature measurements are used to evaluate the area-averaged temperature at the in- and outlet section of the array. 
This information allows the overall heat transfer coefficient $h_{0}$ to be estimated by applying the effectiveness-and-number-of-transfer-units ($\epsilon$-NTU) method, or the logarithmic mean temperature difference (LMTD) approach (Refs.~\onlinecite{kays1984compact,hu1995prandtl,wang2000data,dong2007air}). 
However, both methods rely on two critical assumptions: i) the temperature distribution is assumed to be one-dimensional, and ii) the overall heat transfer coefficient is assumed to be spatially constant. 
Also, a simplified expression for the fin temperature effectiveness is used with these methods. 

\begin{figure}[ht!]
\includegraphics[scale = 1.0]{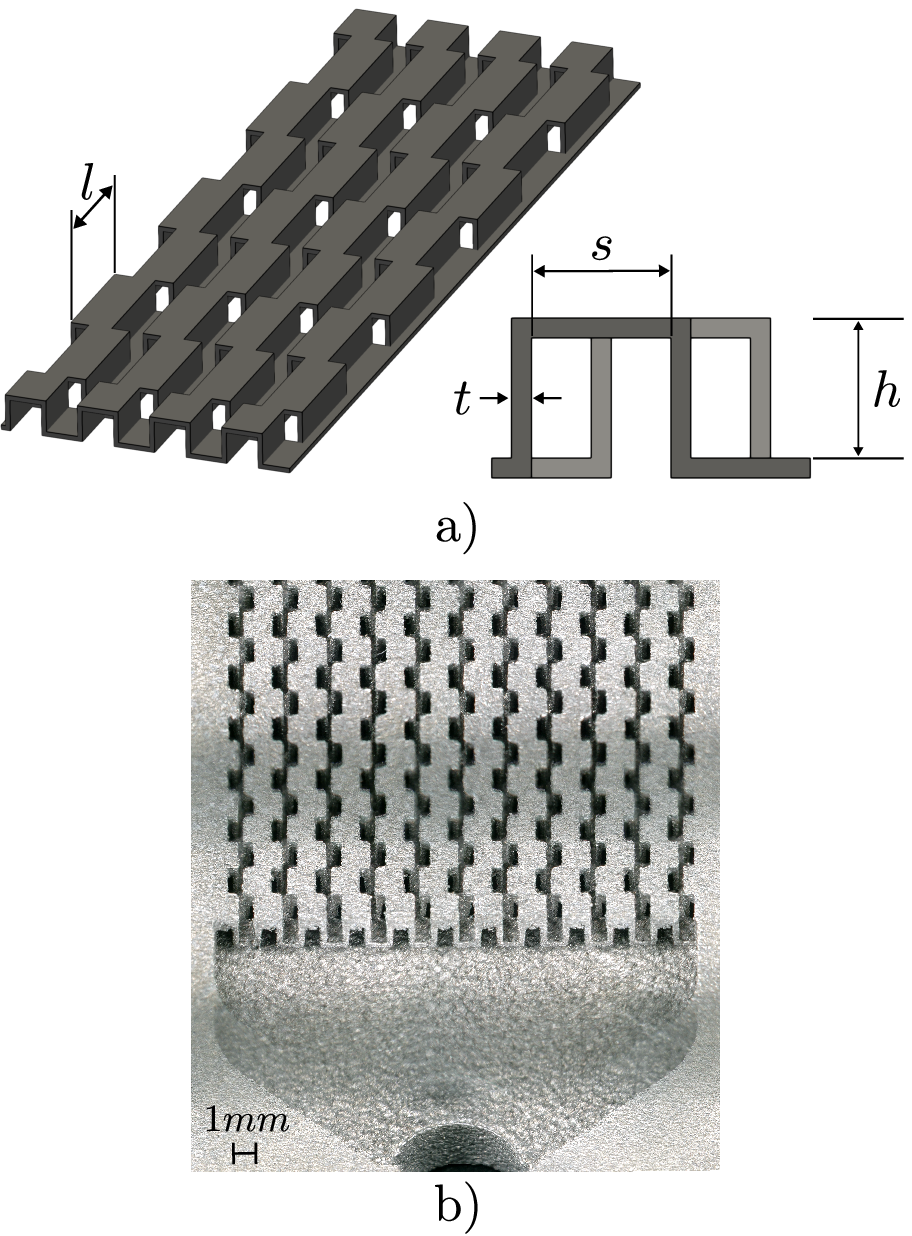}
\caption{\label{fig:osf} (a) Geometrical parameters of an array of periodic offset strip fins, actual mini-channel with an array of periodic offset strip fins, (b) produced in-house by the laser powder bed fusion manufacturing technique}
\end{figure}

The relation between the overall heat transfer coefficient and the flow rate is usually expressed by means of a correlation between the Stanton number $St \triangleq \frac{Nu}{Re_{D_h} Pr_f}$ or the Colburn j-factor $j \triangleq \frac{Nu}{Re_{D_h} Pr_f^{1/3}}$, based on the Nusselt number $Nu \triangleq h_{0} D_h/k_f$, and the Reynolds number $Re_{D_h}\triangleq \rho_f U_{\text{ref}} D_h/\mu_f$. 
These dimensionless numbers depend on the hydraulic diameter $D_h$, the conductivity $k_f$, the density $\rho_f$, and the dynamic viscosity $\mu_f$ of the fluid, as well as some reference flow speed $U_{\text{ref}}$. 
The heat transfer correlations for offset strip fin arrays reported in the literature are chronologically listed in Tables \ref{tab:literature} and \ref{tab:literature2}. 
For each study, the tables include the corresponding definition of the hydraulic diameter $D_h$, which is derived from the geometrical parameters of the offset strip fin array. 
These are the fin length $l$, the fin height $h$, the lateral fin pitch $s$ and the fin thickness $t$, as shown in Figure \ref{fig:osf}. 
Figure \ref{fig:osf} also displays an actual offset strip fin mini-channel which has been fabricated by the additive manufacturing technique of laser powder bed fusion (Refs.~\onlinecite{jadhav2021laser}). 
In addition, the considered ranges of the Reynolds number $Re_{D_h}$, the fin height-to-length ratio $h/l$ and the fluid Prandtl number $Pr_f$ in each study are given in Tables \ref{tab:literature} and \ref{tab:literature2}. 
In the study of Joshi and Webb, the reference velocity $U_{\text{ref}}$ is defined as the average bulk velocity through the cross-section area $(s-t)h$, whereas in the study of Bhowmik and Lee, it is based on the cross-section area  $(s+t)h$. 
In the remaining studies listed in Tables \ref{tab:literature} and \ref{tab:literature2}, $U_{\text{ref}}$ is defined with respect to the flow passage area $sh$. 
Next, the correlations from Tables \ref{tab:literature} and \ref{tab:literature2} are briefly reviewed. \\

\begin{table*}[ht!]
\caption{\label{tab:literature}Chronological list of heat transfer correlations for offset strip fin channels}
\begin{ruledtabular}
\begin{tabular}{llccc}
Researchers & Correlation & $Re_{D_h}$ & $h/l$ & $Pr_f$ \\
\hline\hline
\makecell[l]{Manson \\ (1950)} & \makecell[l]{
    $j = 
    \begin{cases}
        0.6 (l/D_h)^{-0.5} Re_{D_h}^{-0.5} & \text{for } l/D_h \leqslant 3.5 \\
        0.6 (3.5)^{-0.5}   Re_{D_h}^{-0.5} & \text{for } l/D_h > 3.5
    \end{cases} $\\
    where \quad $D_h = 2sh/(s+h)$}
    & 370-9020 & 4.8-7.9 & 0.7\\
\hline
\makecell[l]{Wieting \\ (1975)} & \makecell[l]{
    For $Re_{D_h} \leqslant 1000$ \\ 
    \quad $j = 0.483 (l/D_h)^{-0.162} (s/h)^{-0.184} Re_{D_h}^{-0.536}$ \\
    For $Re_{D_h} > 2000$ \\ 
    \quad $j = 0.242 (l/D_h)^{-0.322} (t/D_h)^{0.089} Re_{D_h}^{-0.368}$ \\
    where $D_h = 2sh/(s+h)$}
    & 120-50000 & 0.23-5.1 & 0.7\\
\hline
\makecell[l]{Joshi \& Webb \\ (1987)} & \makecell[l]{
    For $Re_{D_h} < Re^{*}$ \\ 
    \quad $j = 0.53 (l/D_h)^{-0.15} (s/h)^{-0.14} Re_{D_h}^{-0.50}$ \\
    For $Re_{D_h} > Re^{*} + 1000$ \\ 
    \quad $j = 0.21 (l/D_h)^{-0.24} (t/D_h)^{0.02} Re_{D_h}^{-0.40}$ \\
    where \\
    \quad $Re^{*} = 257 (l/s)^{1.23} (t/l)^{0.58} D_h$ \\
    \quad \quad $\times \left[ t + 1.328 \left( Re_{D_h} / (l D_h) \right)^{-0.5}  \right]^{-1} $ \\
    \quad $D_h = 2(s-t)h/((s+h)+ht/l))$}
    & 120-50000 & 0.23-5.1 & 0.7\\
\hline
\makecell[l]{Manglik \& Bergles \\ (1995)} & \makecell[l]{
    $j = 0.6522 (s/h)^{-0.1541} (t/l)^{0.1499} (t/s)^{-0.0678} Re_{D_h}^{-0.5403}$ \\
    \quad $\times \left[ 1 + 5.269 \cdot 10^{-5} (s/h)^{0.504} (t/l)^{0.456} (t/s)^{-1.055} Re_{D_h}^{1.340} \right]^{0.1} $ \\
    where $D_h = 4shl/(2(sl+hl+th)+ts)$}
    & 120-10000 & 0.23-5.1 & 0.7\\
\hline
\makecell[l]{Dong et al. \\ (2007)} & \makecell[l]{
    $j = 0.101 (s/h)^{-0.488} (t/l)^{-0.297} (t/s)^{0.479} Re_{D_h}^{-0.189} (L/l)^{-0.315}$ \\
    where \\
    \quad $L$ is the flow length \\
    \quad $D_h = 2sh/(s+h)$}
    & 500-7500 & 0.91-2.3 & 0.7\\
\end{tabular}
\end{ruledtabular}
\end{table*}

\begin{table*}[ht!]
\caption{\label{tab:literature2}Chronological list of heat transfer correlations for offset strip fin channels (continued)}
\begin{ruledtabular}
\begin{tabular}{llccc}
Researchers & Correlation & $Re_{D_h}$ & $h/l$ & $Pr_f$ \\
\hline\hline
\makecell[l]{Guo et al. \\ (2008)} & \makecell[l]{
    $j = 0.8452 (2(s+t)/D_h)^{-0.5476 \alpha + 0.2326} (h/D_h)^{0.1819 \alpha - 1.4207}$ \\
    \quad $\times (l/D_h)^{0.2558 \alpha - 0.7685} Re_{D_h}^{-0.3958} \alpha^{0.05855}$ \\
    where \\
    \quad $\alpha$ is the flow angle of attack \\
    \quad $D_h = 4V_{\text{free}}/A_{\text{surface}}$ \\
    \quad $V_{\text{free}}$ is the free flow volume \\
    \quad $A_{\text{surface}}$ is the wetted surface area}
    & 20-400 & 1.9-2.9 & >100\\ 
\hline
\makecell[l]{Bhowmik \\ \& Lee (2009)} & \makecell[l]{
    $j = 0.489 Re_{D_h}^{-0.445}$ \\
    where $D_h = 2(s-t)h/(s+h+th/l)$}
    & 10-3500 & 1 & 7\\
\hline
\makecell[l]{Kim et al. \\ (2011)} & \makecell[l]{
    For $\beta < 0.2$ \\
    \quad $j = exp(1.96) (s/h)^{-0.098} (t/l)^{0.235} (t/s)^{-0.154}$ \\
    \quad \quad $\times  Re_{D_h}^{0.0634 ln (Re_{D_h}) - 1.3} Pr_f^{0.00348}$ \\
    For $0.2 \leqslant \beta < 0.25$ \\
    \quad $j = 1.06 (s/h)^{-0.1} (t/l)^{0.131} (t/s)^{-0.08}$ \\
    \quad \quad $\times  Re_{D_h}^{0.0323 ln (Re_{D_h}) - 0.856} Pr_f^{0.0532}$ \\
    For $0.25 \leqslant \beta < 0.3$ \\
    \quad $j = exp(1.3) (s/h)^{0.004} (t/l)^{0.251} (t/s)^{0.031}$ \\
    \quad \quad $\times  Re_{D_h}^{0.0507 ln (Re_{D_h}) - 1.07} Pr_f^{0.051}$ \\
    For $0.3 \leqslant \beta < 0.35$ \\
    \quad $j = 0.2 (s/h)^{-0.125} (t/l)^{0.21} (t/s)^{-0.069}$ \\
    \quad \quad $\times  Re_{D_h}^{0.0005 ln (Re_{D_h}) - 0.338} Pr_f^{0.0549}$ \\
    where \\
    \quad $\beta = ((2s + 2t)(h + t) - 2sh)/((2s + 2t)(h + t))$\\
    \quad $D_h = 4shl/(2(sl+hl+th)+ts)$}
    & 100-6000 & 0.046-10 & 0.72-50\\
\end{tabular}
\end{ruledtabular}
\end{table*}

\newpage
\clearpage

Manson developed the first Colburn j-factor correlations from temperature measurements related to airflows over three different offset strip fin array geometries, as well as channels with other fin geometries, such as louvered fins and flat finned tubes (Refs.~\onlinecite{manson1950correlations}). 
Manson's study mainly deals with the transitional and turbulent flow regime in larger conventional channels, as apparent from the range of both $Re_{D_h}$ and $h/l$ reported in Table \ref{tab:literature}. 
Moreover, it did not report under which thermal boundary condition the heat transfer was established in the offset strip fin array. 
Therefore, the applicability of Manson's correlations to micro- and mini-channels remains limited. 

The temperature measurements of Kays and London were employed to construct multiple heat transfer correlations (Refs.~\onlinecite{kays1984compact}). 
However, the experimental data is limited to only 50 operating points for the laminar flow regime, and the experimental study was conducted with a uniform channel wall temperature. 
This condition was established by circulating saturated steam on the outer surface of the air channels, the so-called control side of the setup. 

Wieting (Refs.~\onlinecite{wieting1975empirical}) presented one of the first correlations based on the data of Kays and London. 
One of the main limitations of Wieting's work is that it is unclear whether a consistent definition of the hydraulic diameter was used throughout the study (Refs.~\onlinecite{manglik1995heat}). 

Wieting's correlation was adapted by Joshi and Webb (Refs.~\onlinecite{joshi1987heat}). 
They re-calibrated a Colburn j-factor correlation based on a data set which included, besides the data of Kays and London, measurements from London and Shah for one offset strip fin geometry (Refs.~\onlinecite{shah1967offset}) and measurements from Walters for two other geometries (Refs.~\onlinecite{walters1969hypersonic}). 
Similarly to Kays and London, London and Shah used a condensing steam flow to impose a uniform channel wall temperature. The same experimental condition was achieved in the study of Walters through a large cooling water flow rate. 
The correlation of Joshi and Webb exhibits a discontinuity when the Reynolds number approaches a critical value $Re_{D_h}^{*}$ at the transition to a turbulent flow regime. 

Manglik and Bergles (Refs.~\onlinecite{manglik1995heat}) analyzed the same data sets of Joshi and Webb. 
Their Colburn j-factor correlation is one of the most referenced in the literature since it features an accuracy of 20\% without a discontinuous dependence on the Reynolds number over the transitional regime. 
Still, due to the considered experimental data sets, the correlation is only applicable to larger conventional offset strip fin channels. 

In order to account for entrance effects in relatively short offset strip fin arrays, Dong et al. (Refs.~\onlinecite{dong2007air}) developed a Colburn j-factor correlation which includes a dependence on the flow length $L$ (see Table \ref{tab:literature}). 
This correlation was calibrated based on the experimental data for 16 offset strip fin channel geometries with a relatively large height and a flow length ranging from 5 to 14 fin lengths. 
The experiments were carried out with air and Reynolds numbers characteristic of the transitional flow regime. 
A high flow rate of hot water was imposed on the control side to approximately achieve a uniform channel wall temperature. 
Given the geometries considered in the experiments, also the correlation of Dong et al. has a limited applicability to offset strip fins in micro- and mini-channels. 

All the previously mentioned correlations were derived from tests carried out with air as the working fluid. 
Only a limited amount of studies considered other fluids. 
Guo et al. (Refs.~\onlinecite{guo2008lubricant}) studied the heat transfer performance of lubrication oil with a Prandtl number above 100. 
They aimed to study novel designs of oil coolers containing inclined offset strip fin arrays with respect to the flow direction. 
The authors developed a Colburn j-factor correlation based on the experimental data for 16 different geometries with $h/l > 1$ and array inclination angles $\alpha$ between 0$^{\circ}$ and 90$^{\circ}$. 
This data was gathered for the laminar flow regime, as it can be seen in Table \ref{tab:literature2}, and a uniform heat flux was imposed to the channel using an electrical film heater. 
Nevertheless, since the data set encompasses different Prandtl numbers and relatively large fin heights, the correlation is not accurate enough for the micro- and mini-channel applications considered in this work. 
Moreover, its use is also hampered by the lack of an explicit definition for the hydraulic diameter and a missing fin thickness value. 

In the work of Bhowmik and Lee (Refs.~\onlinecite{bhowmik2009analysis}), the working fluid is water with a Prandtl number of 7. 
Moreover, as it can be seen from the range of Reynolds numbers in Table \ref{tab:literature2}, their $j$-correlation covers both the laminar and turbulent flow regime. 
Their correlation was fitted to numerical data from simulations for a single offset strip fin geometry with $h/l = 1$. 
Therefore, its validity for different geometries is limited. 
The simulations were performed on a three-dimensional domain containing one fin row spanning 20 fin lengths in the streamwise direction under the assumption of lateral flow and temperature periodicity. 
All material properties were assumed constant. 
Furthermore, a uniform temperature was imposed as the thermal boundary condition on the top and bottom walls of the channel. 
As mentioned before, this boundary condition is not representative of the majority of micro- and mini-channel applications where offset strip fins are employed. \\

The extension of the correlations presented above to fluids with a different Prandtl number is generally obtained by assuming the Colburn j-factor to be independent of the Prandtl number (Refs.~\onlinecite{kays1984compact,manglik1995heat,bhowmik2009analysis}). 
More specifically, if $j$ is independent of $Pr_f$, this implies that the Nusselt number scales according to $Nu \sim Pr_f^{1/3}$. 
In the work of Kays and London (Refs.~\onlinecite{kays1984compact}), it is mentioned that this scaling law applies to fluids with a Prandtl number between 0.5 and 15. 
However, they note that for viscous fluids with a high value of $Pr_f$, the relation $Nu \sim Pr_f^{0.4}$ is more appropriate. 
The influence of the Prandtl number on the Colburn j-factor for offset strip fin channels was explicitly analyzed by Kim et al. (Refs.~\onlinecite{kim2011correlations}). 
In their work, to cover a range of working fluids varying from air and water to diesel fuel, a Colburn j-factor correlation is presented for Prandtl numbers ranging from 0.72 to 50. 
Their correlation was fitted to heat transfer data obtained from numerical simulations on a three-dimensional fin row of 68 fin lengths in the streamwise direction, with a periodic boundary condition in the lateral direction. 
A wide range of values for the so-called blockage ratio $\beta$, as defined in Table \ref{tab:literature2}, and consequently also the relative fin height $h/l$, has been included in the data set by analyzing 39 different geometries. 
Yet, the numerical study of Kim et al. focuses on the turbulent flow regime and is based on the assumption of a uniform temperature in the entire solid domain of the channel. 
The accuracy of their correlation for the laminar flow regime thus remains questionable. 
The same consideration applies to the investigated range of Prandtl numbers, as the used numerical data are not available. 

Also, other studies have examined the influence of the Prandtl number on the heat transfer in offset strip fin channels. 
The most notable references are the works of Tinaut et al. (Refs.~\onlinecite{tinaut1992correlations}) and Hu and Herold (Refs.~\onlinecite{hu1995prandtl}). 
Tinaut et al. (Refs.~\onlinecite{tinaut1992correlations}) experimentally confirmed the independence of the Colburn j-factor with respect to the Prandtl number for $Pr_f \in \left(1, 100 \right)$. 
Their measurements were performed for both laminar and turbulent flows on a single offset strip fin geometry, cooled by a large water flow rate on the control side. 
Although the authors fitted an empirical Colburn j-factor correlation to their experimental data, it has little relevance to the present study since their work reports no explicit definition of the hydraulic diameter nor a value of the fin length $l$ and fin thickness $t$. 
Conversely, Hu and Herold (Refs.~\onlinecite{hu1995prandtl}) tried to capture the influence of the fluid Prandtl number on the Nusselt number through an analytical model for Prandtl numbers between 0.7 and 150. 
The model used was based on simplified heat transfer relations for flow through rectangular channels and is therefore expected to have limited validity for offset strip fin arrays. 
Their results show that the correlations calibrated on experimental data for air overestimate the Colburn j-factor with a factor of two in the case of a liquid working fluid. 
This demonstrates the limitations of assuming $j$ to be independent of $Pr_f$ and therefore contradicts the conclusion from Tinaut et al. (Refs.~\onlinecite{tinaut1992correlations}). \\

It must be emphasized that, due to their empirical nature, the discussed correlations do not provide an explicit definition of the heat transfer coefficient in the Nusselt number. 
However, the heat transfer coefficients for periodic fin arrays are commonly defined on the basis of a cross-sectional averaged or bulk-temperature difference, or the one-dimensional reference temperature which appears in the $\epsilon-NTU$ or LMTD methods (Refs.~\onlinecite{kays1984compact}), similar to the theoretically derived heat transfer coefficients for straight channels without fins. 
As such, they are all spatially dependent along the channel in case of a periodic fin array.
Nevertheless, their space dependence has been ignored in the correlations from the literature, so that the existing Nusselt number correlations are not exact and challenging to interpret from a theoretical viewpoint (Refs.~\onlinecite{buckinx2015macro,buckinx2016macro}).
For that reason, in this work, we adopt the definition of the heat transfer coefficient proposed by Buckinx and Baelmans (Refs.~\onlinecite{buckinx2015macro,buckinx2016macro}). 
The latter is spatially constant and has a clear physical meaning in the macro-scale description of the periodically developed heat transfer regimes (Refs.~\onlinecite{buckinx2015macro,buckinx2016macro}).
Its definition is also consistent with that of the interfacial heat transfer coefficient in porous media (Refs.~\onlinecite{quintard1997two,degroot2011closure,penha2012fully}), apart from some geometric scaling factors.\\



The previous literature review shows that a thorough analysis of the laminar heat transfer regime in micro- and mini-channels with offset strip fin arrays is still required. 
After all, it was found that only 50 temperature measurement points are available for the laminar flow regime (Refs.~\onlinecite{vangeffelen2021friction}). 
Therefore, our study aims to investigate the Nusselt number in micro- and mini-channels for a steady laminar flow of water and air. 
In particular, we present the first study on the Nusselt number in the periodically developed conjugate heat transfer regime, in which the channel wall is subject to a uniform heat flux. 
This heat transfer regime is of interest, since it is expected to occur after a short development length from the channel inlet in the specified applications (Refs.~\onlinecite{lee2005investigation,lee2006thermally}), based on previous flow observations (Refs.~\onlinecite{vangeffelen2021friction}). 
Furthermore, the boundary condition of a uniform heat flux, instead of a constant wall temperature, is more suitable for the analysis of many micro- and mini-channel applications. 
Finally, in contrast to the existing correlations in the literature, the Nusselt number correlations in this work are based on an exact heat transfer coefficient which is spatially constant and represents the macro-scale temperature difference between the fluid and the fins. 

This paper is organized as follows. 
Section \ref{sec:period} presents the numerical model, from the geometry of the unit cell to the periodic temperature equations in the periodically developed regime, and the numerical procedure. 
Afterwards, in Section \ref{sec:aligned}, the influence of the Reynolds number, the Prandtl number and the geometrical parameters on the Nusselt number are discussed in detail. 
The analysis covers a wide range of values for each of the geometrical parameters of the offset strip fins, as well as for the Reynolds number, which is varied between 1 and 600. 
Section \ref{sec:correlation} presents the final Nusselt number correlation for periodically developed flow and heat transfer in micro- and mini-channels with an offset strip fin array. 
Some final remarks on the influence of the thermal boundary condition and the reference temperature difference are given in Section \ref{sec:final remarks}.


\section{\label{sec:period}Unit cell geometry and periodically developed heat transfer equations}

\subsection{Geometry}

\begin{figure}[ht]
\includegraphics[scale = 1.0]{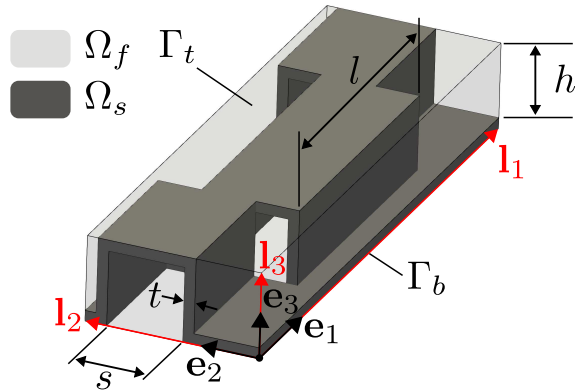}
\caption{\label{fig:UnitCellDomains} Unit cell domain of an offset strip fin array}
\end{figure}

Figure \ref{fig:UnitCellDomains} illustrates the three-dimensional unit cell $\Omega_{\text{unit}}$ (Refs.~\onlinecite{vangeffelen2021friction}) in which the periodically developed heat transfer regime is simulated. 
The unit cell consists of a fluid domain $\Omega_{f}$ and a solid domain $\Omega_{s}$, divided by a fluid-solid interface $\Gamma_{fs}$. 
The top surface $\Gamma_{t}$ and bottom surface $\Gamma_{b}$, which are a part of the unit cell's exterior boundary $\Gamma = \partial \Omega_{\text{unit}}$, coincide with the solid walls at the top and bottom of the channel. 
The unit cell is spanned by the three lattice vectors $\textbf{l}_{1} = l_{1} \textbf{e}_{1} = 2l \textbf{e}_{1}$, $\textbf{l}_{2} = l_{2} \textbf{e}_{2} = 2(s+t) \textbf{e}_{2}$ and $\textbf{l}_{3} = l_{3} \textbf{e}_{3} = (h+t) \textbf{e}_{3}$, with respect to the normalized Cartesian vector basis $\{ \textbf{e}_{j} \}_{j=1,2,3}$. 
Furthermore, the geometry of the unit cell is uniquely determined by the non-dimensional geometrical parameters $h/l$, $s/l$, and $t/l$, where the fin length $l$ represents the reference length. 
From these three non-dimensional parameters, the porosity of the unit cell can be computed as
\begin{equation}
    \epsilon = \frac{(h/l)(s/l)}{[(h/l)+(t/l)][(s/l)+(t/l)]}. 
\end{equation}

\vfill

\subsection{Periodically developed conjugate heat transfer equations}

In the steady periodically developed heat transfer regime driven by an imposed uniform heat flux, the temperature field can be decomposed into a component that varies linearly with a spatially constant gradient $\mathrm{\nabla{T}}$ and a spatially periodic component $T^{*}$ (Refs.~\onlinecite{patankar1977fully,penha2012fully}). 
Therefore, the periodic temperature field $T^{*}$, which equals $T_f^{*}$ in $\Omega_f$ and $T_s^{*}$ in $\Omega_s$, is governed by the following energy conservation equations, if viscous dissipation is negligible, and other heat sources are absent (Refs.~\onlinecite{buckinx2016macro}): 
\begin{equation}
\begin{aligned}
  \rho_{f} c_{f} \nabla \cdot \left( \boldsymbol{u} T^{*}_{f} \right) &= - \rho_{f} c_{f} \boldsymbol{u} \cdot \mathrm{\nabla{T}} + k_{f} \nabla^2 T^{*}_{f} & \text{in } \Omega_{f}, \\
  0 &= k_{s} \nabla^2 T^{*}_{s} & \text{in } \Omega_{s}.
\end{aligned}
\label{eq:Temperature}
\end{equation}
The applicable periodicity conditions and boundary conditions for the conjugate heat transfer problem are given by
\begin{equation}
\begin{aligned}
  T^{*} \left( \textbf{x} + \textbf{l}_{j}  \right) &= T^{*} \left( \textbf{x}  \right) & &\text{in } \Omega_{f} \cup \Omega_{s}, \\
  T^{*}_{f} &= T^{*}_{s} & &\text{in } \Gamma_{fs}, \\
  - \textbf{n}_{fs} \cdot k_{f} \left( \mathrm{\nabla{T^{*}_{f}}} + \mathrm{\nabla{T}} \right) &= - \textbf{n}_{fs} \cdot k_{s} \left( \mathrm{\nabla{T^{*}_{s}}} + \mathrm{\nabla{T}} \right) & &\text{in } \Gamma_{fs}, \\
  - \textbf{n} \cdot k_{f} \left( \mathrm{\nabla{T^{*}_{f}}} + \mathrm{\nabla{T}} \right) &= q_{b} & &\text{in } \Gamma_{bf}, \\
  - \textbf{n} \cdot k_{s} \left( \mathrm{\nabla{T^{*}_{s}}} + \mathrm{\nabla{T}} \right) &= q_{b} & &\text{in } \Gamma_{bs}, \\
  \langle T^{*} \rangle &= \text{constant}, &  \\
\end{aligned}
\label{eq:BoundaryConditionsHeat}
\end{equation}
with $j=\{ 1,2 \}$. 
In the former equations, the thermal conductivity of the fluid and solid, $k_f$ and $k_s$, as well as the densities $\rho_f$ and $\rho_s$, are assumed to be constant, just like the dynamic viscosity $\mu_f$ and specific heat capacity $c_f$ of the fluid.
The temperature field $T^{*}$ is spatially periodic over the unit cell along the lattice vectors $\textbf{l}_{1}$ and $\textbf{l}_{2}$, similarly to the periodically developed velocity field $\boldsymbol{u}$, which follows from the periodic flow equations given in (Refs.~\onlinecite{vangeffelen2021friction}). 
At the fluid-solid interface $\Gamma_{fs}$, the continuity of the temperature field and heat flux is required. 
In addition, at the bottom surface of the unit cell, which contains parts that belong to the fluid and solid ($\Gamma_{b} = \Gamma_{bf} \cup \Gamma_{bs}$), a uniform heat flux $q_{b}$ is imposed in the form of a Neumann boundary condition. 
We remark that the unit normal vector $\textbf{n}_{fs}$ at $\Gamma_{fs}$ points from the fluid domain $\Omega_{f}$ towards the solid domain $\Omega_{s}$, while the unit normal vector $\textbf{n}$ at $\Gamma$ points outward of the unit cell $\Omega_{\text{unit}}$, such that the heat flux $q_{b}$ is negative when directed towards $\Omega_{\text{unit}}$. 
Finally, in order to have a unique solution for the periodic temperature field, its volume-averaged value over the unit cell domain $\langle T^{*} \rangle$ is imposed. 
This value depends on the development of the temperature field in the channel but does not affect the heat transfer coefficient. 

For a specified volume-averaged velocity vector $\langle \boldsymbol{u} \rangle$ over the unit cell, the temperature field $T^{*}$ can be computed from equations (\ref{eq:Temperature})-(\ref{eq:BoundaryConditionsHeat}), since the constant temperature gradient $\mathrm{\nabla{T}}$ is determined by the imposed heat flux $q_{b}$ (Refs.~\onlinecite{penha2012fully,buckinx2016macro}): 
\begin{equation}
    \mathrm{\nabla{T}} = \frac{ \langle q_{b} \delta_b \rangle }{ \rho_{f} c_{f} \|\langle \boldsymbol{u} \rangle\| } \textbf{e}_{s} \,. 
\label{eq:temperaturegradient}
\end{equation}
Here, the Dirac surface indicator $\delta_b$ associated with the bottom interface $\Gamma_{b}$ is defined such that $\langle \delta_b \rangle=1/l_3$ represents the integral area of bottom surface per unit cell volume. 
In the right hand side of (\ref{eq:temperaturegradient}), the unit vector $\textbf{e}_{s} \triangleq \langle \textbf{u} \rangle / \|\langle \textbf{u} \rangle\|$ indicates the direction of the volume-averaged velocity, and  $\| \, \|$ denotes the Euclidean vector norm. 
Lastly, the volume-averaged value of any physical quantity $\phi$ is defined as 
\begin{align}
  \langle \phi \rangle  &\triangleq \frac{1}{V_{\text{unit}}} \int_{\textbf{r} \in \Omega_{\text{unit}} \left( \textbf{x} \right)} \phi \left( \textbf{r} \right) \,d\Omega \left( \textbf{r} \right), \\
  V_{\text{unit}} & = \textbf{l}_{1} \cdot \left( \textbf{l}_{2} \times \textbf{l}_{3} \right). 
\end{align}

The temperature field $T^{*}$ in the unit cell determines the following heat transfer coefficient between the fluid and solid:
\begin{equation}
    h_{\text{unit}} \triangleq \epsilon_{f}^{-1} \frac{ \langle q_{b} \delta_b \rangle }{ \langle T^{*} \rangle^{f} - \langle T^{*} \rangle^{s} } \,. 
\label{eq:heattransfercoefficient}
\end{equation}
This heat transfer coefficient characterizes the difference between the intrinsic volume-averaged temperatures of the fluid and solid in the unit cell, which are defined by $\langle T^{*} \rangle^{f} \triangleq \epsilon_{f}^{-1} \langle T^{*} \gamma_{f} \rangle$ and $\langle T^{*} \rangle^{s} \triangleq \epsilon_{s}^{-1} \langle T^{*} \gamma_{s} \rangle$, with $\epsilon_{f} \triangleq \langle \gamma_{f} \rangle = \epsilon$ and $\epsilon_{s} \triangleq \langle \gamma_{s} \rangle = 1 - \epsilon$. 
Here, we have used  $\gamma_{f}$ and $\gamma_{s}$ to denote the fluid indicator and solid indicator respectively: $\gamma_{f}(\boldsymbol{x}) = 1 \leftrightarrow \boldsymbol{x} \in \Omega_{f}$, $\gamma_{f}(\boldsymbol{x}) = 0 \leftrightarrow \boldsymbol{x} \notin \Omega_{f}$ and $\gamma_{s}(\boldsymbol{x}) = 0 \leftrightarrow \boldsymbol{x} \in \Omega_{f}$, $\gamma_{s}(\boldsymbol{x}) = 1 \leftrightarrow \boldsymbol{x} \notin \Omega_{f}$. 
The heat transfer coefficient $h_{\text{unit}}$ in equation (\ref{eq:heattransfercoefficient}) has been introduced in the macro-scale descriptions of periodically developed heat transfer by Buckinx and Baelmans (Refs.~\onlinecite{buckinx2015macro,buckinx2016macro}). 
In their macro-scale descriptions, the macro-scale variables are obtained through a double volume-averaging operation $\langle \, \rangle_{m}$ so that the interfacial heat transfer coefficient $h_{fs} \triangleq \epsilon_{fm}^{-1} \langle q_{fs} \delta_{fs} \rangle_m / \left( \langle T \rangle_{m}^{f} - \langle T \rangle_{m}^{s} \right)$ becomes spatially constant in the periodically developed heat transfer regime, and identical to $h_{\text{unit}}$. 
Therefore, $h_{\text{unit}}$ is the exact and physically meaningful heat transfer coefficient that relates the constant macro-scale heat transfer rate $\langle q_{fs} \delta_{fs} \rangle_m$ between the fluid and solid to the constant macro-scale temperature difference $\langle T \rangle_{m}^{f} - \langle T \rangle_{m}^{s} = \langle T^{*} \rangle^{f} - \langle T^{*} \rangle^{s}$ (Refs.~\onlinecite{buckinx2016macro}).

The relationship between the heat transfer coefficient $h_{\text{unit}}$ and the volume-averaged velocity $\langle \boldsymbol{u} \rangle$ will further be expressed as a non-dimensional relationship between the Nusselt number, \begin{equation}
    Nu_{\text{unit}} \triangleq \frac{ h_{\text{unit}} l^{2} }{k_f} \,, 
\label{eq:nusseltnumber}
\end{equation}
and the Reynolds number, 
\begin{equation}
    Re_l \triangleq \frac{\rho_{f} \|\langle \boldsymbol{u} \rangle\| l}{\mu_{f}}. 
\label{eq:reynoldsnumber}
\end{equation}
To be consistent with our previous work (Refs.~\onlinecite{vangeffelen2021friction}), the fin length $l$ is thus chosen as the reference length for the Nusselt number and the Reynolds number. 
Contrary to the hydraulic diameter $D_{h}$, the fin length $l$ simplifies the interpretation of the Nusselt number and Reynolds number, as $l$ is a single geometrical parameter instead of a combination of all the geometrical parameters.
Besides, the actual definition of the hydraulic diameter implies that the latter varies in the streamwise direction along the fin, since the wetted area is different at each cross section of the channel. 

When the periodically developed heat transfer regime extends over the largest part of the channel with offset strip fins, the Nusselt number $Nu_{\text{unit}}$ will give an accurate indication of the overall temperature difference over the channel, for a constant heat flux at the channel wall. 
Furthermore, the volume-averaged velocity and thus the Reynolds number $Re_l$ will correspond directly to the bulk velocity through the channel, as long as the flow retardation near the side walls of the channels does not significantly impact the total mass flow rate through the channel.

\subsection{Numerical procedure}

The periodic temperature equations (\ref{eq:Temperature})-(\ref{eq:BoundaryConditionsHeat}) have been solved for periodic velocity fields at different Reynolds numbers, and for different material properties, using a finite-element discretization, in which the discretized temperature field was represented by continuous Galerkin tetrahedral elements of the second order. 
The same structured mesh as in (Refs.~\onlinecite{vangeffelen2021friction}) was employed for the spatial discretization of the unit cell.
For the finite-element formulation and the numerically parallelized solution of the discretized temperature equations, we use the software package FEniCSLab.
This package was developed by G. Buckinx in the finite-elements-based computing platform FEniCS (Refs.~\onlinecite{AlnaesBlechta2015a}). 

To validate the employed numerical discretization, a mesh-independence study was performed at the two largest Reynolds numbers, for the four lowest unit cell porosities. 
Through the Richardson extrapolation and grid convergence index (Refs.~\onlinecite{richardson1911ix,roache1994perspective}), the relative discretization error on the computed Nusselt number was estimated to remain below 1\%. 

\begin{figure}
\includegraphics[scale = 1.0]{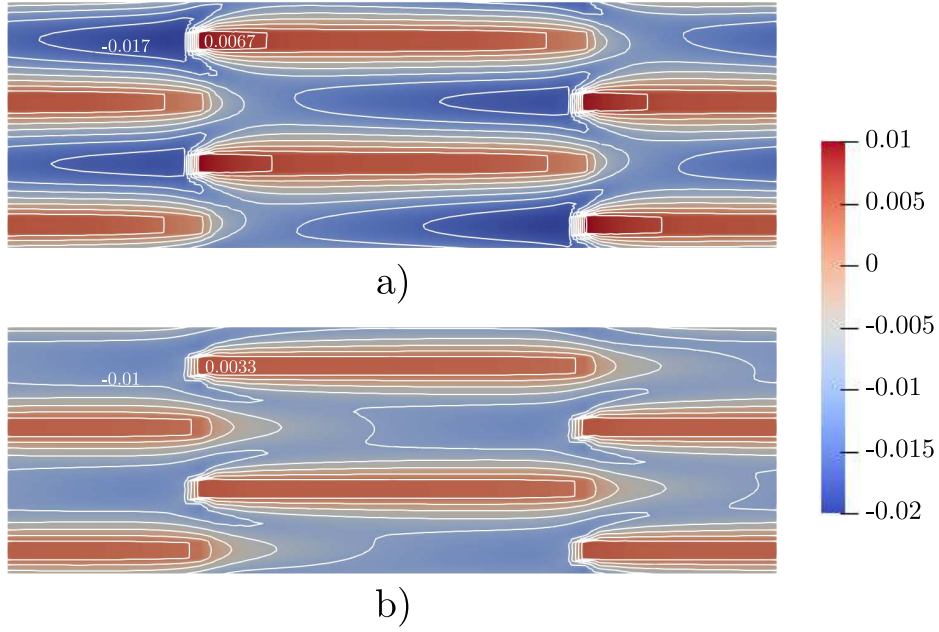}
\caption{\label{fig:isothermsPR07RE100_300T4H48S28} The non-dimensional periodic temperature field $ \left( T^{*} - \langle T^{*} \rangle \right) k_f / \left( q_{b} l \right) $, visualized by its iso-lines in the unit cell's mid-plane spanned by $\textbf{l}_{1}$ and $\textbf{l}_{2}$ for (a) $Re_{l}=100$ (b) and $Re_{l}=300$ when $Pr_f=0.7$, $k_{s}/k_{f}=10^4$, $t/l=0.04$, $h/l=0.48$, $s/l=0.28$}
\end{figure}

In Figure \ref{fig:isothermsPR07RE100_300T4H48S28}, the non-dimensional periodic temperature field $ \left( T^{*} - \langle T^{*} \rangle \right) k_f / \left( q_{b} l \right) $ in the mid-plane of the unit cell, spanned by $\textbf{l}_{1}$ and $\textbf{l}_{2}$, is visualized through its iso-lines at a Reynolds number of 100 and 300 for a fluid Prandtl number of 0.7. 
From the depicted iso-line values, it can be seen that the temperature difference between the fluid and solid domain decreases when the Reynolds number $Re_{l}$ is higher, which indicates an increased heat transfer coefficient. 

Finally, we remark that due to the linearity of the periodic temperature equations (\ref{eq:Temperature})-(\ref{eq:BoundaryConditionsHeat}) and the top-down symmetry of the fin geometry, the obtained heat transfer coefficient $h_{\text{unit}}$ for an imposed heat flux $q_{b}$ at the bottom surface of the unit cell is identical to $h_{\text{unit}}$ when a heat flux $q_{b}$ is imposed at both the top and bottom surface. 

\newpage
\clearpage

\section{\label{sec:aligned} Nusselt number for periodically developed heat transfer}

Considering the case where the volume-averaged velocity is aligned with the lattice vector $\textbf{e}_{1}$, we have determined the Nusselt number $Nu_{\text{unit}}$ for a large set of Reynolds numbers and geometrical parameters, relevant to micro- and mini-channel applications (Refs.~\onlinecite{bapat2006thermohydraulic,yang2007advanced, hong2009three,do2016experimental,nagasaki2003conceptual,yang2017heat,jiang2019thermal,yang2014design,pottler1999optimized}). 
The Nusselt number has been determined for two fluids, air and water, with a Prandtl number of $Pr_f=0.7$ and $Pr_f=7$ respectively (Ref.~\onlinecite{shah2003fundamentals}).
The thermal conductivity ratio has been chosen accordingly as $k_{s}/k_{f}=500$ and $k_{s}/k_{f}=10^4$, since these values are representative for copper and air, and copper and water. 
All the geometrical parameters and material properties selected for our study are listed in Table \ref{tab:UCgeom}. 
The periodically developed heat transfer equations were solved for 197 different offset strip fin geometries. 
In total, 1168 data points for the Nusselt number were obtained for $Pr_f=0.7$ and $k_{s}/k_{f}=10^4$, while 1114 data points were collected for $Pr_f=7$ and $k_{s}/k_{f}=500$. 
The entire data set is tabulated in Appendix \ref{sec:alignedflowdata}. 
In the remainder of this section, the influence of the Reynolds number and the geometrical parameters on the unit cell's Nusselt number will be discussed in detail. \\

\begin{table}[h]
\caption{\label{tab:UCgeom} Values for the Reynolds number, the geometrical parameters, the Prandtl number, and thermal conductivity ratio, considered in the numerical study.}
\begin{tabular}{l|l}
\hline\hline
$Re_{l}$ \quad & \quad 1, 10, 15, 25, 35, 50, 75, 100, 150, 200, 300, 400, 600\\
$h/l$    \quad & \quad 0.12, 0.16, 0.20, 0.24, 0.28, 0.32, 0.40, 0.48, 0.56, 0.68, 1.00\\
$s/l$    \quad & \quad 0.12, 0.16, 0.20, 0.24, 0.28, 0.32, 0.40, 0.48\\
$t/l$    \quad & \quad 0.01, 0.02, 0.04, 0.06\\
\hline\hline
\end{tabular}
\quad\quad
\begin{tabular}{l|cc}
\hline\hline
 & Air & Water \\
 \hline
 $Pr_f$ & 0.7 & 7 \\
 $k_{s}/k_{f}$ & $10^4$ & 500 \\
\hline\hline
\end{tabular}
\end{table}

\subsection{\label{sec:influence re}The influence of the Reynolds number $Re_l$ on the Nusselt number}

\begin{figure}[ht]
\centering
\begin{minipage}{.475\textwidth}
\raggedleft
\includegraphics[scale = 1.00]{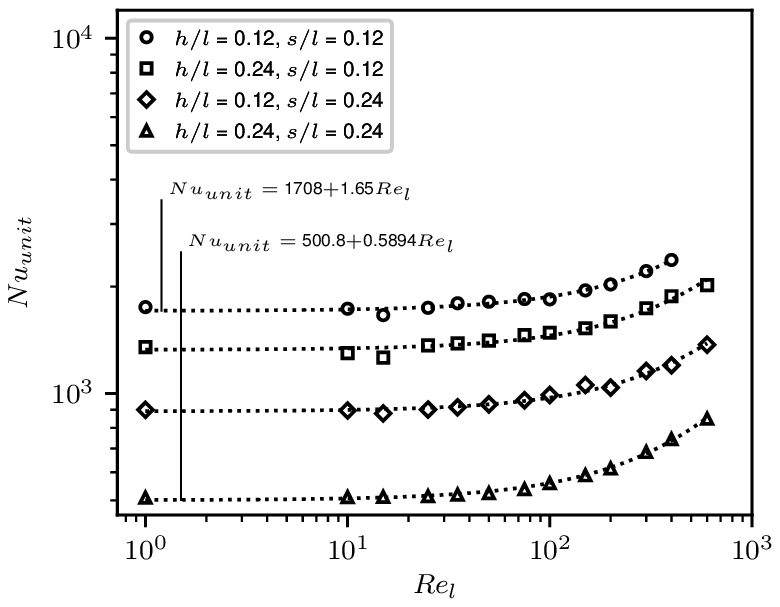}
\caption{\label{fig:NU_REdataP07T4} Influence of the Reynolds number on the Nusselt number for steady periodically developed heat transfer, when $Pr_f=0.7$, $k_{s}/k_{f}=10^4$, $t/l=0.04$}
\end{minipage}
\hfill
\begin{minipage}{.475\textwidth}
\raggedright
\includegraphics[scale = 1.00]{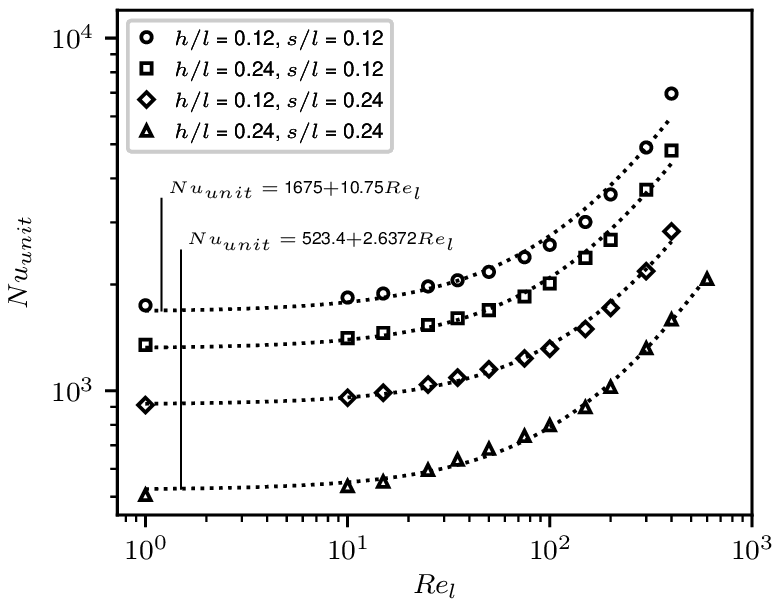}
\caption{\label{fig:NU_REdataP7T4} Influence of the Reynolds number on the Nusselt number for steady periodically developed heat transfer, when $Pr_f=7$, $k_{s}/k_{f}=500$, $t/l=0.04$}
\end{minipage}
\end{figure}

The dependence of the Nusselt number on the Reynolds number is illustrated in Figures \ref{fig:NU_REdataP07T4} and \ref{fig:NU_REdataP7T4} for various offset strip fin geometries. 
It can be observed that a linear relationship of the form
\begin{equation}
Nu_{\text{unit}} \simeq A + B Re_l,
\label{eq:nu_re}
\end{equation}
accurately captures the data from our work. 
In this linear relationship, the parameters $A$ and $B$ are both functions of the geometrical parameters, the Prandtl number and the thermal conductivity ratio.
For all the parameters in Table \ref{tab:UCgeom}, the linear correlation (\ref{eq:nu_re}) captures the dependence of the Nusselt number on the Reynolds number with an average and maximum relative error of 2\% and 15\%, respectively. 
A log-linear regression analysis confirmed that when the exponent of the Reynolds number equals 1, the relative error with respect to the data from this work is minimized, and the standard deviation on this exponent is 0.3. 
At the same time, the constant $A$ exhibits no significant dependence on $Pr_{f}$ and $k_{s}/k_{f}$.
This can also be observed from the fitted functions in Figures \ref{fig:NU_REdataP07T4} and \ref{fig:NU_REdataP7T4}. 
The influence of the Prandtl number and the thermal conductivity ratio on $Nu_{\text{unit}}$ will be discussed in more detail in Section \ref{sec:influence prkskf}.

Figures \ref{fig:NU_REdataP07T4} and \ref{fig:NU_REdataP7T4} clearly show that for lower Reynolds numbers, i.e., when $Re_l < 10$, the Nusselt number becomes virtually independent of $Re_l$, as it is primarily determined by the constant $A$. 
At higher Reynolds numbers, when $Re_l > 100$, the influence of the Reynolds number $Re_l$ on the Nusselt number becomes more significant through the linear term $B Re_l$. 

The observed linear relationship between $Nu_{\text{unit}}$ and $Re_l$ is supported by the findings in various experimental and numerical performance studies for micro- and mini-channels applications in the literature (Refs.~\onlinecite{bapat2006thermohydraulic,nagasaki2003conceptual,yang2017heat,jiang2019thermal,pottler1999optimized}). 
In addition, it closely resembles the form of the empirical correlations for the Nusselt number characterizing the convective heat transfer coefficient in porous media and arrays of circular and square cylinders: $Nu = A + B Re^{b}$ (Refs.~\onlinecite{wakao1978effect,hwang1994heat,martin1998frictional,kuwahara2001numerical,mandhani2002forced,saito2006correlation,gamrat2008numerical,alshare2010modeling,lu2019effect}). 
Nevertheless, in these studies, the reported exponent $b$ lies between 0.5 and 0.9. 
Moreover, many studies have suggested that $Nu \sim Re^{b}$ with $b < 1$, since they focused on the transitional regime in porous media and cylinder arrays.
Indeed, the relation between the Nusselt number $Nu$ and the Reynolds number $Re$, and thus the specific value of the exponent $b$, depends on the nature of the flow regime through the porous medium, as clarified in the study of Lu and Zhao (Refs.~\onlinecite{lu2019effect}). 
Therefore, the influence of the flow regime on the precise relationship between $Nu_{\text{unit}}$ and $Re_l$ in arrays of offset strip fins is examined in more detail in Section \ref{sec:critical}.
Here, we first show that the linear relationship (\ref{eq:nu_re}) found for  $b=1$ is more accurate with respect to our data than the available correlations from the literature with $b<1$.

\begin{figure}[ht]
\centering
\begin{minipage}{.475\textwidth}
\raggedleft
\includegraphics[scale = 1.00]{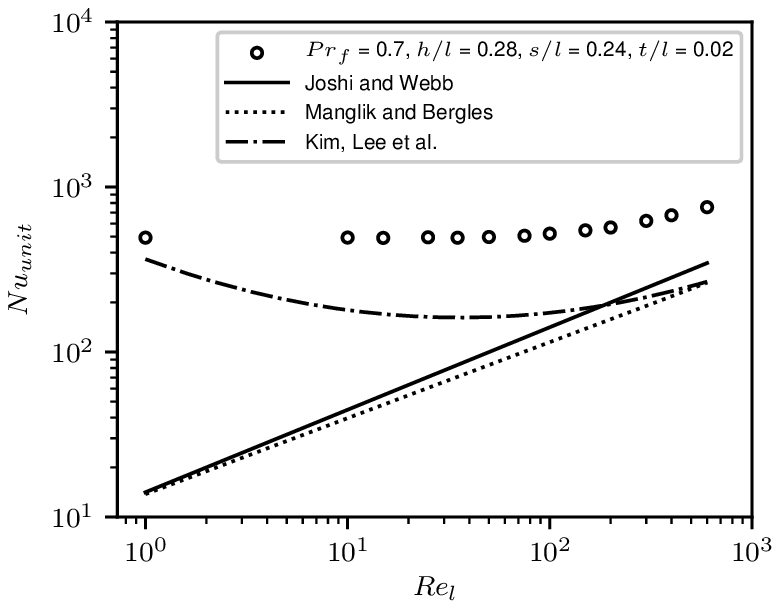}
\caption{\label{fig:litcomp_NU_PR07T2H28S24_nofit} A comparison between the Nusselt number correlations from the literature with the data from this work for steady periodically developed heat transfer at low Reynolds numbers, when $Pr_f=0.7$, $t/l=0.02$, $h/l=0.28$, $s/l=0.24$}
\end{minipage}
\hfill
\begin{minipage}{.475\textwidth}
\raggedright
\includegraphics[scale = 1.00]{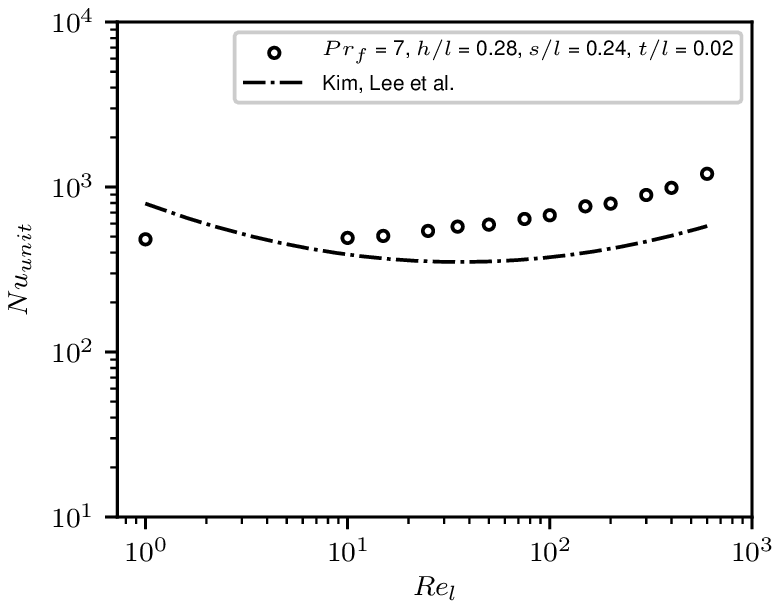}
\caption{\label{fig:litcomp_NU_PR7T2H28S24_nofit} A comparison between the Nusselt number correlations from the literature with the data from this work for steady periodically developed heat transfer at low Reynolds numbers, when $Pr_f=7$, $t/l=0.02$, $h/l=0.28$, $s/l=0.24$}
\end{minipage}
\end{figure}

In figures \ref{fig:litcomp_NU_PR07T2H28S24_nofit} and \ref{fig:litcomp_NU_PR7T2H28S24_nofit}, the various Nusselt number correlations from the literature, listed in Tables \ref{tab:literature} and \ref{tab:literature2}, are compared with  the data from this work, for a single representative geometry. 
For the comparison, the different definitions of $Nu$ and $Re$ in the correlations from the literature, given in Section \ref{sec:intro}, are converted to our definitions of $Nu_{\text{unit}}$ and $Re_{l}$. 
Note that only an approximate conversion of the heat transfer coefficient is possible, due to the approximations made in the $\epsilon-NTU$ and LMTD methods. 
This underlines again the importance of using an exact and physically meaningful definition of $h_{\text{unit}}$, as in equation (\ref{eq:heattransfercoefficient}). 

As it can be seen in Figure \ref{fig:litcomp_NU_PR07T2H28S24_nofit}, the correlations of Wieting (Refs.~\onlinecite{wieting1975empirical}), Joshi and Webb (Refs.~\onlinecite{joshi1987heat}), and Manglik and Bergles (Refs.~\onlinecite{manglik1995heat}), all deviate significantly from our data, both in their predicted values and scaling with $Re_l$. 
They all predict a trend $Nu \sim Re^{0.5}$ in the laminar flow regime, while the trend in our data is $Nu \sim Re$.
The reason is likely the fact that these correlations were fitted mainly to data points pertaining to the transitional and turbulent flow regime. 
As such, these correlations underestimate the Nusselt number data determined in this work with an average relative error of 70\% and a maximum relative error of 90\%, which occurs for $Re_{l} \simeq 1$. 

The correlation from Dong et al. (Refs.~\onlinecite{dong2007air}), which predicts a trend $Nu \sim Re^{0.8}$, results in an even more considerable underestimation of our Nusselt number data, although an illustrative comparison has been omitted here. 
Again, this is most likely due to the fact that the latter authors only considered Reynolds numbers above 500, as shown in Table \ref{tab:literature}.
Also the correlation of Kim et al. (Ref.~\onlinecite{kim2011correlations}),
which is illustrated in Figures \ref{fig:litcomp_NU_PR07T2H28S24_nofit} and \ref{fig:litcomp_NU_PR7T2H28S24_nofit}, does not accurately capture the correct dependence of the Nusselt number on the Reynolds number for $Re_{l} < 100$ when $Pr_f=0.7$ and $Pr_f=7$, despite the fact it was obtained from a data set which did include lower Reynolds numbers. 
For $Re_{l} > 100$, it underestimates our Nusselt number data with an error similar to that of the correlations of Joshi and Webb, and Manglik and Bergles. 
It thus appears that the highly non-linear trend $Nu \sim Re^{0.05 ln (Re) - 0.1}$ proposed by Kim et al. does not apply to the steady laminar regime. 

From the previous considerations, we conclude that the correlations from the literature cannot accurately capture the Reynolds number dependence of the Nusselt number for periodically developed flow and heat transfer in micro- and mini-channels with offset strip fins subject to an imposed heat flux. \\

\subsection{\label{sec:influence h/l}The influence of the fin height-to-length ratio $h/l$ on the Nusselt number}

\begin{figure}[ht]
\centering
\begin{minipage}{.475\textwidth}
\raggedleft
\includegraphics[scale = 1.00]{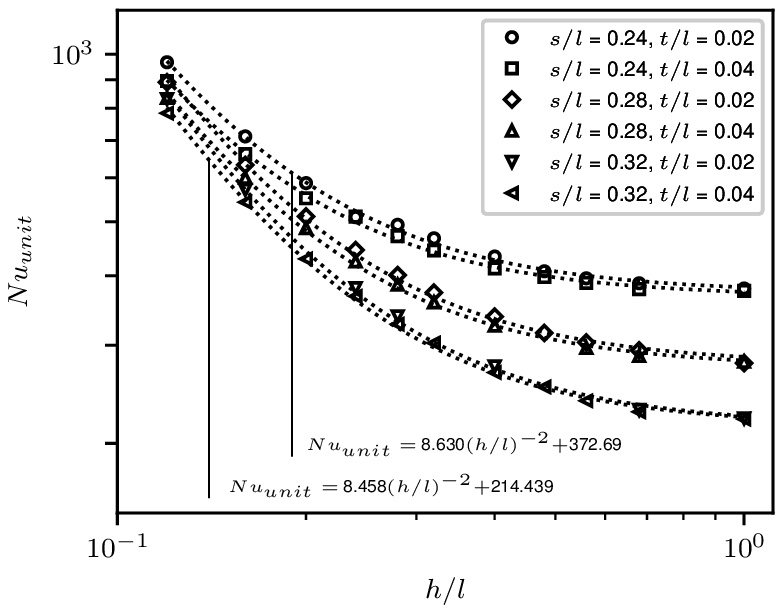}
\caption{\label{fig:NU_Hdata} Influence of the fin height-to-length ratio on the Nusselt number for steady periodically developed heat transfer, when $Re_{l}=10$, $Pr_{f}=0.7$, $k_{s}/k_{f}=10^4$ \newline}
\end{minipage}
\hfill
\begin{minipage}{.475\textwidth}
\raggedright
\includegraphics[scale = 1.00]{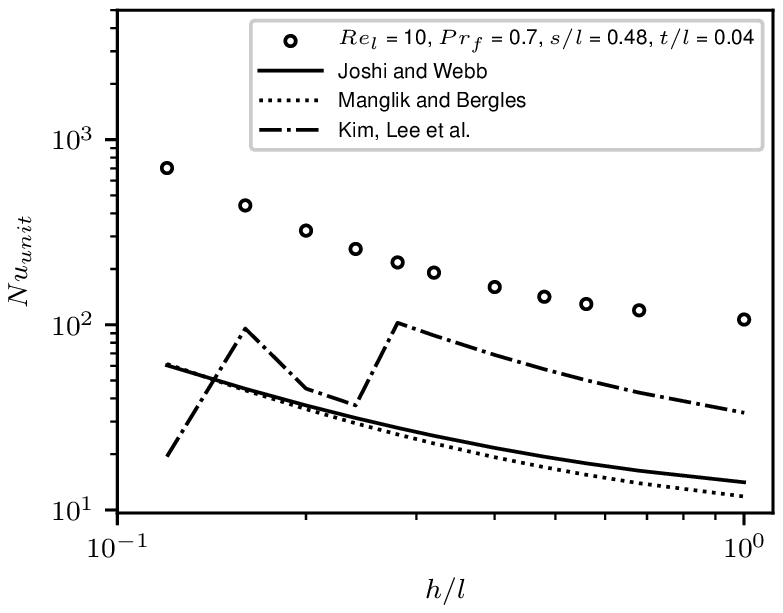}
\caption{\label{fig:litcomp_NU_PR07R10T4S48_nofit} A comparison between the Nusselt number correlations from the literature with the data from this work for steady periodically developed heat transfer at small fin heights, when $Re_{l}=10$, $Pr_{f}=0.7$, $s/l=0.48$, $t/l=0.04$}
\end{minipage}
\end{figure}

The influence of the fin height-to-length ratio $h/l$ on the Nusselt number in the steady periodically developed heat transfer regime is illustrated in Figure \ref{fig:NU_Hdata}. 
From our data set, it can be concluded that the Nusselt number $Nu_{\text{unit}}$ becomes proportional to $(h/l)^{-2}$ for low values of $h/l$, in analogy to what is observed for developed heat transfer between two parallel plates. 
Notice that, in that case, the commonly used Nusselt number based on the channel height $h$ is a constant (Refs.~\onlinecite{shah1978laminar}). 
This is consistent with our findings, as the Nusselt number in this work, $Nu_{\text{unit}}$, is based on the reference length $l$ instead of $h$, and the heat transfer coefficient $h_{\text{unit}}$ includes the area of heat transfer surface per unit cell volume, given by $\langle \delta_b \rangle \simeq 1/h$. 
For developed heat transfer between two parallel plates with an imposed heat flux, this results in the analytical relation $Nu_{\text{unit}} = 10 (h/l)^{-2}$. 
The asymptotic trend $Nu_{\text{unit}} \sim (h/l)^{-2}$ for $h/l \rightarrow 0$, thus can be explained by similarity with developed flow and heat transfer between two parallel plates. 
Indeed, as $h/l$ decreases, the top and bottom plates become the main heat transfer surfaces, and the flow field resembles more and more the developed flow field between two parallel plates (Refs.~\onlinecite{vangeffelen2021friction}). 

On the other hand, when the fin height-to-length ratio is sufficiently large, the Nusselt number becomes independent of $h/l$, as Figure \ref{fig:NU_Hdata} shows. 
This asymptotic behaviour results from the fact that when $h/l$ increases, the fins, but no longer the plates, become the main heat exchanging surface. 
For $h/l \rightarrow \infty$, the flow becomes more two-dimensional (Refs.~\onlinecite{vangeffelen2021friction}). 
As a consequence, the temperature gradients at the fin sides will vary less along the direction $\textbf{e}_{3}$ and therefore become independent of $h/l$. 

The occurrence of the two asymptotic trends $Nu_{\text{unit}} \sim (h/l)^{-2} $ and $Nu_{\text{unit}} \sim (h/l)^{0}$ is greatly influenced by the fin pitch-to-length ratio $s/l$, and thus the aspect ratio $s/h$, as illustrated in Figure \ref{fig:NU_Hdata}. 
More specifically, as $s/h$ decreases, the contribution of the fin sides to the total transferred heat increases, so that the trend $Nu_{\text{unit}} \sim (h/l)^{0}$ starts to prevail at relatively lower $h/l$-ratios.

As a consequence of the former asymptotic trends, the influence of the fin height-to-length ratio on the Nusselt number is accurately described by 
\begin{equation}
\label{eq: correlation form h/l}
Nu_{\text{unit}} \simeq E \left(\frac{h}{l}\right)^{-2} + F. 
\end{equation}
Here, the parameters $E$ and $F$ depend on $Re_{l}$, $Pr_{f}$ and $k_{s}/k_{f}$, as well as the geometrical parameters $s/l$ and $t/l$. 
When $E$ and $F$ are determined by a least-squares fitting, the correlation (\ref{eq: correlation form h/l}) can predict all the Nusselt number data from this work with a mean relative error of 0.5\%, and a maximum relative error of 8\%.

In the literature, the influence of the fin height-to-length ratio on the Nusselt number has been taken into account by a factor of the form $(h/l)^{c}$, hence by means of a single exponent $c$. 
However, the existing correlations are also implicitly affected by the ratio $h/l$ through the Reynolds number, via the specific definition of the hydraulic diameter. 
For the studies in Tables \ref{tab:literature} and \ref{tab:literature2}, this leads to an effective exponent $C$, such that $Nu_{\text{unit}} \sim (h/l)^{C}$, which lies between -1.3 to -1.8 for small fin heights. 
As a result, the correlations from the literature, which mainly focus on conventional offset strip fin channels with larger fin height-to-length ratios, underestimate the Nusselt number for micro- and mini-channels. 
This statement is supported by Figure \ref{fig:litcomp_NU_PR07R10T4S48_nofit}. 
Even when the least-square differences with our data is minimized by rescaling the correlations from the literature with a constant to account for any incorrect scaling with the other parameters, our data is still underestimated with a mean relative error of 40\%, and a maximum relative error of 85\% near $h/l\simeq 0.1$. 
Furthermore, Figure \ref{fig:litcomp_NU_PR07R10T4S48_nofit} reveals that the correlation presented in the numerical study of Kim et al. (Refs.~\onlinecite{kim2011correlations}) does not even predict a continuous trend of the Nusselt number with respect to the height-to-length ratio, since the latter is based on a piecewise function of the porosity (see Table \ref{tab:literature2}). 

Finally, all the available correlations from the literature for offset strip fins do not recover a constant Nusselt number $Nu_{\text{unit}}$ for $h/l \rightarrow \infty$ as it can be seen in Figure \ref{fig:litcomp_NU_PR07R10T4S48_nofit}, but instead predict a scaling close to $Nu_{\text{unit}} \sim (h/l)^{0.2}$ for $h/l \rightarrow \infty$.

\subsection{\label{sec:influence s/l}The influence of the fin pitch-to-length ratio $s/l$ on the Nusselt number}

\begin{figure}[ht]
\centering
\begin{minipage}{.475\textwidth}
\raggedleft
\includegraphics[scale = 1.00]{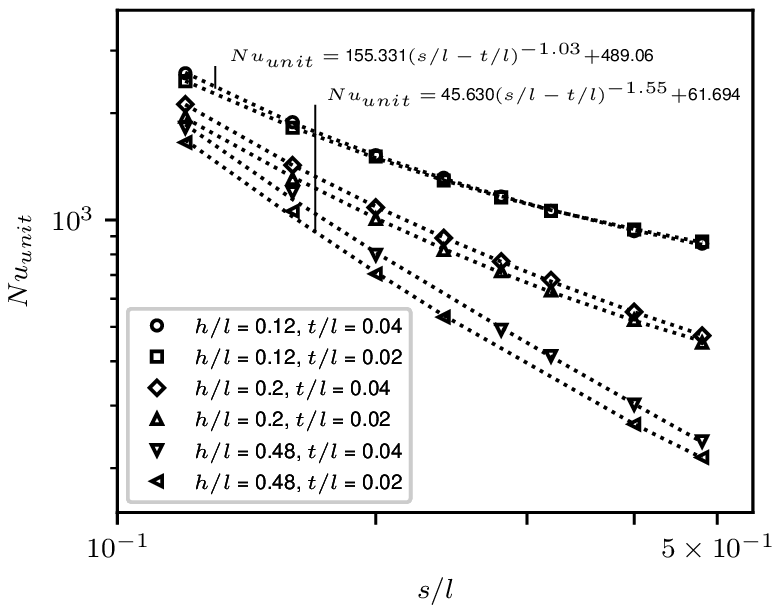}
\caption{\label{fig:NU_Sdata} Influence of the fin pitch-to-length ratio on the Nusselt number for steady periodically developed heat transfer, when $Re_{l}=100$, $Pr_{f}=7$, $k_{s}/k_{f}=500$ \newline}
\end{minipage}
\hfill
\begin{minipage}{.475\textwidth}
\raggedright
\includegraphics[scale = 1.00]{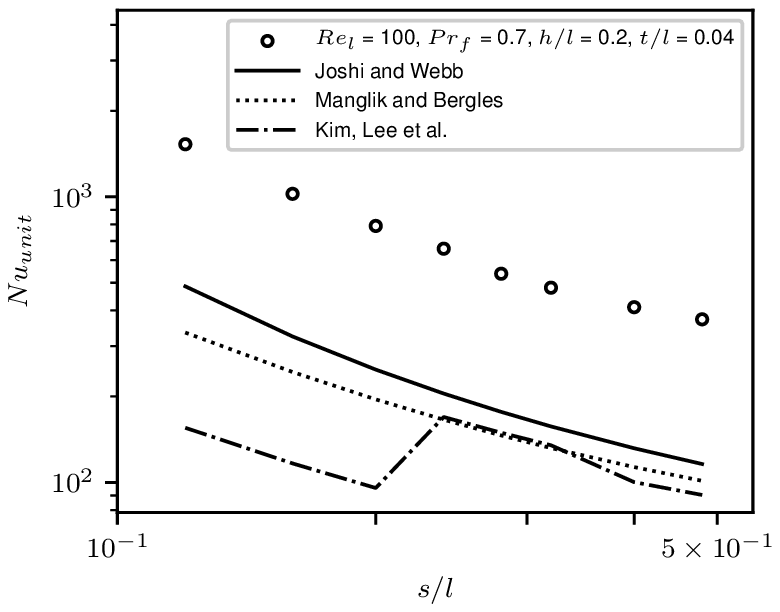}
\caption{\label{fig:litcomp_NU_PR07R100T4H2_nofit} A comparison between the Nusselt number correlations from the literature with the data from this work for steady periodically developed heat transfer at small fin heights, when $Re_{l}=100$, $Pr_{f}=0.7$, $t/l=0.04$, $h/l=0.20$}
\end{minipage}
\end{figure}

Figure \ref{fig:NU_Sdata} illustrates the variation of the Nusselt number with the fin pitch-to-length ratio $s/l$. 
According to our data, the Nusselt number becomes infinite as $s/l$ approaches a value close to that of the relative fin thickness $t/l$. 
In the limit $s=t$, the channel becomes fully blocked. 
This implies that the fluid flows ever more slowly past the fins as $s \rightarrow t$.
Consequently, the average temperature difference between the fluid and solid domain progressively decreases as $s \rightarrow t$, giving rise to an ever larger value of the Nusselt number. 
Therefore, we propose to capture the variation of the Nusselt number with the fin pitch-to-length ratio by means of a correlation of the form 
\begin{equation}
\label{eq: correlation form s/l}
Nu_{\text{unit}} \simeq G\left(\frac{s}{l} - \frac{t}{l}\right)^{d} + H \,. 
\end{equation} 
Here, the parameters $G$, $H$ and $d$ are a function of $Re_{l}$, $Pr_{f}$ and $k_{s}/k_{f}$, and the remaining geometrical parameters $h/l$ and $t/l$. 

As it is shown in Figure \ref{fig:NU_Sdata}, the correlation form (\ref{eq: correlation form s/l}) results in a quite accurate fit of our Nusselt number data. 
However, the negative exponent $d$ cannot be assumed to be constant over the entire data set. 
Nevertheless, the exponent $d$ can be approximated by a constant over distinct ranges of the Reynolds number and the geometrical parameters. 
For example, for $Re_l \in (150, 600) $ and $h/l \in (0.3, 1) $, the correlation (\ref{eq: correlation form s/l}) results in a relative error below 5\% with respect to our data, if $d=-1.29$, for all Prandtl numbers, thermal conductivity ratios and fin thickness-to-length ratios studied in this work. 

Besides the asymptotic behaviour $Nu_{\text{unit}} \rightarrow \infty$ for $s/l \rightarrow t/l$, the relationship (\ref{eq: correlation form s/l}) is in line with our observation from Figure \ref{fig:NU_Sdata} that the Nusselt number becomes independent of the relative fin pitch for high values of $s/l$.
This asymptotic trend for $s/l \rightarrow \infty$ corresponds to the case where less heat is transferred by the fin sides in comparison to the top and bottom plate. 
Therefore, the value of $s/l$ after which the Nusselt number becomes independent of $s/l$ depends significantly on how large the fin sides are compared to the plate surface. 
Such a characteristic is well represented by the fin aspect ratio $s/h$. 
More specifically, for lower values of $s/h$ or smaller fin heights, the fin sides will inevitably contribute less to the total heat transfer surface area than the bottom and top plate. 
Thus, the Nusselt number will become independent of the relative fin pitch $s/l$ at lower $s/l$-values. 

The correlations from the literature typically express the influence of the fin pitch-to-length ratio on the Nusselt number as $Nu \sim (s/l)^{k}$ with $k<0$. 
Of all correlations listed in Tables \ref{tab:literature} and \ref{tab:literature2}, practically only the one proposed by Joshi and Webb (Refs.~\onlinecite{joshi1987heat}) complies with the limit of $Nu_{\text{unit}} \rightarrow \infty$ for $s/l \rightarrow t/l$. 
Nevertheless, this limit is incorporated implicitly through the definition of the hydraulic diameter $D_{h} \sim (s-t)$ and reference velocity $U_{\text{ref}} \sim (s-t)^{-1}$, so that the correlation becomes equivalent to the scaling law $Nu \sim (s/l-t/l)^{-0.85}$ when $s/l \rightarrow t/l$. 
Such a dependence leads to an overestimation of our Nusselt number data up to 50\% for $s/l < 0.25$. 
Strictly speaking, also  the correlation of Bhowmik (Refs.~\onlinecite{bhowmik2009analysis}) respects the limit of $Nu_{\text{unit}} \rightarrow \infty$ for $s/l \rightarrow t/l$, since the author used a definition of the hydraulic diameter identical to that of Joshi and Webb. 
However, the correlation of Bhowmik is only applicable to just a single offset strip fin geometry. 
The remaining correlations in Tables \ref{tab:literature} and \ref{tab:literature2} typically understimate our Nusselt number data by as much as $40\%$ for $s/l < 0.25$, as we have illustrated in Figure \ref{fig:litcomp_NU_PR07R100T4H2_nofit}. 
The correlation of Kim et al. (Refs.~\onlinecite{kim2011correlations}) in particular, shows the largest and most inconsistent deviations with our data, due to its discontinuous dependence on the fin porosity and thus the fin pitch-to-length ratio. 
We remark that, in the preceding discussion, all the reported errors have been computed after rescaling the correlations with a constant, to minimize the least-square differences with our data and compensate for any incorrect scaling with the other parameters. 
Lastly, we notice that the correlations from the literature fail to predict the correct limit of a constant Nusselt number  for $s/l \rightarrow \infty$.
On the contrary, they 
still show a dependence on $s/l$ of the form $(s/l)^{k}$ with $k \in (-0.2, 0.1)$ for larger values of the fin pitch-to-length ratio, \\

\subsection{\label{sec:influence t/l}The influence of the fin thickness-to-length ratio $t/l$ on the Nusselt number}

\begin{figure}[ht]
\centering
\begin{minipage}{.475\textwidth}
\raggedleft
\includegraphics[scale = 1.00]{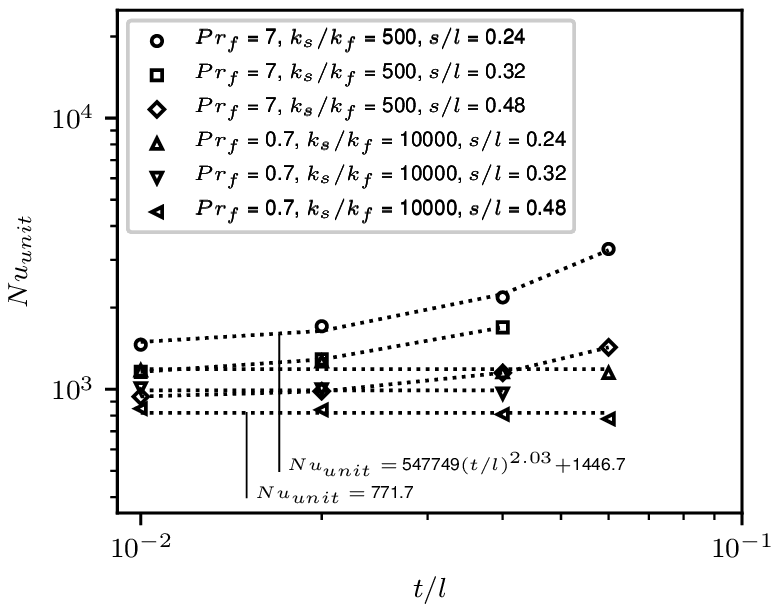}
\caption{\label{fig:NU_Tdata} Influence of the fin thickness-to-length ratio on the Nusselt number for steady periodically developed heat transfer, when $Re_{l}=300$, $h/l=0.12$ \newline}
\end{minipage}
\hfill
\begin{minipage}{.475\textwidth}
\raggedright
\includegraphics[scale = 1.00]{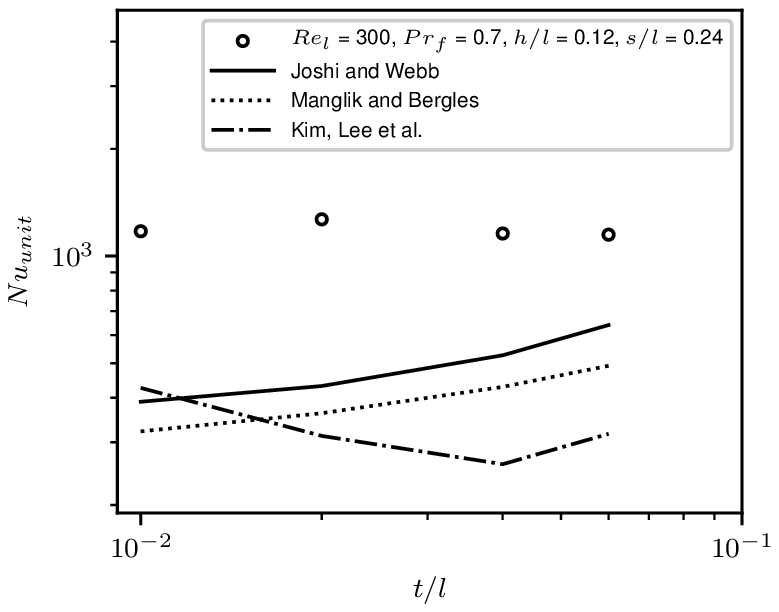}
\caption{\label{fig:litcomp_NU_PR07R300H12S24_nofit} A comparison between the Nusselt number correlations from the literature with the data from this work for steady periodically developed heat transfer at small fin heights, when $Re_{l}=300$, $Pr_{f}=0.7$, $h/l=0.12$, $s/l=0.24$}
\end{minipage}
\end{figure}

The third and final geometrical parameter whose influence on the Nusselt number is assessed is the fin thickness-to-length ratio $t/l$. 
In Figure \ref{fig:NU_Tdata} we observe that, as the fin thickness decreases, the Nusselt number becomes independent of the fin thickness-to-length ratio $t/l$. 
This can be explained by the fact that the heat transfer at the fin sides normal to the main flow direction $\boldsymbol{e}_1$ and parallel to $\boldsymbol{e}_2$, becomes negligible once the fin thickness $t$, and thus the width of these sides, is sufficiently small. 
This implies that the relative fin thickness $t/l$ will no longer affect the heat transfer coefficient.
We remark that the heat transfer at the fin sides mainly occurs at the leading edge of the fin, as little heat is transferred near the wake at the trailing edge of the fin.
This is also visible from the closely spaced iso-lines at the leading edge in Figure \ref{fig:isothermsPR07RE100_300T4H48S28}, which indicate the presence of large temperature gradients. 
Additionally, we observe again in Figure \ref{fig:NU_Tdata} that the Nusselt number becomes infinite when $t/l$ approaches the fin pitch-to-length ratio $s/l$, as discussed in the preceding subsection. 

Based on our data, the dependence of the Nusselt number on the fin height-to-length ratio is accurately captured by a correlation of the form 
\begin{equation}
\label{eq: correlation form t/l}
Nu_{\text{unit}} \simeq J\left(\frac{t}{l} \right)^m + K\,, 
\end{equation}
where the parameters $J$, $K$ and $m$ depend on $Re_l$, $Pr_{f}$, $k_{s}/k_{f}$, as well as the other geometrical parameters $s/l$ and $t/l$. 
It appears from Figure \ref{fig:NU_Tdata} that the exponent $m$ cannot be treated as a constant over all our Nusselt number data. 
We found that $m=0$ for $Pr_{f}=0.7$ and $k_{s}/k_{f}=10^4$, as $Nu_{\text{unit}}$ appears to be virtually independent of the fin thickness-to-length ratio for all $t/l$-values considered in this work.
On the other hand, for $Pr_{f}=7$ and $k_{s}/k_{f}=500$, the best fitting is obtained for $m > 0$. At the same time, $m$ can be taken constant over wide intervals of the fin pitch-to-length ratio. 
For instance, the correlation form (\ref{eq: correlation form t/l}) holds within a relative error of 5\% in the case that $Pr_{f}=7$, $k_{s}/k_{f}=500$, if $m=0.50$, as long as $Re_l \in (1,600)$, $h/l \in (0.1,1)$ and $s/l \in (0.3,0.5)$. 

The correlations of Joshi and Webb (Refs.~\onlinecite{joshi1987heat}) and Bhowmik (Refs.~\onlinecite{bhowmik2009analysis}) comply with  the observation that $Nu_{\text{unit}}$ becomes independent of the fin thickness-to-length ratio for small values of $t/l$. 
However, the other correlations from the literature express the influence of the fin thickness on the Nusselt number through a scaling law of the form 
$Nu_{\text{unit}} \sim (t/l)^{m}$ with $m \in (0.02, 0.2)$ for $t/l \rightarrow 0$. 
As a result, they underestimate the Nusselt number data from this work with a relative error up to 40\% for $t/l<0.02$, even after they are rescaled by a constant factor to minimize the least-square differences between each correlation and our data, and to compensate for their incorrect scaling with other parameters. 
Moreover, the correlation of Kim et al. (Refs.~\onlinecite{kim2011correlations}) does not predict a continuous relationship between $Nu$ and $t/l$ as illustrated in Figure \ref{fig:litcomp_NU_PR07R300H12S24_nofit}, for the same reason as discussed previously. \\

\subsection{\label{sec:critical} The influence of the critical Reynolds number}

\begin{figure}[ht]
\includegraphics[scale = 1.00]{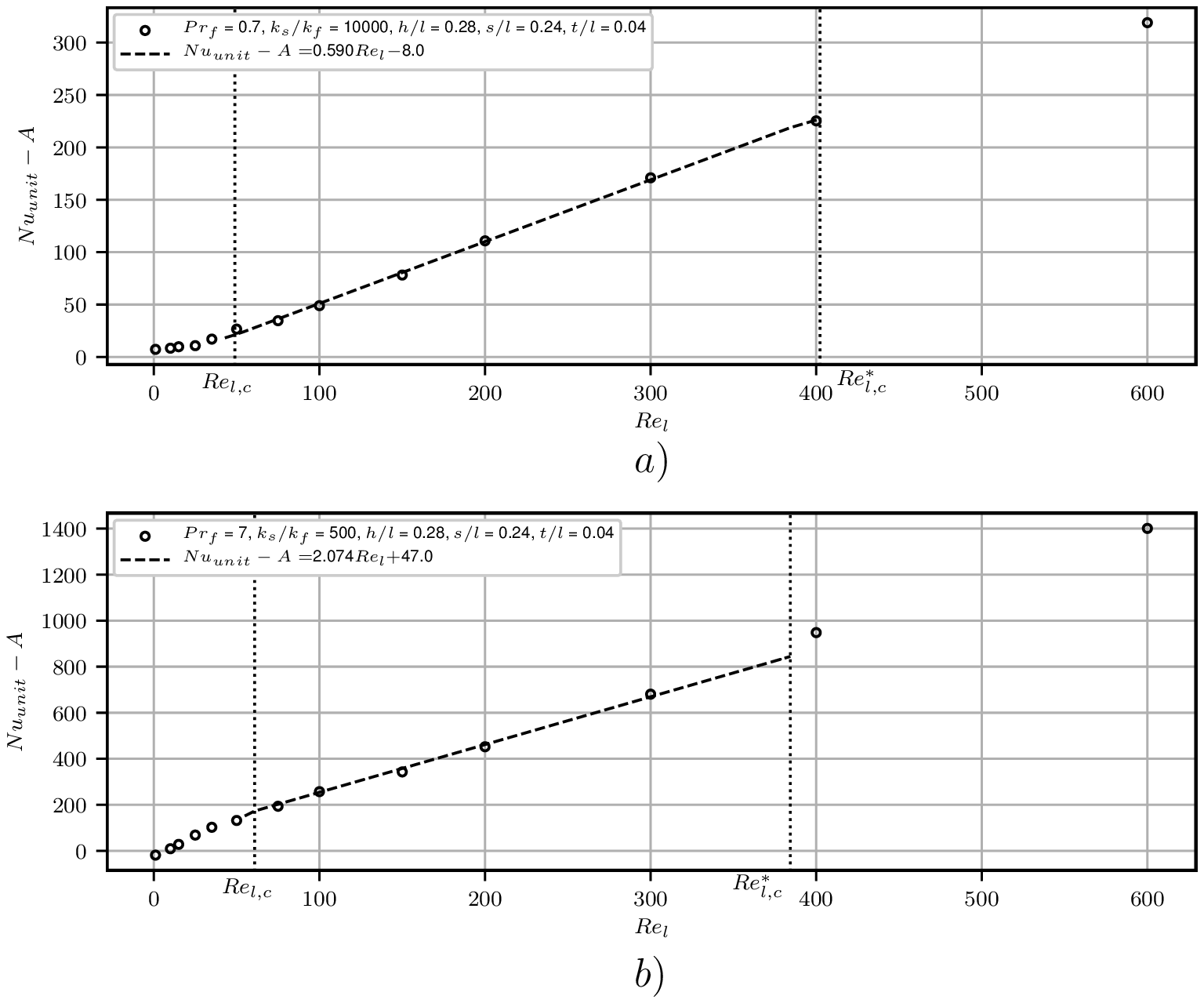}
\caption{\label{fig:Nu_correction_fit} Scaling of the correction term $Nu_{\text{unit}}-A$ with $Re_{l}$ for steady periodically developed heat transfer with (a) $Pr_f=0.7$ and $k_{s}/k_{f}=10^4$, and (b) $Pr_f=7$ and $k_{s}/k_{f}=500$, when $h/l = 0.28$, $s/l = 0.24$, $t/l = 0.04$}
\end{figure}

Despite the fact that our entire Nusselt number data set can be accurately captured by a linear relationship of the form $Nu_{\text{unit}} \simeq A + B Re_l$, a detailed analysis reveals a more complex scaling of $Nu_{\text{unit}}$ with $Re_{l}$.
Although the Nusselt number approximately becomes constant for $Re_l \rightarrow 0$, so $Nu_{\text{unit}} \rightarrow A$, the correction term $Nu_{\text{unit}} - A$ only approximately behaves as $B Re_l$ for some constant $B$. 
In Figure \ref{fig:Nu_correction_fit}, it is illustrated that the correction term $Nu_{\text{unit}} - A$ can only be considered to be linearly dependent on $Re_{l}$ with a constant slope $B$ over a specific Reynolds number range $Re_l \in \left(Re_{l,c} , Re_{l,c}^{*}\right)$.
In this figure, the constant term $A$ has been determined by fitting 
the Nusselt number in the lower Reynolds range, such that the fit $Nu_{\text{unit}} = A$ has a relative error below 1\% over this $Re_{l}$-range. 
Furthermore, the critical Reynolds numbers $Re_{l,c}$ and $Re_{l,c}^{*}$ have been determined as the values of $Re_{l}$ for which the correction term $Nu_{\text{unit}} - A$ deviates 1\% from the linear fit $\left( Nu_{\text{unit}} - A \right) = B Re_{l} + D$. 
It should be noted that the offset $D$ was added to ensure the robustness of the fitting procedure, since there is some numerical uncertainty on $A$. 
However, since $D \ll A$, the interval $Re_l \in \left(Re_{l,c} , Re_{l,c}^{*}\right)$ actually indicates the validity of the linear relationship $Nu_{\text{unit}} \simeq A + B Re_l$. 

In our previous work (Ref.~\onlinecite{vangeffelen2021friction}), two critical Reynolds numbers $Re_{l,ws}$ and $Re_{l,st}$ were introduced to indicate the transition from the weak inertia regime to the strong inertia regime and the transition from the strong inertia regime to the transitional regime, respectively. 
For the weak and strong inertia regime, the pressure drop deviates from Darcy's law with a term that is cubic and quadratic in the volume-averaged velocity, respectively (Refs.~\onlinecite{lasseux2011stationary,vangeffelen2021friction}). 
This term is typically known as Forchheimer's correction. 
A comparison of the critical Reynolds numbers for these flow regimes, $Re_{l,ws}$ and $Re_{l,st}$, with the critical Reynolds numbers for the heat transfer regime, $Re_{l,c}$ and $Re_{l,c}^{*}$, indicates a strong correspondence between these quantities. 
More specifically, $Re_{l,c}$ and $Re_{l,ws}$ have been found to virtually coincide for all unit cell geometries, Prandtl numbers and thermal conductivity ratios investigated in this work, as their average and maximum relative difference are 1\% and 10\%, respectively. 
The same conclusion can be drawn for $Re_{l,c}^{*}$ and $Re_{l,st}$, whose average and maximum relative difference equal 3\% and 10\%, respectively. 
As a consequence, we can deduce that the critical Reynolds numbers $Re_{l,c}$ and $Re_{l,c}^{*}$ depend on the geometrical parameters in the same way as $Re_{l,ws}$ and $Re_{l,st}$, as discussed in (Refs.~\onlinecite{vangeffelen2021friction}). 

The coincidence of the critical Reynolds numbers suggests that distinguishable heat transfer regimes are directly linked to the underlying flow regimes.
In the literature, similar observations have been made for the heat transfer regimes in porous media (Refs.~\onlinecite{hwang1994heat,martin1998frictional,mandhani2002forced,alshare2010modeling,lu2019effect}). 
For instance, in the work of Lu et al. (Refs.~\onlinecite{lu2019effect}), it was empirically assessed that the scaling of the Nusselt number with the Reynolds number is given by $Nu_{\text{unit}} \sim Re_l^{b}$, where the exponent $b$ can be considered a constant for each different flow regime. 
Besides the weak and strong inertia regime, they also considered the so-called pre-Darcy regime, which is not considered in this work. 
Nevertheless, as no analytical studies have been conducted on the theoretical scaling laws for the Nusselt number in porous media and fin arrays, the precise connection between the different flow and heat transfer regimes is currently not fully understood. 

Finally, we remark that Figure \ref{fig:Nu_correction_fit} confirms that any deviation from the trends $Nu_{\text{unit}}=A$ and $Nu_{\text{unit}}=A + B Re_l$ remains small. 
This implies that, from a practical perspective, the linear correlation (\ref{eq:nu_re}) is sufficiently accurate to model periodically developed heat transfer in offset strip fin arrays in micro- and mini-channels subject to a constant heat flux. \\

\subsection{\label{sec:influence prkskf}The influence of the Prandtl number $Pr_f$ and the thermal conductivity ratio $k_{s}/k_{f}$ on the Nusselt number}

So far, the Nusselt number was only investigated for two Prandtl numbers: $Pr_f=0.7$ (air) and $Pr_f=7$ (water), while the conductivity ratio was kept fixed to $k_{s}/k_{f}=10^4$ and $k_{s}/k_{f}=500$. 
Therefore, the influence of these material properties is now examined more thoroughly. 

\begin{figure}[ht]
\centering
\begin{minipage}{.475\textwidth}
\raggedleft
\includegraphics[scale = 1.00]{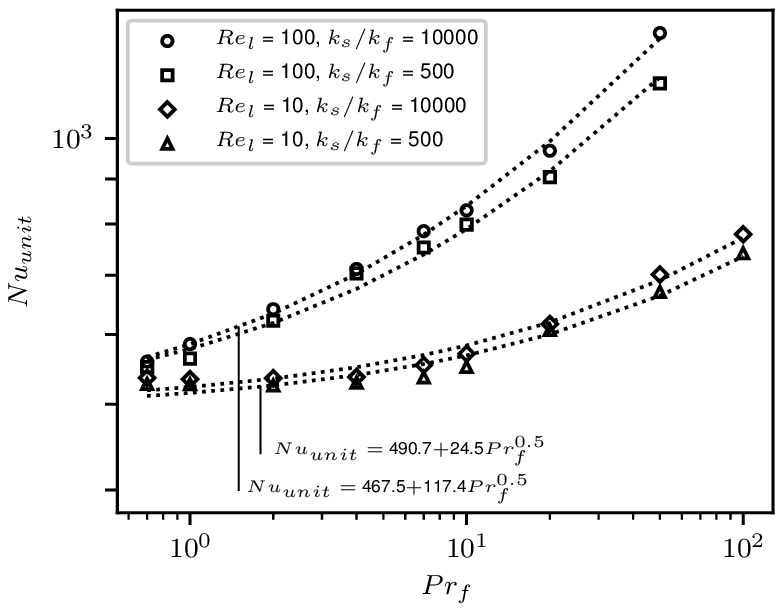}
\caption{\label{fig:NU_Pdata} Influence of the Prandtl number on the Nusselt number for steady periodically developed heat transfer, when $h/l=0.24$, $s/l=0.24$, $t/l=0.02$}
\end{minipage}
\hfill
\begin{minipage}{.475\textwidth}
\raggedright
\includegraphics[scale = 1.00]{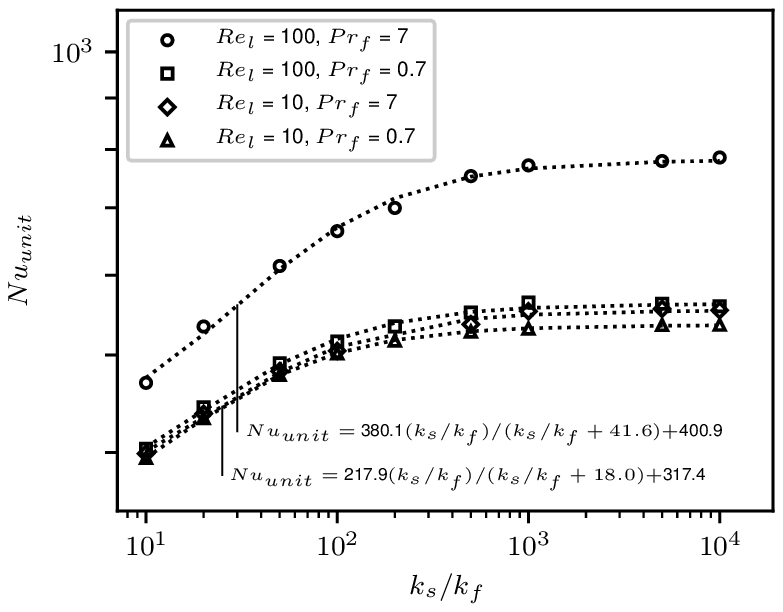}
\caption{\label{fig:NU_Kdata} Influence of the thermal conductivity ratio on the Nusselt number for steady periodically developed heat transfer, when $h/l=0.24$, $s/l=0.24$, $t/l=0.02$}
\end{minipage}
\end{figure}

In Figure \ref{fig:NU_Pdata}, the influence of the Prandtl number $Pr_{f}$ is illustrated for a single unit cell geometry, for two conductivity ratios, $k_s/k_f=500$ and $k_s/k_f=10^4$, and two Reynolds numbers, $Re_{l}=10$ and $Re_{l}=100$. 
The range of Prandtl numbers displayed in this figure is relevant for thermal oils used in heat recuperators (Refs.~\onlinecite{yih2020experimental}). 
It can be observed that for the selected parameters, the Nusselt number correlates well with the Prandtl number according to a scaling law of the type
\begin{equation}
\label{eq: correlation form pr}
Nu_{\text{unit}} \simeq M + N Pr_{f}^{0.5}\,, 
\end{equation}
where the standard deviation on the exponent of the Prandtl number equals 0.1.
As a matter of fact, the correlation above predicts all the data points for the Nusselt number in Figure \ref{fig:NU_Pdata} with an average error of 2\%, and a maximum relative error of 4\%.
It clearly reflects that for low Prandtl numbers, the Nusselt number becomes independent of $Pr_{f}$. 
In general, the parameters $M$ and $N$ in correlation (\ref{eq: correlation form pr}) vary, of course, with the Reynolds number, the thermal conductivity ratio, and the offset strip fin geometry. 

Figure \ref{fig:NU_Pdata} shows that the Nusselt number remains constant over a wider range of Prandtl numbers, when the Reynolds number decreases. 
This observation is predicted by the empirical correlations for convective heat transfer in porous media and cylinder arrays, which relate the Nusselt number to the Reynolds number and Prandtl number through a relationship of the form $Nu = M + N Re_{l}^{b} Pr_f^{n}$.
Commonly, the coefficients $M$ and $N$ in this relationship depend solely on the geometry, and the exponents satisfy $b \in \left(0.5, 0.9 \right)$ and $n \in \left(0.2, 0.4 \right)$ (Refs.~\onlinecite{wakao1978effect,hwang1994heat,kuwahara2001numerical,mandhani2002forced,saito2006correlation,gamrat2008numerical}). 
The finding that our data from Figure \ref{fig:NU_Pdata} indicates a larger exponent ($n=0.5$) can be attributed to the different geometry of offset strip fins and the fact that we considered a limited number of data points for the Prandtl number. 

It is worth pointing out that the scaling law (\ref{eq: correlation form pr}) observed in Figure \ref{fig:NU_Pdata} deviates significantly from the correlations for offset strip fins in the literature.
The latter all predict a trend $Nu \sim Pr_f^{n}$ with $n \in (0.3, 0.4)$ and thus fail to predict a constant limit of the Nusselt number for low Prandtl numbers and low Reynolds numbers, despite its relevance for micro- and mini-channels.
Moreover, these correlations also underestimate our data in Figure \ref{fig:NU_Pdata} for larger Reynolds numbers, due to their lower exponent $n$ for the Prandtl number. 
This mismatch is arguably caused by a different scaling of the Nusselt number with the Prandtl number in the transitional and turbulent flow regime, which most of the data in the literature belong to. 
For example, the correlation of Kim et al. (Refs.~\onlinecite{kim2011correlations}), which was constructed for Prandtl numbers ranging from 0.72 to 50 (see Table \ref{tab:literature2}), underestimates our data in Figure \ref{fig:NU_Pdata} with at least 20\% for $Pr_{f} < 2$ and $Re_{l}=100$, 
as well as $Pr_{f} < 7$ and $Re_{l}=10$. 
Similar discrepancies are observed between our data and the correlations of Manglik and Bergles (Refs.~\onlinecite{manglik1995heat}) and Joshi and Webb (Refs.~\onlinecite{joshi1987heat}).
For those correlations, it is assumed that the dependence of the Nusselt number on the Prandtl number is captured in the definition of the Colburn j-factor, through the factor $Pr^{1/3}$. 
This additional data set is tabulated in Appendix \ref{sec:alignedflowdata_extra}. \\


The influence of the thermal conductivity ratio $k_{s}/k_{f}$ on the Nusselt number is illustrated in Figure \ref{fig:NU_Kdata}. 
For low values of $k_{s}/k_{f}$, the Nusselt number scales linearly with the thermal conductivity ratio $k_{s}/k_{f}$. 
On the contrary, for high values of $k_{s}/k_{f}$, in this case $k_{s}/k_{f} > 500$, the Nusselt number $Nu_{\text{unit}}$ becomes independent of the thermal conductivity ratio. 
When the ratio $k_{s}/k_{f}$ is high, the periodic temperature field $T^{*}$ inside the solid domain becomes approximately uniform with a limited dependence on $k_{s}/k_{f}$, as discussed in (Refs.~\onlinecite{li2016effect}) in the context of conjugate heat transfer in pin-fin arrays. 
This means that the influence of the thermal conductivity ratio $k_{s}/k_{f}$ on the temperature field in the fluid domain, and hence $Nu_{\text{unit}}$, is very weak in that case.
As a result, when considering typical values for the material properties of air and water, namely $Pr_{f} < 10$ and $k_{s}/k_{f} > 500$, both the Prandtl number and the thermal conductivity ratio do not exert any significant influence on the Nusselt number at low Reynolds numbers. 
The same conclusions can be drawn by observing Figures \ref{fig:NU_REdataP07T4} and \ref{fig:NU_REdataP7T4}. 

Based on the two observed trends for low and high $k_{s}/k_{f}$-ratios, the dependence of the Nusselt number on the thermal conductivity ratio is asymptotically equivalent to the form
\begin{equation}
\label{eq: correlation form kskf}
Nu_{\text{unit}} \simeq P \frac{\left( k_{s}/k_{f} \right)}{\left( k_{s}/k_{f} \right) + R} + Q\,.
\end{equation}
This expression results in an average relative error of $0.5\%$ and a maximum relative error of $2\%$ with respect to the data in Figure \ref{fig:NU_Kdata}. 
As discussed in Section \ref{sec:intro}, the influence of $k_{s}/k_{f}$ on the Nusselt number for offset strip fins has not been reported in literature before. 
Therefore, the parameter  $k_{s}/k_{f}$ does not appear in the available correlations from the literature, and a comparison with our data could not be made. 



\section{\label{sec:correlation}Nusselt number correlation}

\subsection{Fitting approach}
To obtain a final correlation for the Nusselt number in the steady periodically developed heat transfer regime in micro- and mini-channels with offset strip fins, we followed the same two-step procedure as proposed for the friction factor in our previous work (Ref.~\onlinecite{vangeffelen2021friction}). First, we determined the optimal parameter values for multiple heuristically chosen candidate correlations, complying with the forms (\ref{eq:nu_re})-(\ref{eq: correlation form t/l}).
Hereto, we used a non-linear least-squares optimization method based on a trust region reflective algorithm (Refs.~\onlinecite{2020SciPy-NMeth}). 
Secondly, we calculated the most likely parameter values for each candidate correlation using the Bayesian approach for parameter estimation and model validation (Refs.~\onlinecite{sivia1993introduction}). 
Through the Bayesian approach, we also obtained for each candidate correlation its log-evidence, which is a statistical measure to quantify the suitability of the correlation to represent the data.  
Finally, we selected the candidate correlation and its fitting parameters that resulted in the highest relative accuracy and the highest log-evidence value for our Nusselt number data. 
We note that both the optimal and most likely parameter values were found to be equal within their significant digits for the final Nusselt number correlations presented next.


\subsection{Fitting result}

The proposed correlation for the Nusselt number in the case of air ($Pr_f=0.7$ and $k_{s}/k_{f}=10^4$) is  
\begin{equation}
    Nu_{\text{unit}} = c_{0} + c_{1} Re_l, 
\nonumber
\end{equation}
with
\begin{equation}
\begin{aligned}
  c_{0} &= 6.44 (h/l)^{-2} + 9.60(h/l)^{-1.24} + 24.4 (s/l)^{-1.85}, \\
  c_{1} &= 0.112 (s/l-t/l)^{-0.61} (h/l)^{-0.48}.\\
\end{aligned}
\label{eq:fit_alignedPr07}
\end{equation}
This correlation results in an average relative error of 3\% with respect to the Nusselt number data in this work. 
The maximum relative error is below 6\%, 8\% and 12\% for, respectively, 90\%, 95\%, 99\% of the data points. 

The proposed Nusselt number correlation in the case of water 
($Pr_f=7$ and $k_{s}/k_{f}=500$), is 
\begin{equation}
    Nu_{\text{unit}} = d_{0} + d_{1} Re_l, 
\nonumber
\end{equation}
with
\begin{equation}
\begin{aligned}
  d_{0} &= 3.84 (h/l)^{-2} + 19.2(h/l)^{-1.39} + 22.3 (s/l)^{-1.87}, \\
  d_{1} &= 1.26 (s/l-t/l)^{-1.07} (t/l)^{0.54} (h/l)^{-0.56}.\\
\end{aligned}
\label{eq:fit_alignedPr7}
\end{equation}
In this case, the average relative error is 4\% with respect to our data, while the maximum relative error remains below 9\%, 11\% and 18\% for, respectively, 90\%, 95\%, 99\% of the data points. 

\begin{figure}[ht]
\centering
\begin{minipage}{.475\textwidth}
\raggedleft
\includegraphics[scale = 1.00]{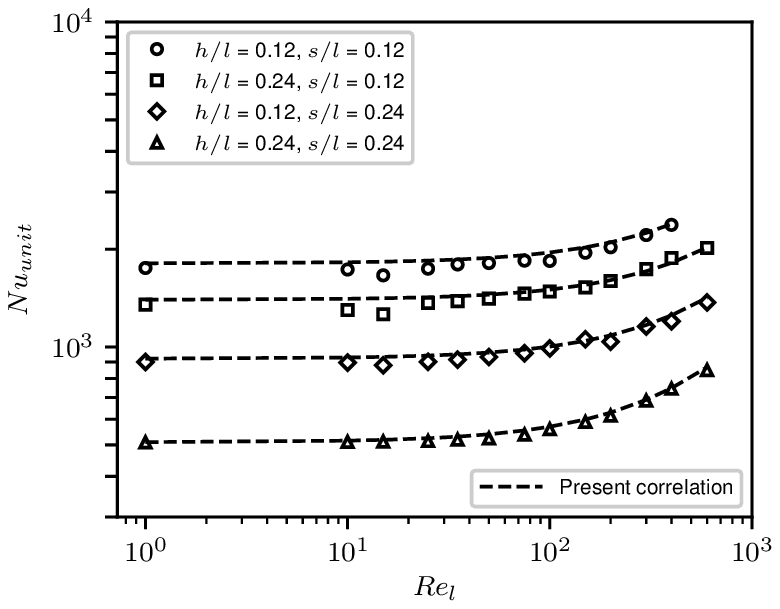}
\caption{\label{fig:NU_fitREdataP07T4} A comparison between our final Nusselt number correlation and the data from this work, at small fin heights, when $Pr_f=0.7$, $k_{s}/k_{f}=10^4$, $t/l=0.04$}
\end{minipage}
\hfill
\begin{minipage}{.475\textwidth}
\raggedright
\includegraphics[scale = 1.00]{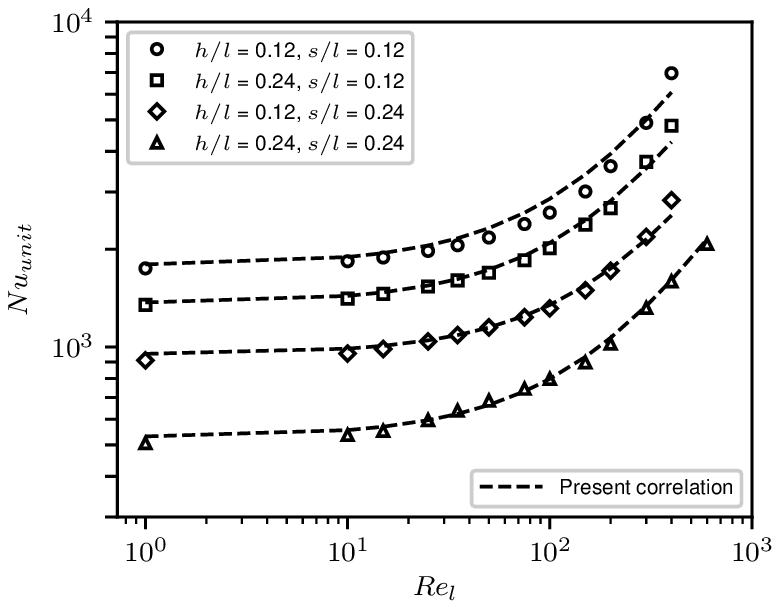}
\caption{\label{fig:NU_fitREdataP7T4} A comparison between our final Nusselt number correlation and the data from this work, for at small fin heights, when $Pr_f=7$, $k_{s}/k_{f}=500$, $t/l=0.04$}
\end{minipage}
\end{figure}

The accuracy of correlations (\ref{eq:fit_alignedPr07}) and (\ref{eq:fit_alignedPr7}) is illustrated in Figures \ref{fig:NU_fitREdataP07T4} and \ref{fig:NU_fitREdataP7T4}. 
For example, all candidate correlations of the form $Nu_{\text{unit}} = A + B Re_l$ resulted in log-evidence values two times larger than those of the form $Nu_{\text{unit}} = B Re_l^{b}$, despite the latter form is most frequently adopted in the literature.

The correlations (\ref{eq:fit_alignedPr07}) and (\ref{eq:fit_alignedPr7}) reflect that a linear relation between the Nusselt and the Reynolds number is very precise for Reynolds numbers $Re_{l}$ ranging from 1 to 600. 
Furthermore, they are consistent with all the discussed trends (\ref{eq: correlation form h/l})-(\ref{eq: correlation form t/l}) which characterize the influence of the geometrical parameters on $Nu_{\text{unit}}$. 

With regard to the influence of the fin height-to-length ratio $h/l$, the final correlations respect the asymptotic trends $Nu_{\text{unit}} \sim (h/l)^{-2}$ for $h/l \rightarrow 0$ and $Nu_{\text{unit}} \sim (h/l)^{0}$ for $h/l \rightarrow \infty$. 
At the same time, they include some additional terms proportional to $(h/l)^{b}$ with $b \neq -2$ to better match our data for intermediate values of $h/l$. 
We remark that their limits $Nu_{\text{unit}}=6.44(h/l)^{-2}$ (for air) and $Nu_{\text{unit}}=3.84(h/l)^{-2}$ (for water), which are found for $h/l \rightarrow 0$, do not correspond to the Nusselt number for fully-developed heat transfer between two parallel plates subject to a constant heat flux (Refs.~\onlinecite{shah1967offset}). 
In contrast, the asymptotic value of the friction factor for offset strip fins in the limit of $h/l \rightarrow 0$ was found to agree well with the one for fully-developed flow between parallel plates (Refs.~\onlinecite{vangeffelen2021friction}). 
This can be explained by the fact that the offset strip fins still thermally connect the top and bottom plates of the channel, even when the fin height-to-length ratio $h/l$ and the fin thickness-to-length ratio $t/l$ approach zero. 
As such, there remains a path of low thermal resistance in the solid fins, along which heat is more easily conducted from the bottom plate to the top plate, than through fluid.
Nevertheless, the asymptotic scaling of $Nu_{\text{unit}}$ with $(h/l)^{-2}$ for $h/l \rightarrow 0$ is in agreement with the Nusselt numbers for fully-developed heat transfer between two parallel plates with any type of thermal boundary condition (Refs.~\onlinecite{shah1967offset}).

Concerning the fin pitch-to-length ratio $s/l$, the final correlations (\ref{eq:fit_alignedPr07}) and (\ref{eq:fit_alignedPr7}) reveal that $Nu_{\text{unit}}$ becomes independent of $s/l$ for  $s/l \rightarrow \infty$.
Besides, they indicate that $Nu_{\text{unit}} \rightarrow \infty$ for $s \rightarrow t$ and that the Nusselt number becomes directly proportional to the Reynolds number when $s \rightarrow t$.
The reason is that the coefficients $c_{1}$ and $d_{1}$ in equations (\ref{eq:fit_alignedPr07}) and (\ref{eq:fit_alignedPr7}) become dominant in that limit. 
This prediction is in line with many empirical correlations from the literature (Refs.~\onlinecite{martin1998frictional,saito2006correlation,lu2019effect}), although it is still not clear whether it is justified from a theoretical perspective.

With respect to the fin thickness-to-length ratio $t/l$, the final correlations show that $t/l$ barely affects the Nusselt number when $t/l$ is small. 
Therefore, the correlation factors $c_0$ and $d_0$ are independent of $t/l$. 
In the case of air ($Pr_f=0.7$ and $k_{s}/k_{f}=10^4$), even the correlation term $c_{1}$ does not directly depend on $t/l$. 
On the contrary, for water ($Pr_f=7$ and $k_{s}/k_{f}=500$), the correlation term $d_{1}$ does scale with $(t/l)^{0.54}$, as it accounts for a more significant influence of the thickness-to-length ratio. 

Finally, it is worth mentioning that the correlations (\ref{eq:fit_alignedPr07}) and (\ref{eq:fit_alignedPr7}) are valid for air at temperatures between 0$ ^\circ C $ and 500$ ^\circ C $, and for water at temperatures between 15$ ^\circ C $ to 25$ ^\circ C $, respectively. 
Yet, from the analysis in Section \ref{sec:influence prkskf}, it is expected that the value of $Nu_{\text{unit}}$, and in particular of the terms $c_{0}$ and $d_{0}$, are not significantly influenced by $Pr_f$ or $k_{s}/k_{f}$, as long as the Reynolds number remains below $10$.

\section{\label{sec:final remarks}Final Remarks}
\begin{figure}[ht!]
\includegraphics[scale = 1.0]{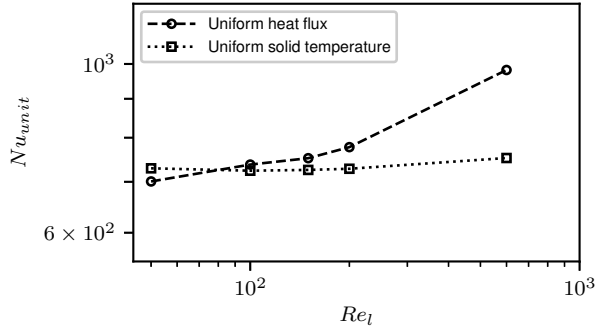}
\caption{\label{fig:BC_comp} Influence of the thermal boundary condition on the Nusselt number for steady periodically developed heat transfer, when $Pr_f=0.7$, $k_{s}/k_{f}=10^4$, $h/l=0.12$, $s/l=0.48$, $t/l=0.04$}
\end{figure}

As the results from this work have been obtained under the condition of a uniform heat flux at the channel wall, while the correlations from the literature apply to the condition of a uniform channel wall (and solid fin) temperature, the influence of the boundary condition deserves a closer inspection.  
Although an investigation of the Nusselt number for a uniform solid temperature falls beyond the scope of the present work, we do present a comparison between our data and a limited data set obtained for a uniform solid temperature. 
This comparison is made in Figure \ref{fig:BC_comp} and shows that 
the Nusselt number for a uniform solid temperature can be up to 23\% smaller
than for a uniform heat flux.
Notably, this difference becomes even larger when the Reynolds number increases. 
The latter observations agree with the findings for straight channels without fins and arrays of pin-fins (Refs.~\onlinecite{shah1978laminar,li2016effect}). 
They suggest that the discrepancies between our Nusselt number correlations and the empirical correlations from the literature are largely, but not wholly, explained by the difference in the considered thermal boundary condition. 
We note that the necessary equations to compute $Nu_{\text{unit}}$ for a uniform solid temperature boundary condition are included in Appendix \ref{sec:thermalBC}. \\

\begin{figure}[ht]
\centering
\begin{minipage}{.475\textwidth}
\raggedleft
\includegraphics[scale = 1.00]{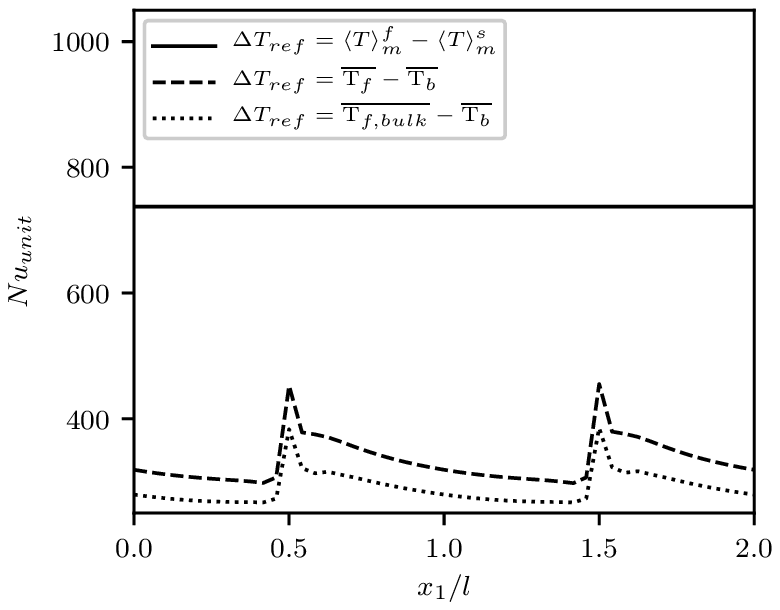}
\caption{\label{fig:Nu_Tref} Influence of the reference temperature difference on the Nusselt number for steady periodically developed heat transfer driven by a uniform heat flux, when $Re_{l}=100$, $Pr_f=0.7$, $k_{s}/k_{f}=10^4$, $h/l=0.12$, $s/l=0.48$, $t/l=0.04$}
\end{minipage}
\hfill
\begin{minipage}{.475\textwidth}
\raggedright
\includegraphics[scale = 1.00]{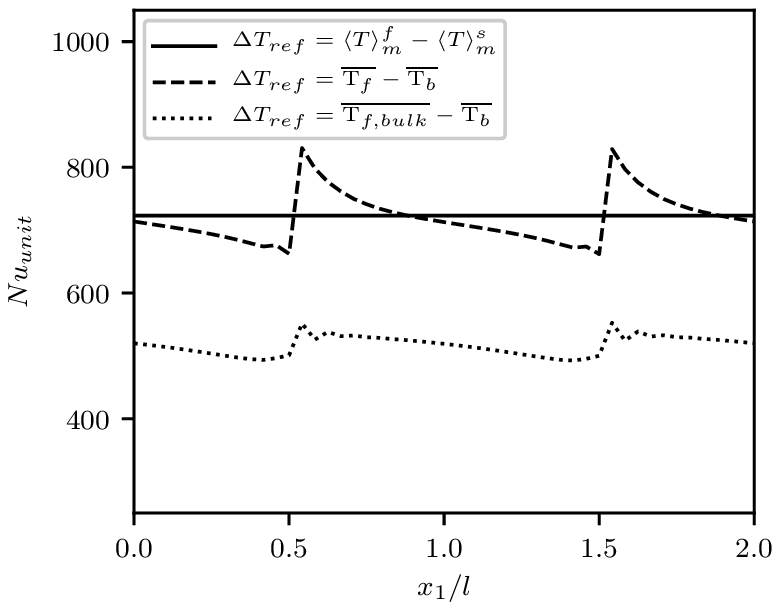}
\caption{\label{fig:Nu_Tref_Ts} Influence of the reference temperature difference on the Nusselt number for steady periodically developed heat transfer driven by a uniform solid temperature, when $Re_{l}=100$, $Pr_f=0.7$, $h/l=0.12$, $s/l=0.48$, $t/l=0.04$}
\end{minipage}
\end{figure}

Another possible source of discrepancies between our correlations and those from the literature is the choice of the temperature difference $\Delta T_{\text{ref}}$ in the definition of the heat transfer coefficient: $h_{\text{unit}} = \langle q_{fs} \delta_{fs} \rangle_{m} / \Delta T_{\text{ref}}$. 
The heat transfer coefficient in the experimental correlations from the literature may be interpreted as a measure for the temperature difference between 
the average fluid temperature and average bottom plate temperature: $\Delta T_{\text{ref}} =  \overline{\mathrm{T}_{f}} - \overline{\mathrm{T}_{b}} $.
However, it may also be interpreted as a measure for the difference in bulk temperatures $\Delta T_{\text{ref}} =\overline{\mathrm{T}_{f, bulk}} - \overline{\mathrm{T}_{b}}$, instead of the difference between the macro-scale temperatures of the fluid and solid, $\Delta T_{\text{ref}} =  \langle T \rangle_{m}^{f} - \langle T \rangle_{m}^{s} $. 
The impact of the chosen temperature difference in the definition of the heat transfer coefficient is illustrated in Figures \ref{fig:Nu_Tref} and \ref{fig:Nu_Tref_Ts}. 
The employed definitions of the various reference temperature differences are further clarified in Appendix \ref{sec:temperaturedef}. 
Both figures show that the Nusselt numbers based on the difference in cross-sectional temperatures and the bulk temperatures exhibit spatial variations up to 20\% within in a single fin unit, whereas the Nusselt numbers based on the macro-scale temperature difference are spatially constant.
Furthermore, the offset between the Nusselt numbers for different reference temperatures can be as high as $50\%$. \\

\section{\label{sec:conclusion}Conclusions}

In this work, the Nusselt number for periodically developed conjugate heat transfer in micro- and mini-channels with offset strip fin arrays has been analyzed under the assumption of a constant heat flux at the channel wall.
An extensive data set for the Nusselt number was collected through 2282 numerical simulations of the periodic temperature field on a unit cell of the array. 

It was shown that the Nusselt number correlations from the literature primarily focus on the heat transfer for airflow through larger conventional offset strip fin arrays, subject to a constant channel wall temperature. 
Consequently, these correlations do not accurately predict the trends and limits for the Nusselt number with respect to the Reynolds number and the geometrical parameters in micro- and mini-channels.
At least, they result in discrepancies as large as 40-90\% in the case of an imposed uniform heat flux.

For this reason, two new Nusselt number correlations have been presented, which apply to air and water in micro- and mini-channels with offset strip fins. 
They result in an average relative error of 4\% with respect to the Nusselt number data from this work. 
The correlations were constructed through a least-squares fitting procedure, after which their suitability was assessed by the Bayesian approach for parameter estimation and model estimation. 

The two new correlations predict a linear relationship between the Nusselt number and the Reynolds number. 
Yet, a more detailed analysis of the Nusselt number correction term has revealed a deviation from this linear scaling outside a specific range of Reynolds numbers bound by two critical Reynolds numbers. 
Within a relative margin of 10\%, these critical Reynolds numbers were shown to correspond to the two critical Reynolds numbers which characterize the transition from the weak inertia regime to the strong inertia regime, and the transition from the strong inertia regime to the transitional regime. 
This result suggests a direct link between the flow regime and the heat transfer regime. Despite the observed deviation from a linear scaling, the linear relationship still accurately captures all the Nusselt number data from this work. 

In addition, the new correlations respect the asymptotic trends observed for each of the geometrical parameters of the offset strip fin array. 
More specifically, the correlations respect the observation that the Nusselt number scales with the inverse square of the relative fin-height-to-length ratio for small fin heights. 
The correlations also comply with the asymptotic limit of the Nusselt number becoming infinite when the fin pitch approaches the value of the fin thickness. 
Furthermore, the observation that the Nusselt number becomes independent of the fin height, fin pitch, and fin thickness for large fin heights, large fin pitches, and small fin thicknesses, respectively, is taken into account as well. 

Finally, the influence of the Prandtl number and thermal conductivity ratio on the Nusselt number has been investigated via 62 additional simulations.
We observed that the Nusselt number does not change with the Prandtl number over the lower Prandtl-number range, in particular at lower Reynolds numbers.
Although this observation is contradicted by the  
available correlations for offset strip fins from the literature, it is in agreement with the empirical correlations for convective heat transfer in porous media and cylinder arrays. 
Furthermore, our simulations indicate that the Nusselt number scales linearly with the thermal conductivity ratio, as long as this ratio remains below 500. 
For larger values, the Nusselt number is no longer affected by the thermal conductivity ratio, as the periodic temperature becomes nearly uniform in the fins. 

\newpage
\clearpage

\section{\label{sec:contributions}Contributions}
The implementation and validation of the computational algorithms and software framework for the periodic flow and temperature equations were performed by G. Buckinx. 
All heat transfer simulations and post-processing calculations were carried out by A. Vangeffelen. 
The results were interpreted by A. Vangeffelen, with input from G. Buckinx regarding the existing literature. 
The paper was written by A. Vangeffelen and G. Buckinx with input from C. De Servi, M. R. Vetrano and M. Baelmans. 

\section{\label{sec:acknowledgements}Acknowledgements}
The work documented in this paper was funded by the Research Foundation — Flanders (FWO) through the post-doctoral project grant 12Y2919N of G. Buckinx, and by the Flemish Institute for Technological Research (VITO) through the Ph.D. grant 1810603 of A. Vangeffelen. 
The resources and services used in this work were provided by the VSC (Flemish Supercomputer Center), funded by the Research Foundation - Flanders (FWO) and the Flemish Government.

\newpage
\clearpage

\appendix

\section{\label{sec:thermalBC}Nusselt number for a uniform solid temperature}


When the solid temperature $T_{s}$ is uniform, the Nusselt number in the steady periodically developed heat transfer regime is defined by the heat transfer coefficient
\begin{equation}
    h_{\text{unit}} \triangleq \epsilon_{fm}^{-1} \frac{ \langle q_{fs} \delta_{fs} \rangle_m }{ \langle T \rangle_{m}^{f} - \langle T \rangle_{m}^{s} } = - \langle \boldsymbol{n}_{fs} \cdot k_{f} \mathrm{\nabla{\theta_{f}}} \delta_{fs} \rangle \,,
\label{eq:heattransfercoefficient_isotherm}
\end{equation}
as the fluid temperature decays exponentially in the main flow direction $\boldsymbol{e}_{s}$ with a spatially periodic amplitude $\theta_{f}$: $T_{f} = T_{0} \theta_{f} \text{exp} \left( \lambda_{T} \boldsymbol{x} \cdot \boldsymbol{e}_s \right) + T_{s}$ for some constant $T_{0}$ (Refs.~\onlinecite{buckinx2015macro}). 
The periodic amplitude $\theta_{f}$ is governed by the following energy conservation equation, if viscous dissipation is negligible, and other heat sources are absent (Refs.~\onlinecite{buckinx2015macro}):
\begin{equation}
  \rho_{f} c_{f} \nabla \cdot \left( \boldsymbol{u} \theta_{f} \right) = k_{f} \nabla^2 \theta_{f} + \sigma_{T} \quad \text{in } \Omega_{f}, 
\label{eq:Temperature_isotherm}
\end{equation}
with
\begin{equation}
\sigma_{T} = \left( 2 k_{f} \mathrm{\nabla{\theta_{f}}} - \rho_{f} c_{f} \boldsymbol{u} \theta_{f} \right) \cdot \boldsymbol{e}_s \lambda_{T} + k_{f} \theta_{f} \lambda_{T}^{2}. 
\label{eq:Temperature_isotherm2}
\end{equation}
The corresponding periodicity and boundary conditions are given by
\begin{equation}
\begin{aligned}
  \theta_{f} \left( \boldsymbol{x} + \boldsymbol{l}_{j}  \right) &= \theta_{f} \left( \boldsymbol{x}  \right) & &\text{in } \Omega_{f}, \\
  \theta_{f} &= 0 & &\text{in } \Gamma_{fs} \cup \Gamma_{t} \cup \Gamma_{b}, \\
  - \boldsymbol{n}_{fs} \cdot k_{f} \mathrm{\nabla{\theta_{f}}} &= 0 & &\text{in } \Gamma_{t} \cup \Gamma_{b}, \\
  \langle \theta \rangle &= 1, &  \\
\end{aligned}
\label{eq:BoundaryConditionsHeat_isotherm}
\end{equation}
with $j=\{ 1,2 \}$. 
The volume-average of the distribution $\theta$, which equals $\theta_{f}$ in $\Omega_{f}$ and $\theta_{s} = 0$ in $\Omega_{s}$ needs to be imposed in order to find a unique solution, but does not affect $Nu_{\text{unit}}$. 
The temperature decay rate $\lambda_{T}$ is the negative solution of the following eigenvalue problem: 
\begin{equation}
\langle \boldsymbol{n}_{fs} \cdot k_{f} \mathrm{\nabla{\theta_{f}}} \delta_{fs} \rangle - \left( \rho_{f} c_{f} \langle \boldsymbol{u} \theta \rangle \cdot \boldsymbol{e}_s \right) \lambda_{T} + k_{f} \lambda_{T}^{2} = 0. 
\label{eq:lambda}
\end{equation}

\newpage
\clearpage

\section{\label{sec:temperaturedef}Reference temperature difference for the Nusselt number}

The heat transfer coefficient for channel flows is commonly defined on the basis of a reference temperature difference $\Delta T_{\text{ref}}$ between cross-sectional averaged temperatures or bulk averaged temperatures. 
Any of these averaged temperatures can be defined as
\begin{equation}
    \overline{\mathrm{T}} \triangleq \frac{\displaystyle\int_{\textbf{r} \in \Gamma^{*} \left( \boldsymbol{x} \right) }  T \left( \boldsymbol{r} \right) w_{\text{ref}} \left( \boldsymbol{r} \right) \,d\Gamma^{*} \left( \boldsymbol{r} \right)}{\displaystyle\int_{\textbf{r} \in \Gamma^{*} \left( \boldsymbol{x} \right) } w_{\text{ref}} \left( \boldsymbol{r} \right) \,d\Gamma^{*} \left( \boldsymbol{r} \right)}, 
\label{eq:area_temperature}
\end{equation}
where $w_{\text{ref}}$ is a weighting function, and $\Gamma^{*}$ the cross-section given by
\begin{equation}
    \Gamma^{*} \left( \boldsymbol{x} \right) \triangleq \left\{ \boldsymbol{r} \: | \: \exists \: c_{j} \in \left( - \frac{1}{2}, \frac{1}{2} \right) \Leftrightarrow \boldsymbol{r} = \boldsymbol{x} + \sum_{j=2}^{3} c_{j} \boldsymbol{l}_{j} \right\}. 
\label{eq:cross_section}
\end{equation}
Often, the reference temperature difference is that between the average fluid temperature and average bottom plate temperature: $\Delta T_{\text{ref}} =  \overline{\mathrm{T}_{f}} - \overline{\mathrm{T}_{b}} $, with $\overline{\mathrm{T}_{f}} \triangleq \left( \overline{\mathrm{T}} \: | \: w_{\text{ref}} = \gamma_{f} \right)$ and $\overline{\mathrm{T}_{b}} \triangleq \left( \overline{\mathrm{T}} \: | \: w_{\text{ref}} = \delta_{b} \right)$. 
On the other hand, the bulk-temperature difference is usually given by $\Delta T_{\text{ref}} =  \overline{\mathrm{T}_{f\text{,bulk}}} - \overline{\mathrm{T}_{b}}$, with $\overline{\mathrm{T}_{f\text{,bulk}}} \triangleq \left( \overline{\mathrm{T}} \: | \: w_{\text{ref}} = \boldsymbol{u} \cdot \boldsymbol{e}_{1} \right)$.

\newpage
\clearpage

\section{\label{sec:alignedflowdata}Periodically developed Nusselt number data for air and water}

\vspace{-5mm}
\hspace{-20mm}
\begin{minipage}{.5\textwidth}
\raggedleft
\includegraphics[scale = 1.00]{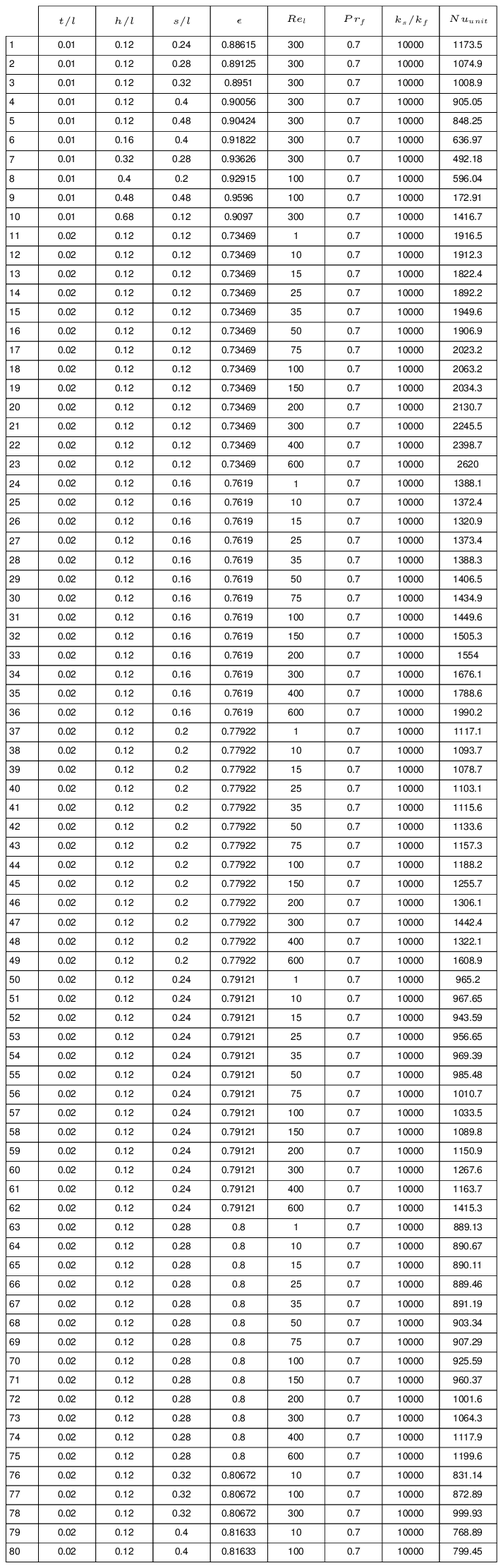}
\end{minipage}
\hfill
\begin{minipage}{.5\textwidth}
\raggedright
\includegraphics[scale = 1.00]{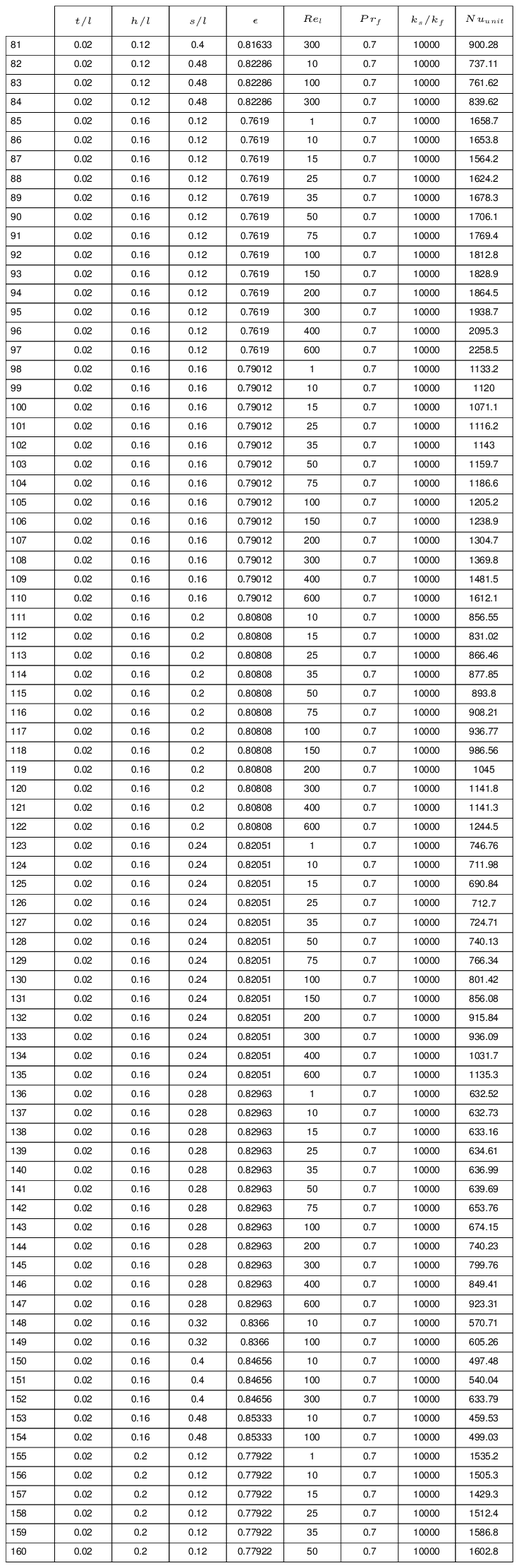}
\end{minipage}

\newpage
\clearpage
\hspace{-20mm}
\begin{minipage}{.5\textwidth}
\raggedleft
\includegraphics[scale = 1.00]{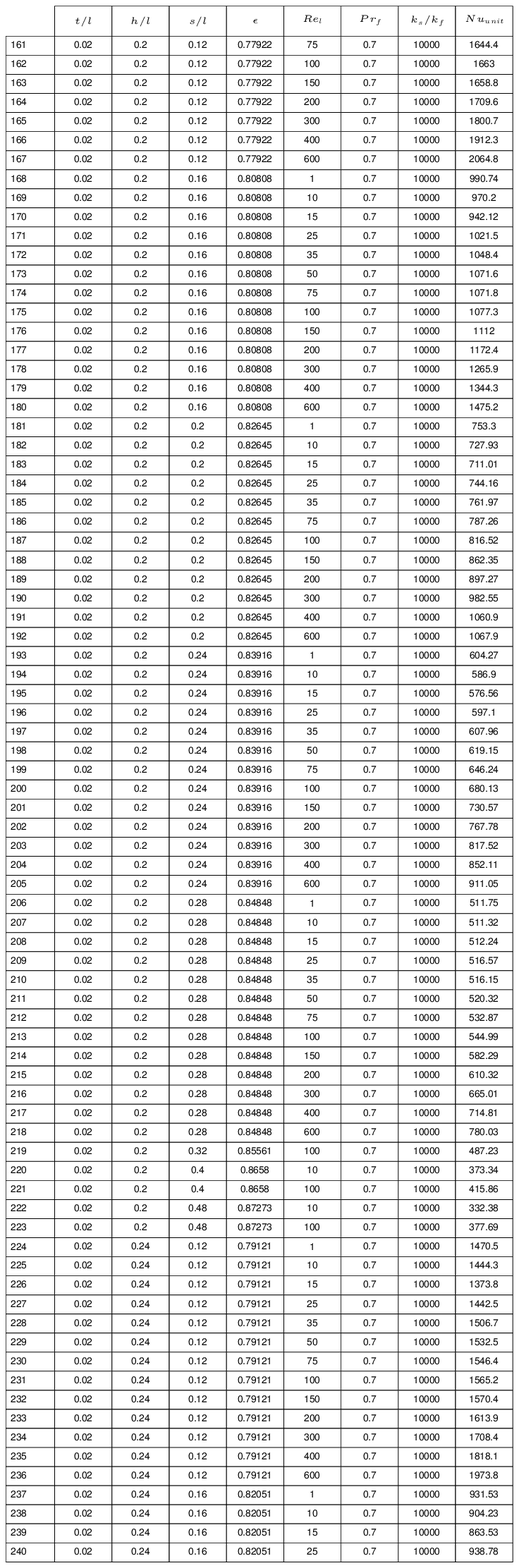}
\end{minipage}
\hfill
\begin{minipage}{.5\textwidth}
\raggedright
\includegraphics[scale = 1.00]{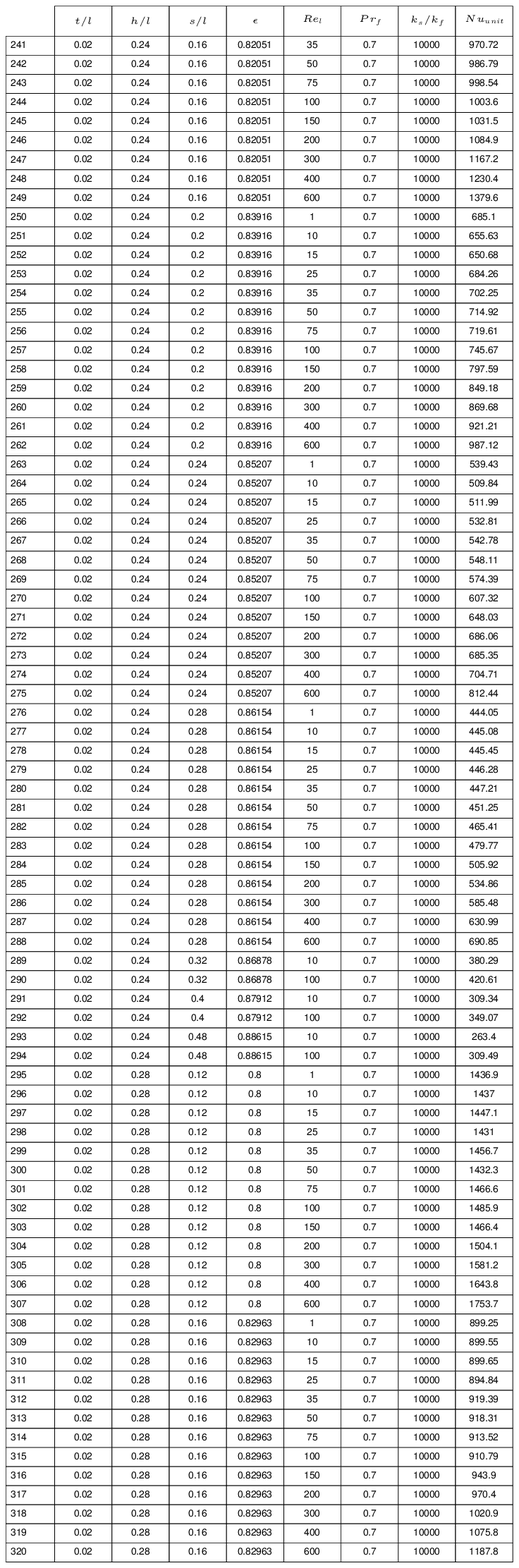}
\end{minipage}

\newpage
\clearpage
\hspace{-20mm}
\begin{minipage}{.5\textwidth}
\raggedleft
\includegraphics[scale = 1.00]{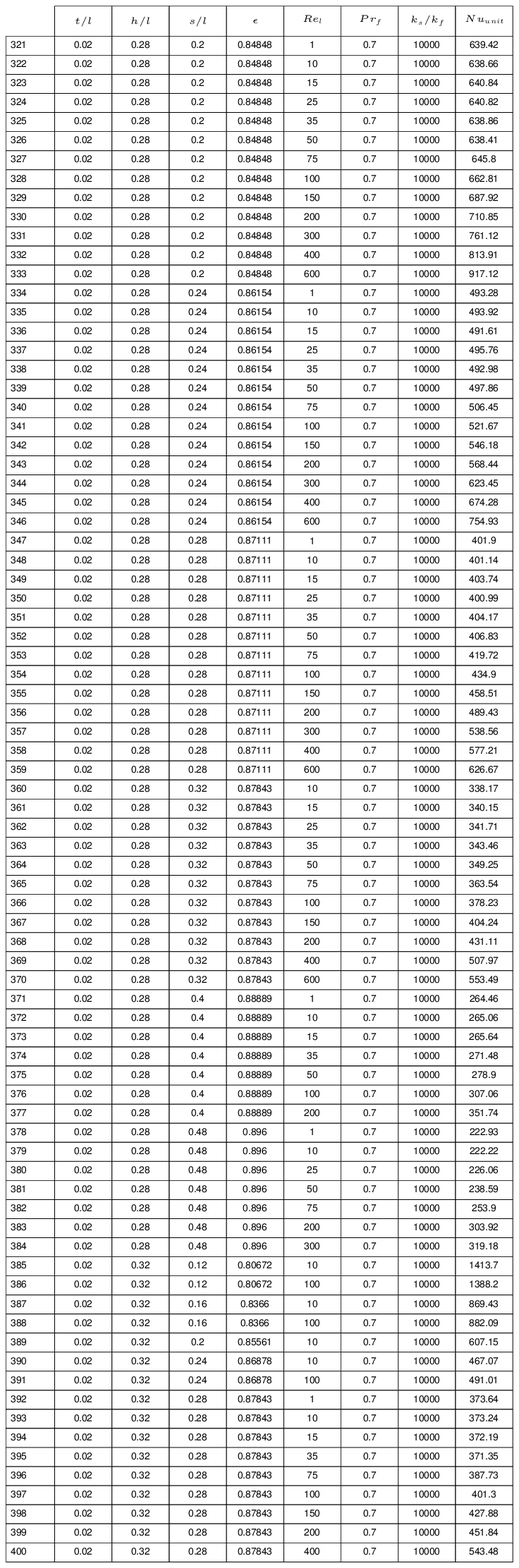}
\end{minipage}
\hfill
\begin{minipage}{.5\textwidth}
\raggedright
\includegraphics[scale = 1.00]{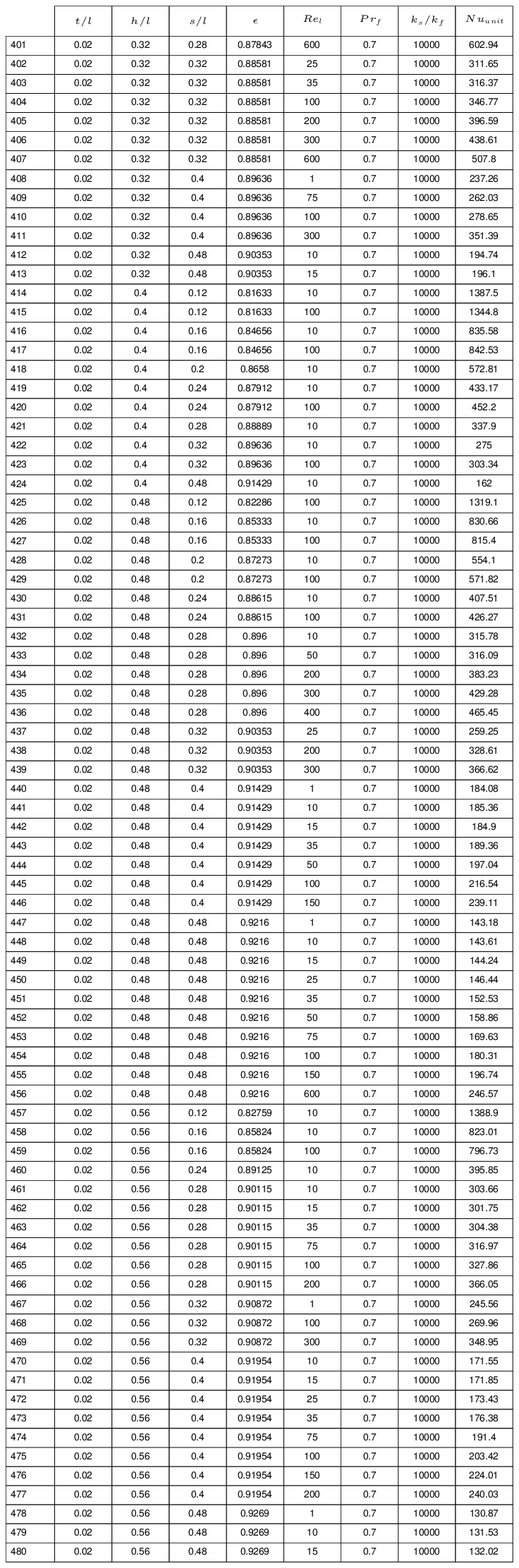}
\end{minipage}

\newpage
\clearpage
\hspace{-20mm}
\begin{minipage}{.5\textwidth}
\raggedleft
\includegraphics[scale = 1.00]{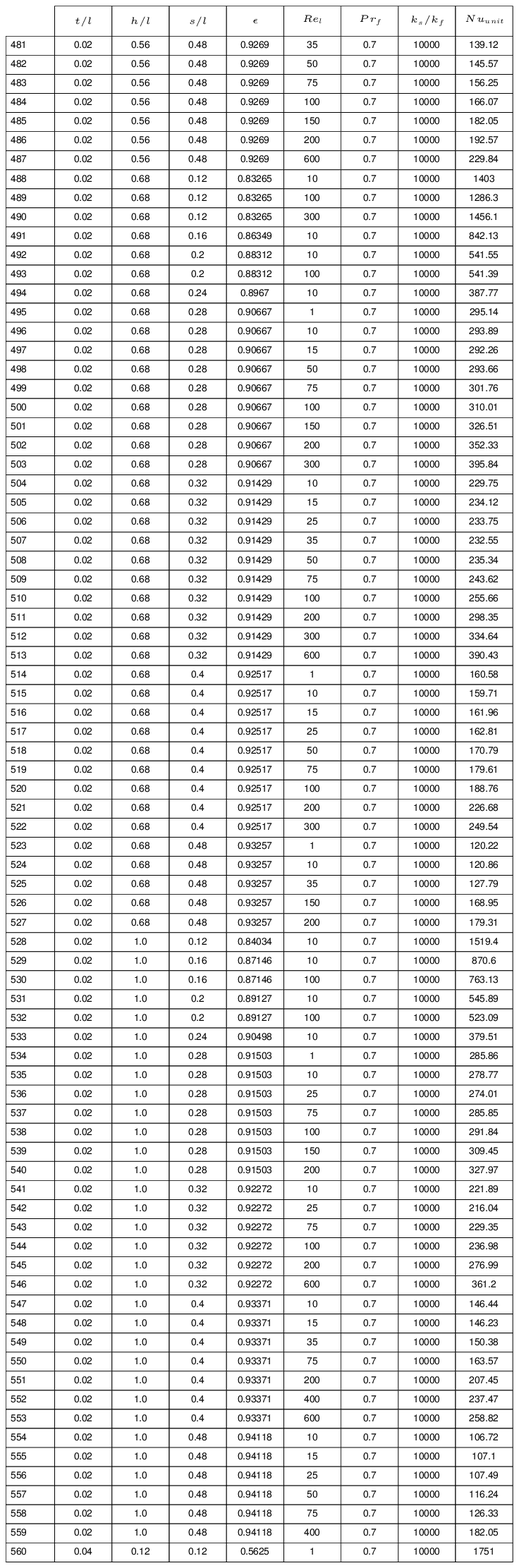}
\end{minipage}
\hfill
\begin{minipage}{.5\textwidth}
\raggedright
\includegraphics[scale = 1.00]{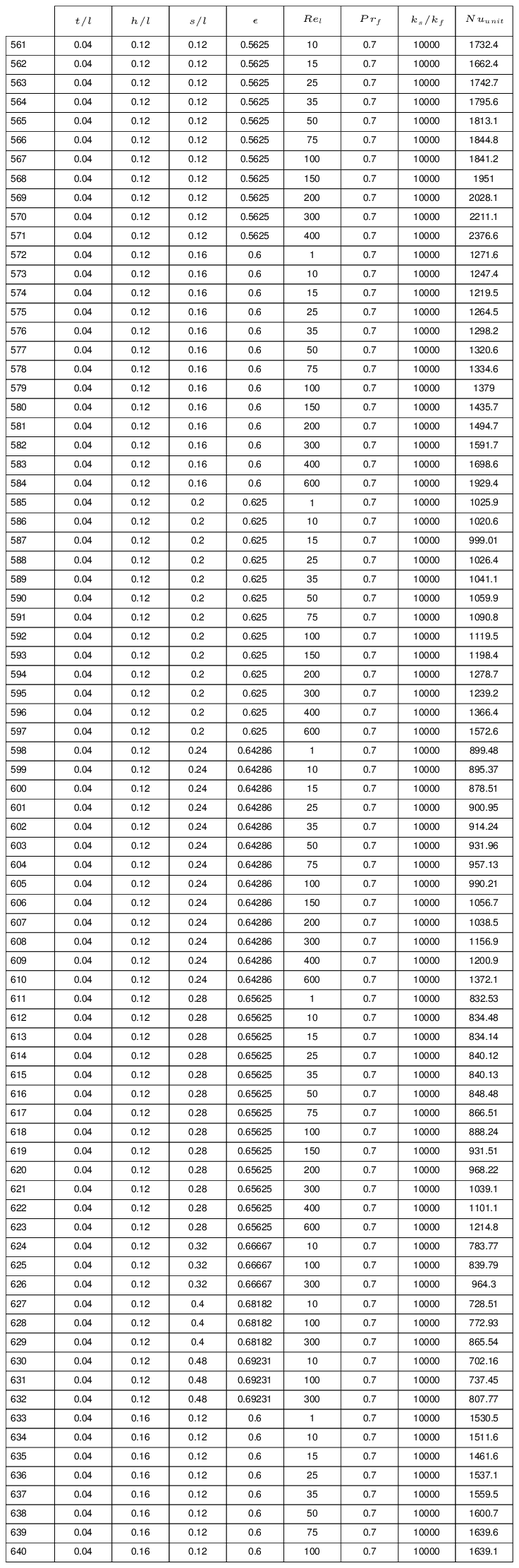}
\end{minipage}

\newpage
\clearpage
\hspace{-20mm}
\begin{minipage}{.5\textwidth}
\raggedleft
\includegraphics[scale = 1.00]{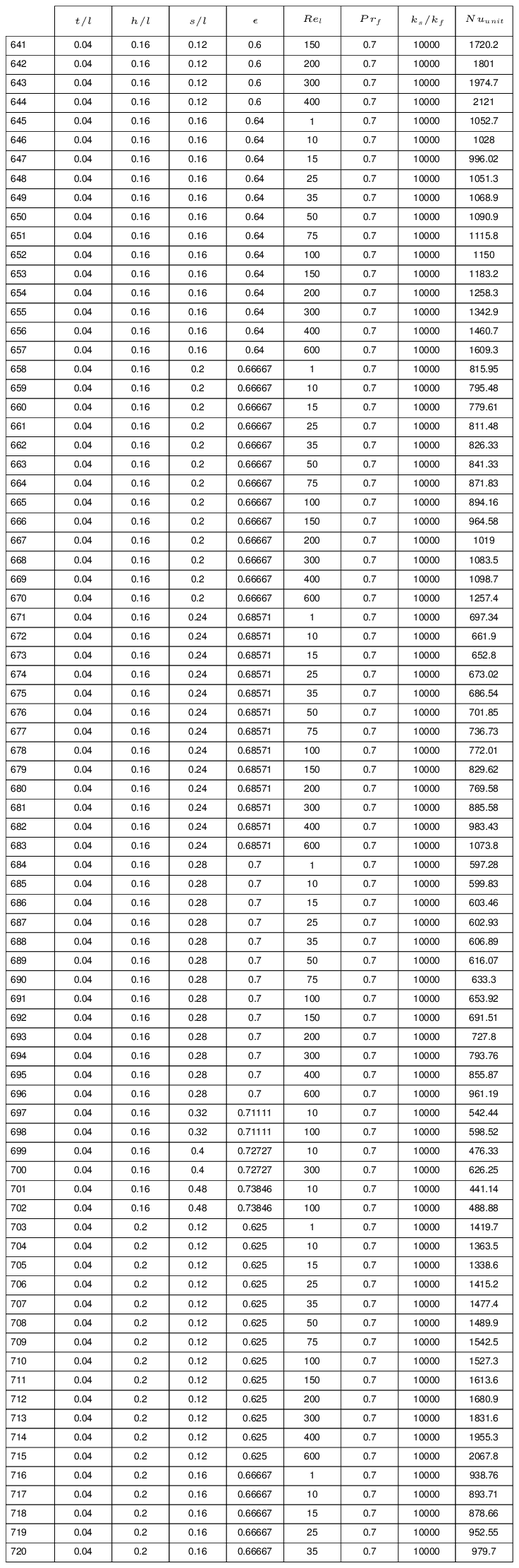}
\end{minipage}
\hfill
\begin{minipage}{.5\textwidth}
\raggedright
\includegraphics[scale = 1.00]{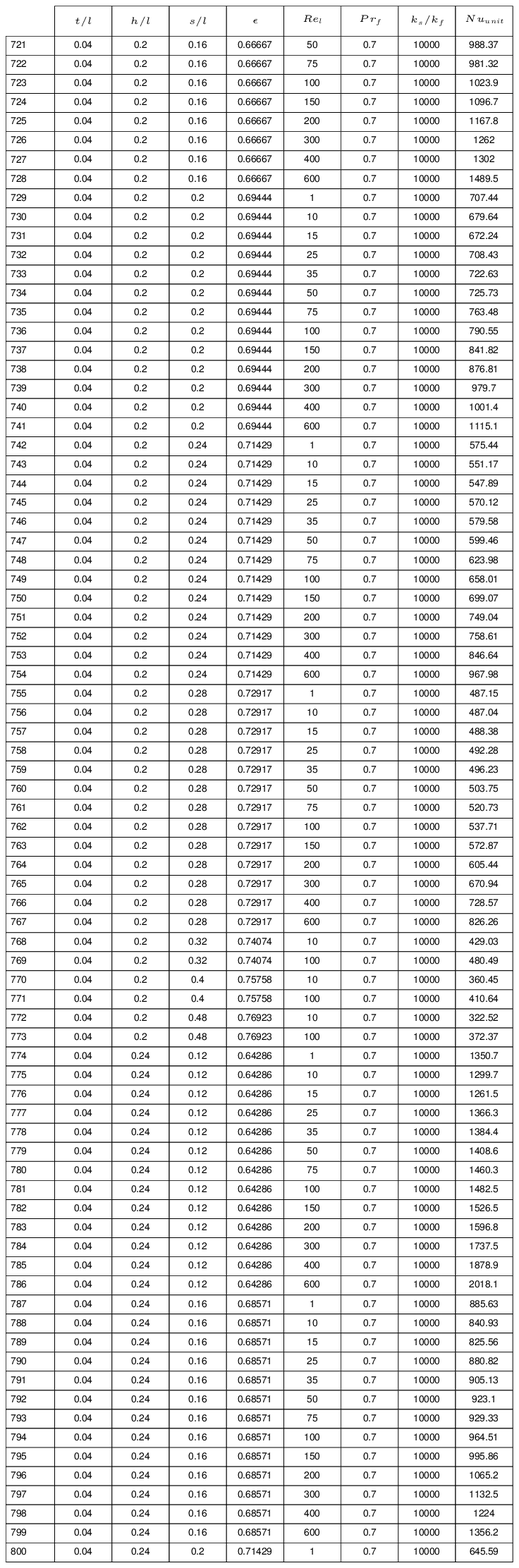}
\end{minipage}

\newpage
\clearpage
\hspace{-20mm}
\begin{minipage}{.5\textwidth}
\raggedleft
\includegraphics[scale = 1.00]{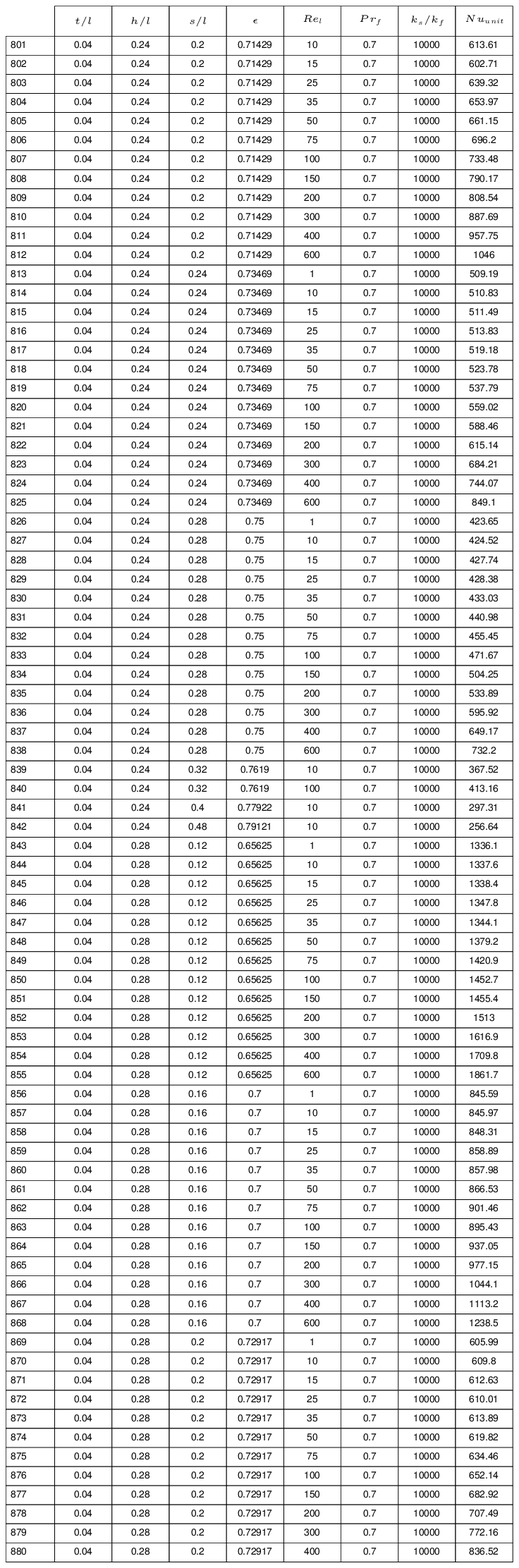}
\end{minipage}
\hfill
\begin{minipage}{.5\textwidth}
\raggedright
\includegraphics[scale = 1.00]{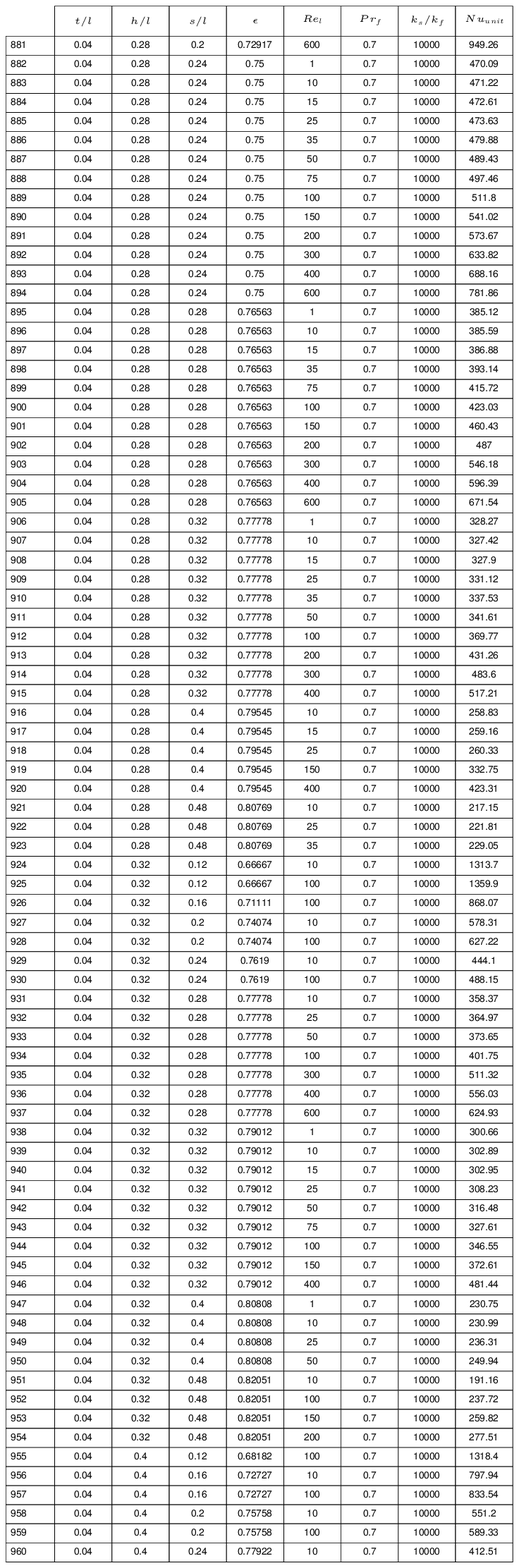}
\end{minipage}

\newpage
\clearpage
\hspace{-20mm}
\begin{minipage}{.5\textwidth}
\raggedleft
\includegraphics[scale = 1.00]{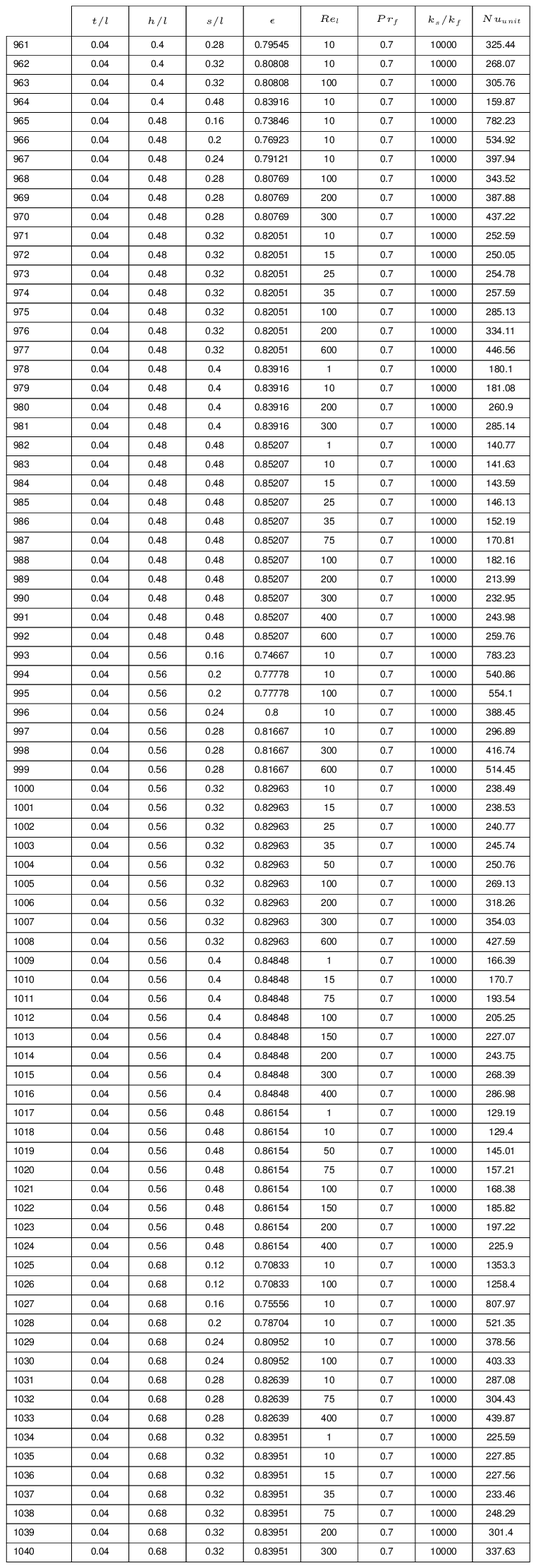}
\end{minipage}
\hfill
\begin{minipage}{.5\textwidth}
\raggedright
\includegraphics[scale = 1.00]{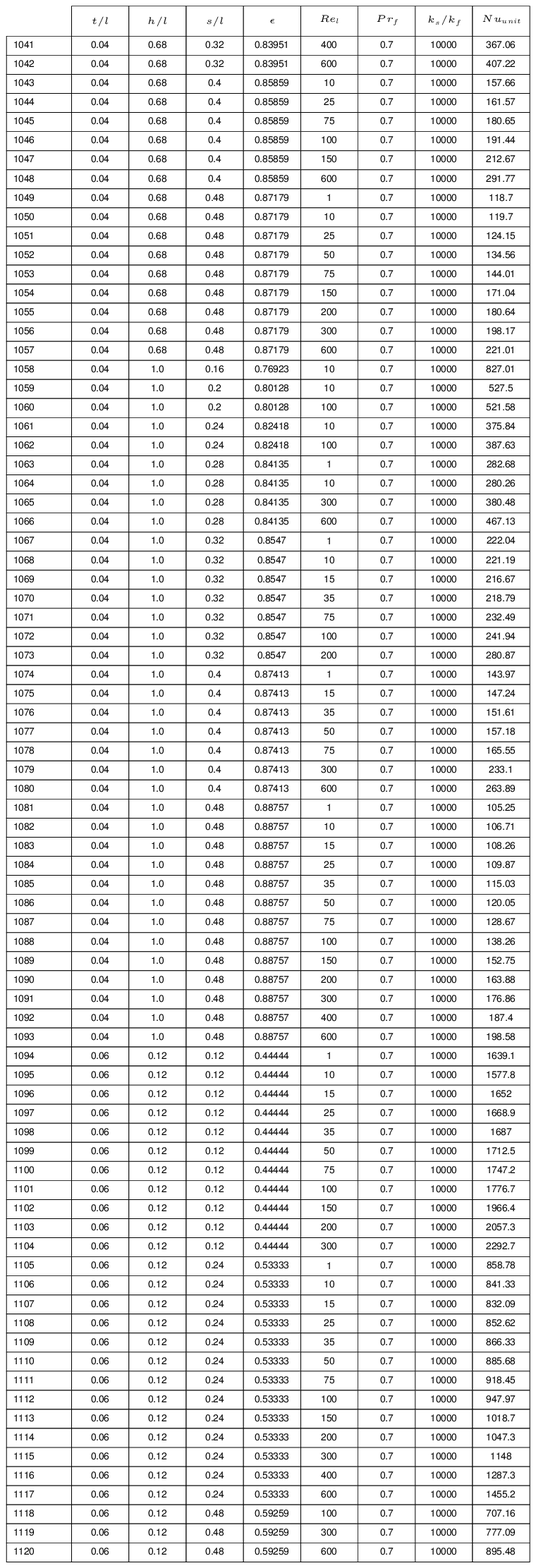}
\end{minipage}

\newpage
\clearpage
\hspace{-20mm}
\begin{minipage}{.5\textwidth}
\raggedleft
\includegraphics[scale = 1.00]{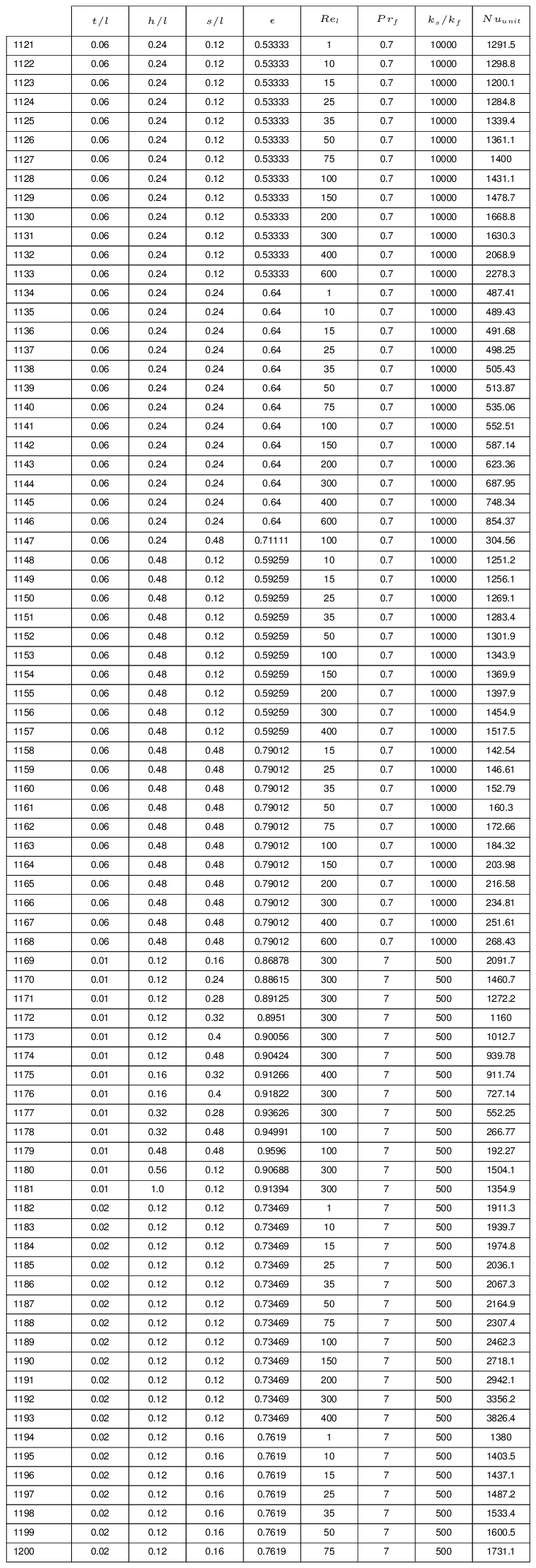}
\end{minipage}
\hfill
\begin{minipage}{.5\textwidth}
\raggedright
\includegraphics[scale = 1.00]{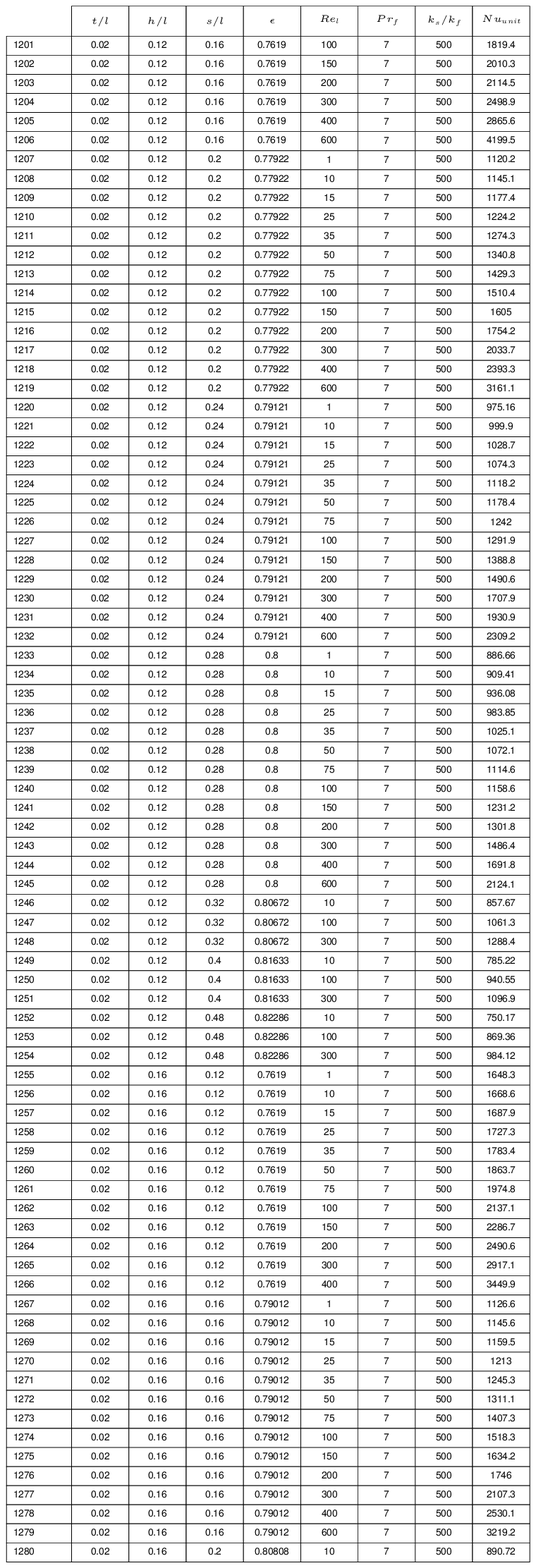}
\end{minipage}

\newpage
\clearpage
\hspace{-20mm}
\begin{minipage}{.5\textwidth}
\raggedleft
\includegraphics[scale = 1.00]{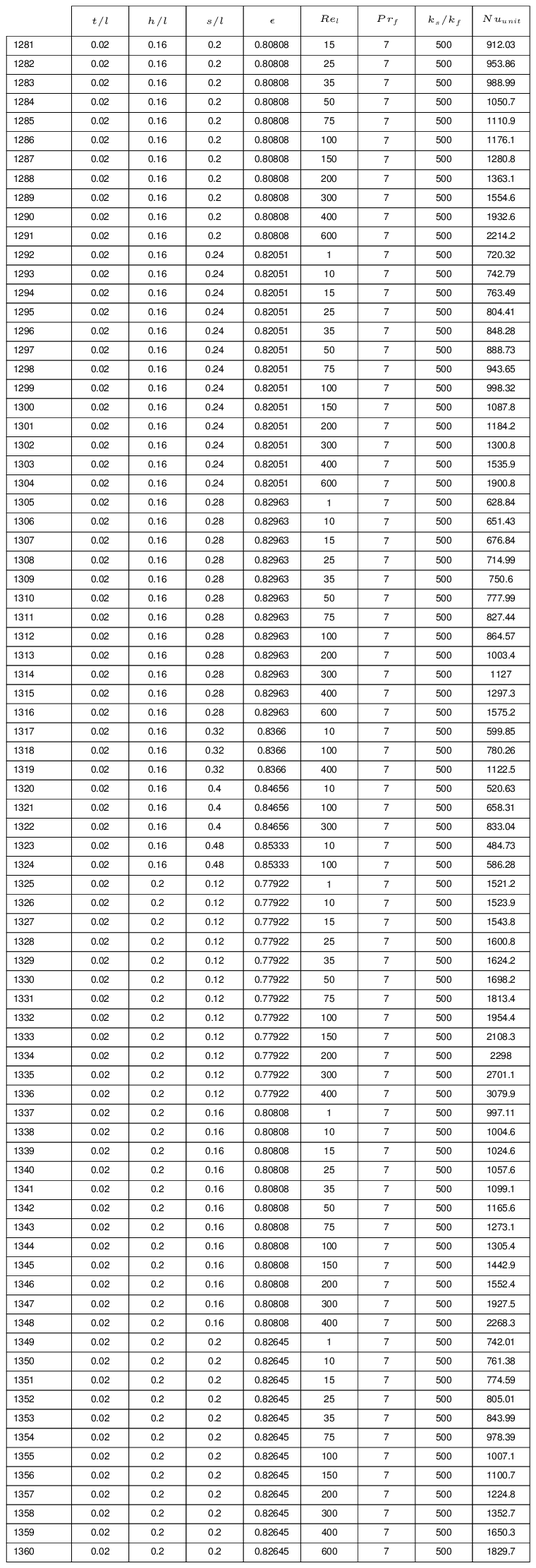}
\end{minipage}
\hfill
\begin{minipage}{.5\textwidth}
\raggedright
\includegraphics[scale = 1.00]{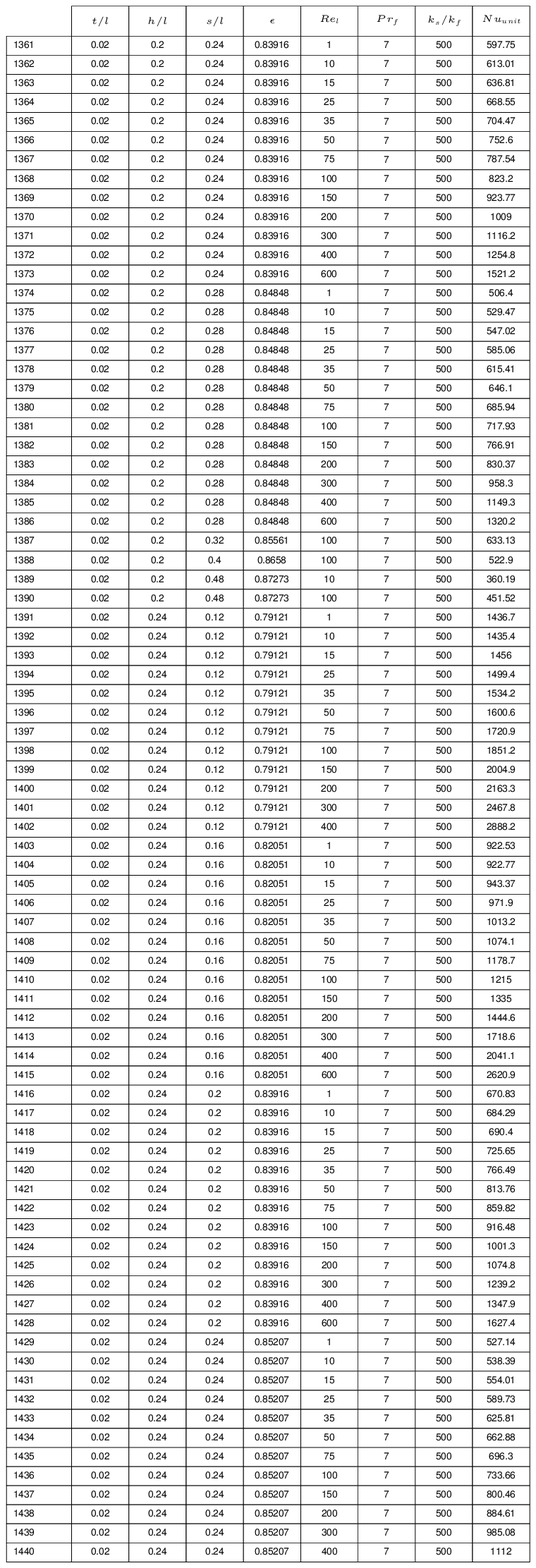}
\end{minipage}

\newpage
\clearpage
\hspace{-20mm}
\begin{minipage}{.5\textwidth}
\raggedleft
\includegraphics[scale = 1.00]{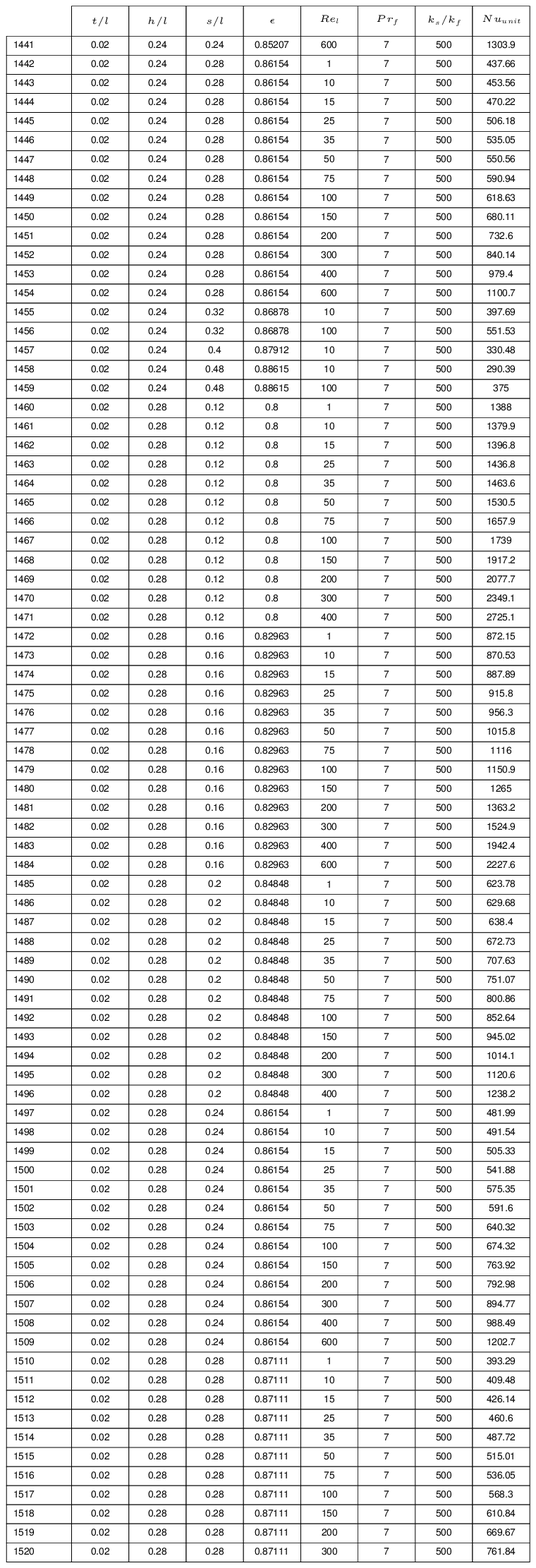}
\end{minipage}
\hfill
\begin{minipage}{.5\textwidth}
\raggedright
\includegraphics[scale = 1.00]{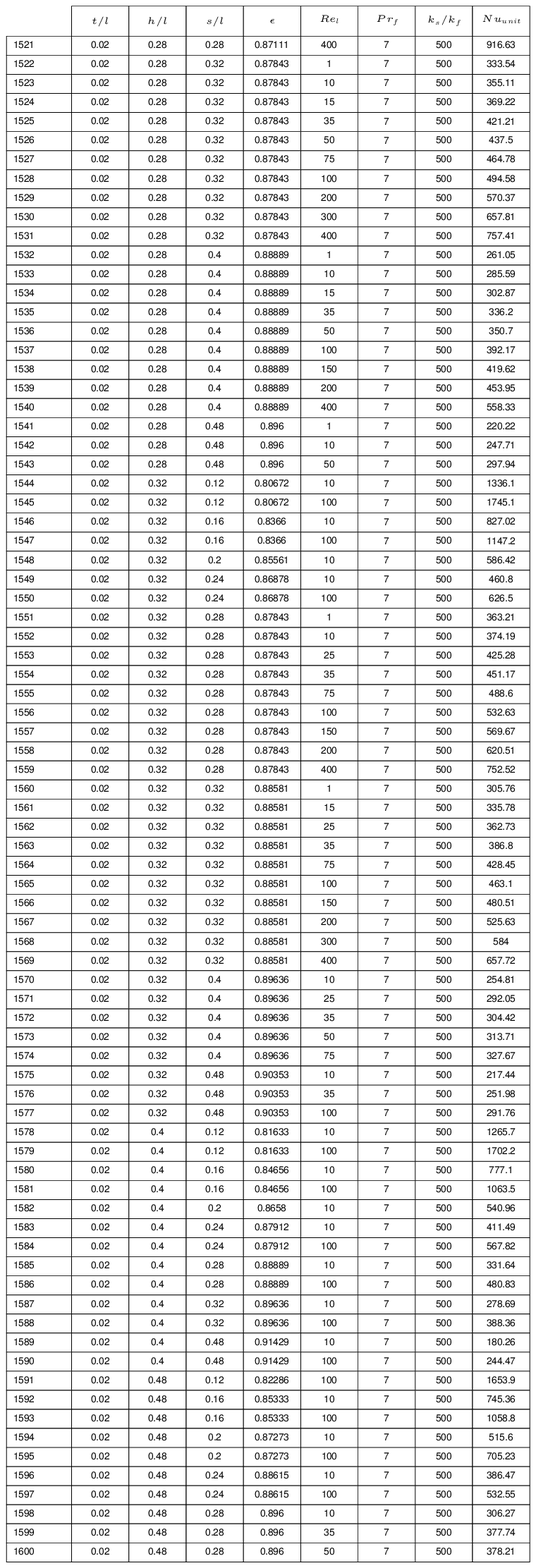}
\end{minipage}

\newpage
\clearpage
\hspace{-20mm}
\begin{minipage}{.5\textwidth}
\raggedleft
\includegraphics[scale = 1.00]{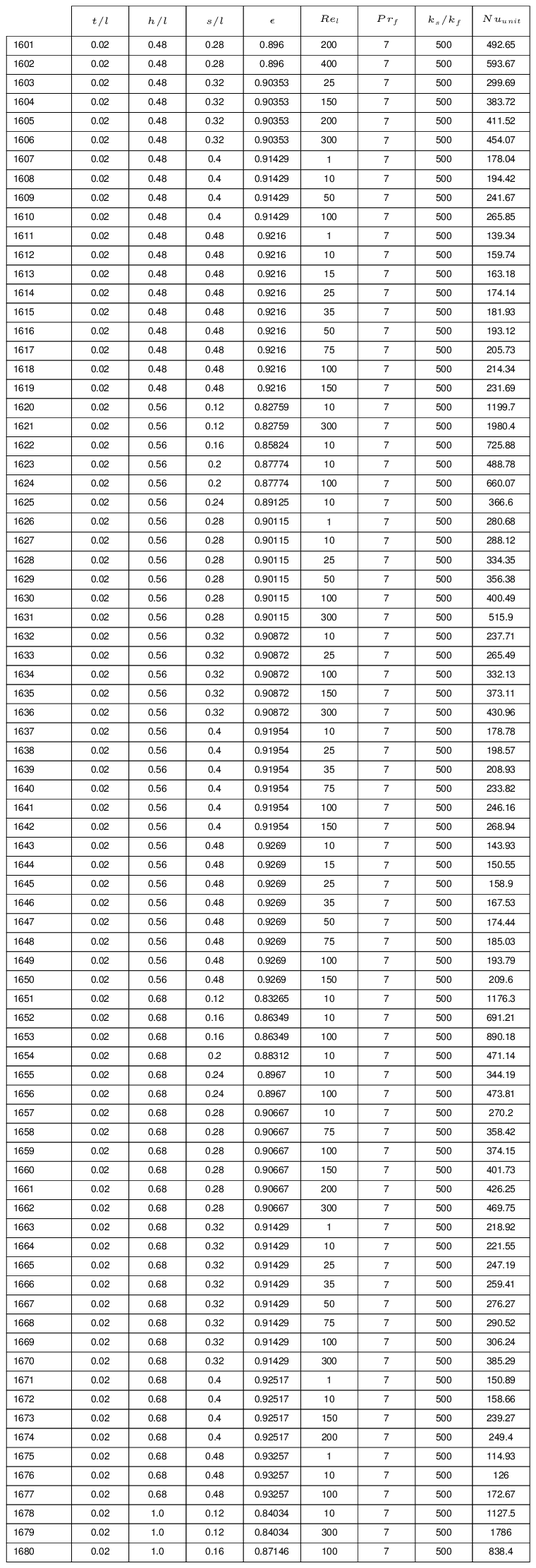}
\end{minipage}
\hfill
\begin{minipage}{.5\textwidth}
\raggedright
\includegraphics[scale = 1.00]{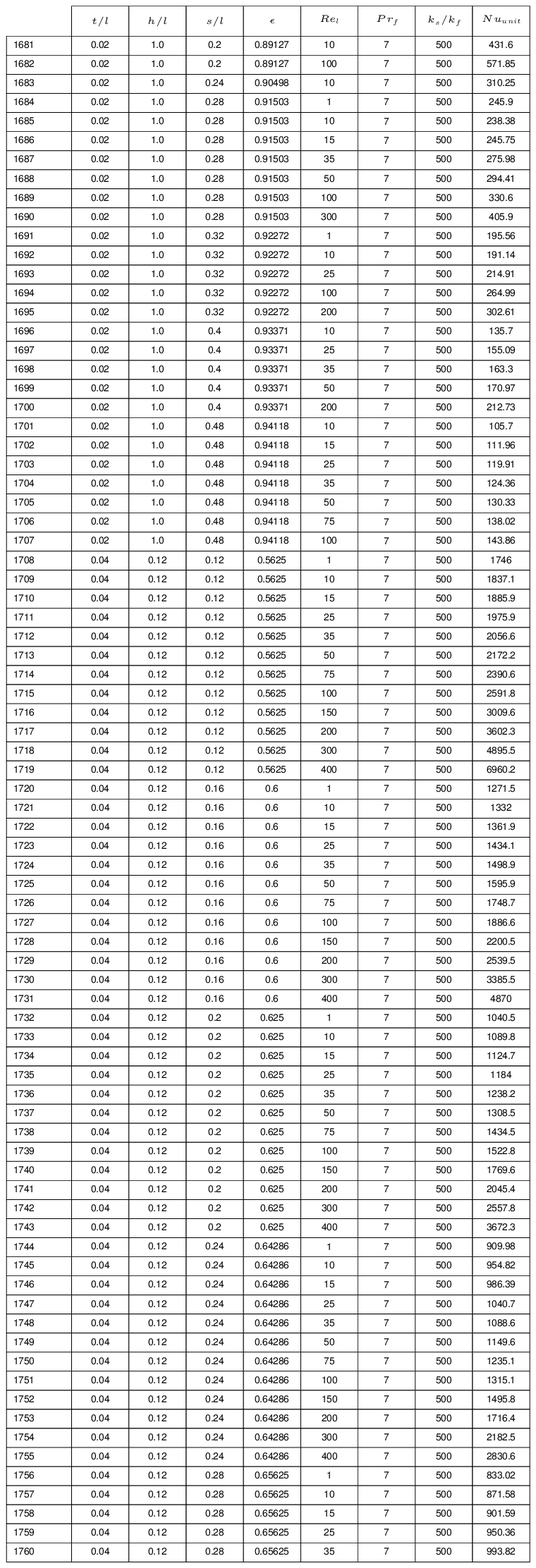}
\end{minipage}

\newpage
\clearpage
\hspace{-20mm}
\begin{minipage}{.5\textwidth}
\raggedleft
\includegraphics[scale = 1.00]{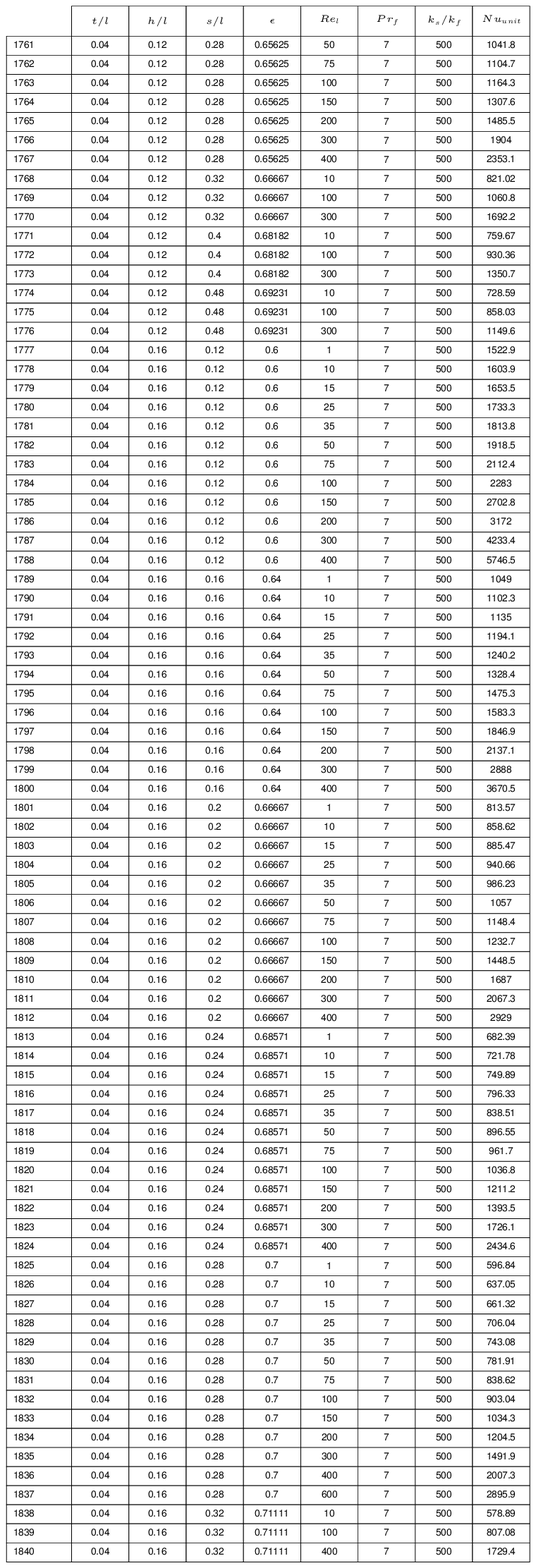}
\end{minipage}
\hfill
\begin{minipage}{.5\textwidth}
\raggedright
\includegraphics[scale = 1.00]{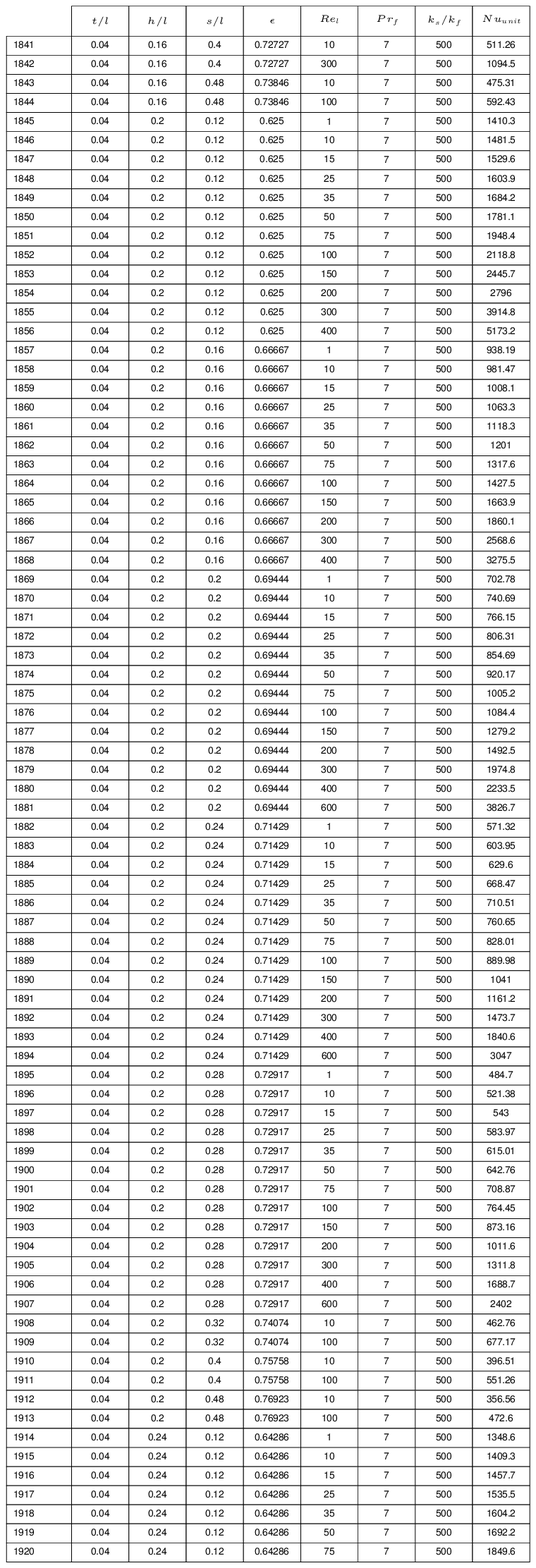}
\end{minipage}

\newpage
\clearpage
\hspace{-20mm}
\begin{minipage}{.5\textwidth}
\raggedleft
\includegraphics[scale = 1.00]{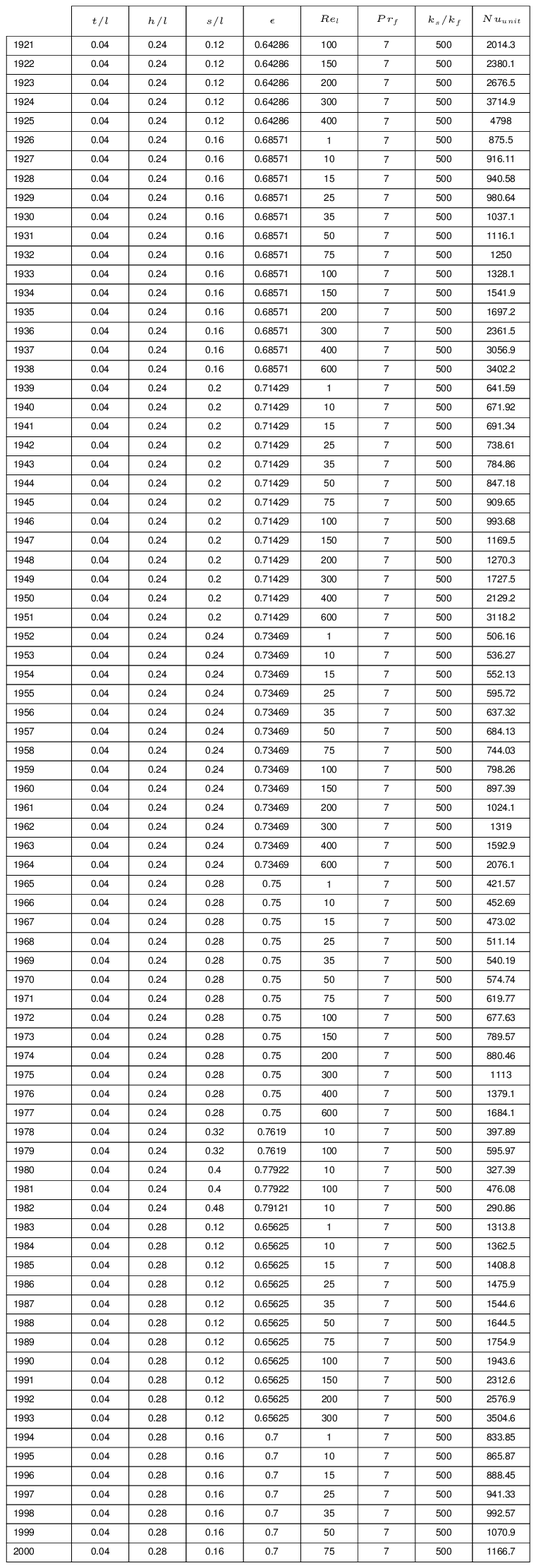}
\end{minipage}
\hfill
\begin{minipage}{.5\textwidth}
\raggedright
\includegraphics[scale = 1.00]{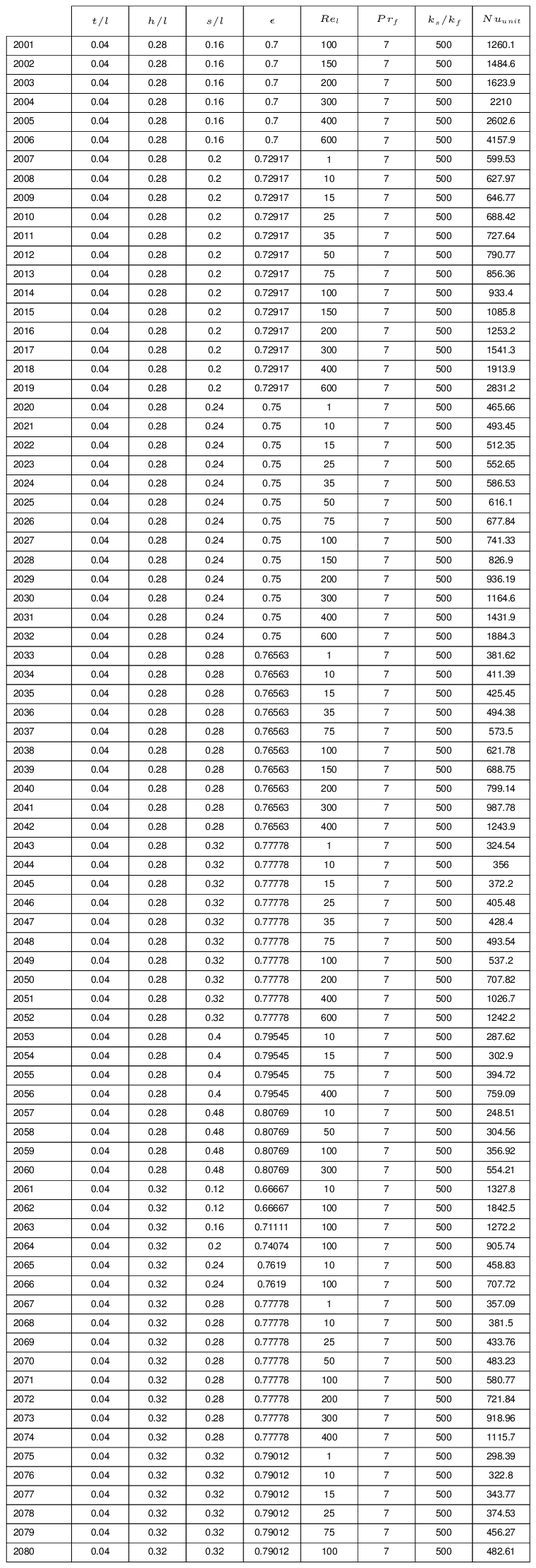}
\end{minipage}

\newpage
\clearpage
\hspace{-20mm}
\begin{minipage}{.5\textwidth}
\raggedleft
\includegraphics[scale = 1.00]{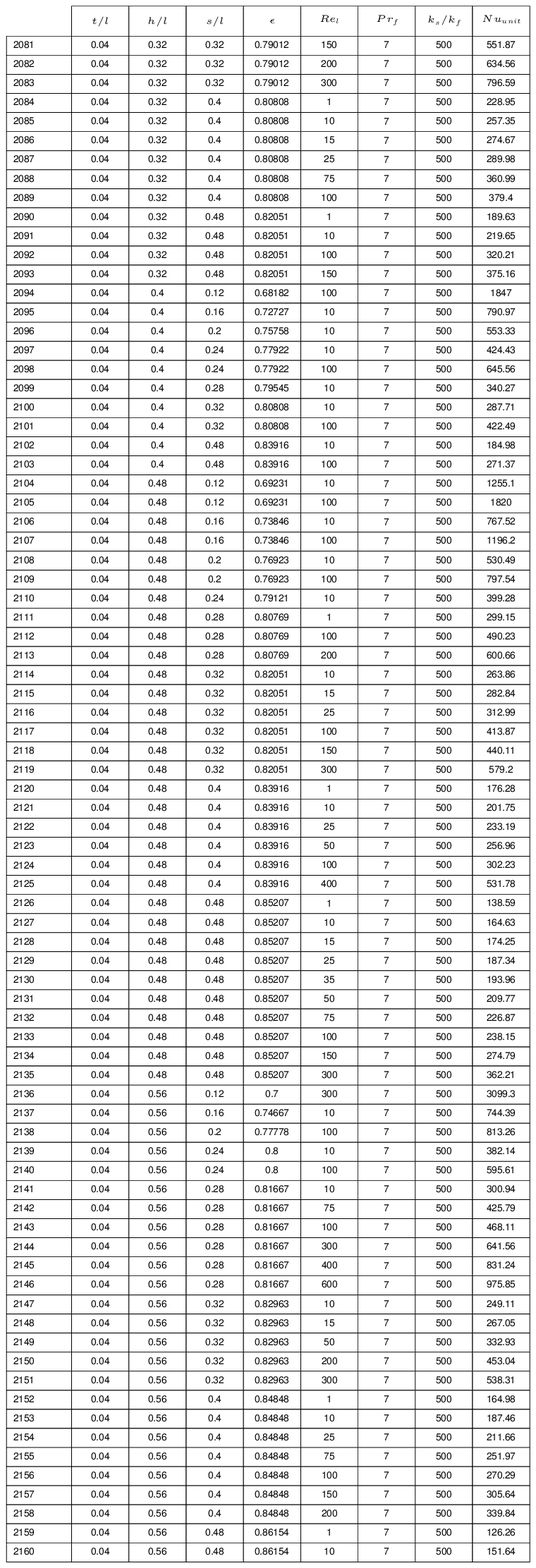}
\end{minipage}
\hfill
\begin{minipage}{.5\textwidth}
\raggedright
\includegraphics[scale = 1.00]{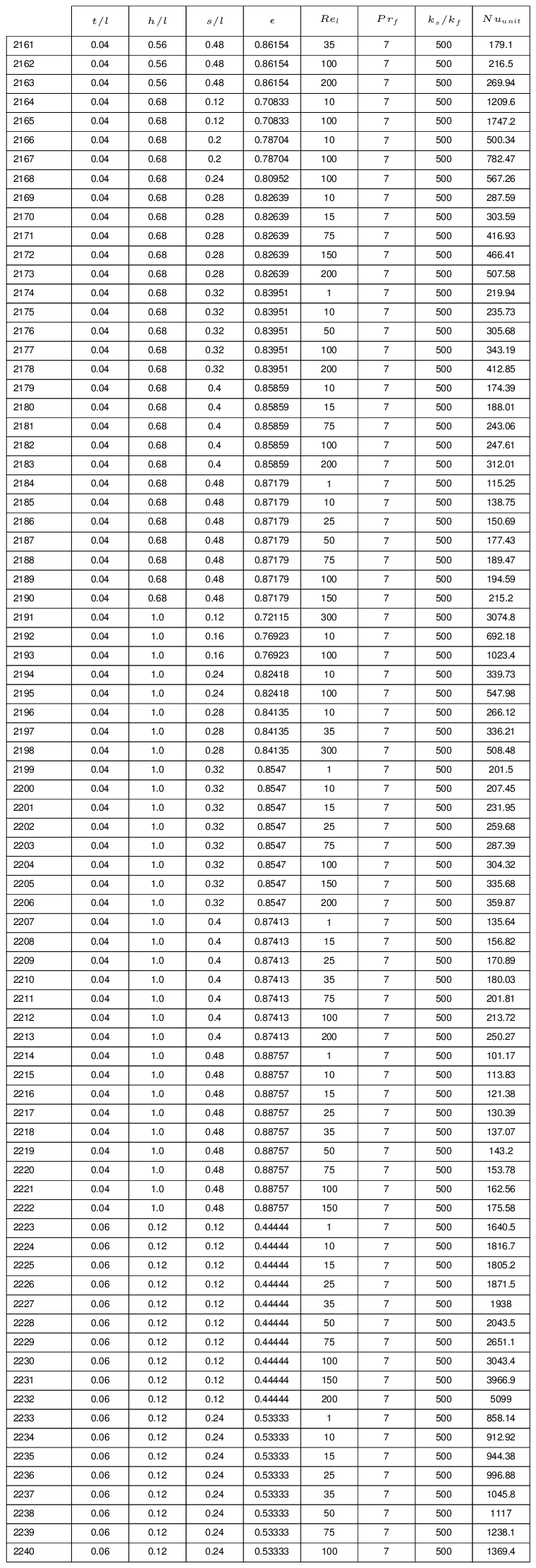}
\end{minipage}

\newpage
\clearpage
\hspace{-20mm}
\begin{minipage}{.5\textwidth}
\raggedleft
\includegraphics[scale = 1.00]{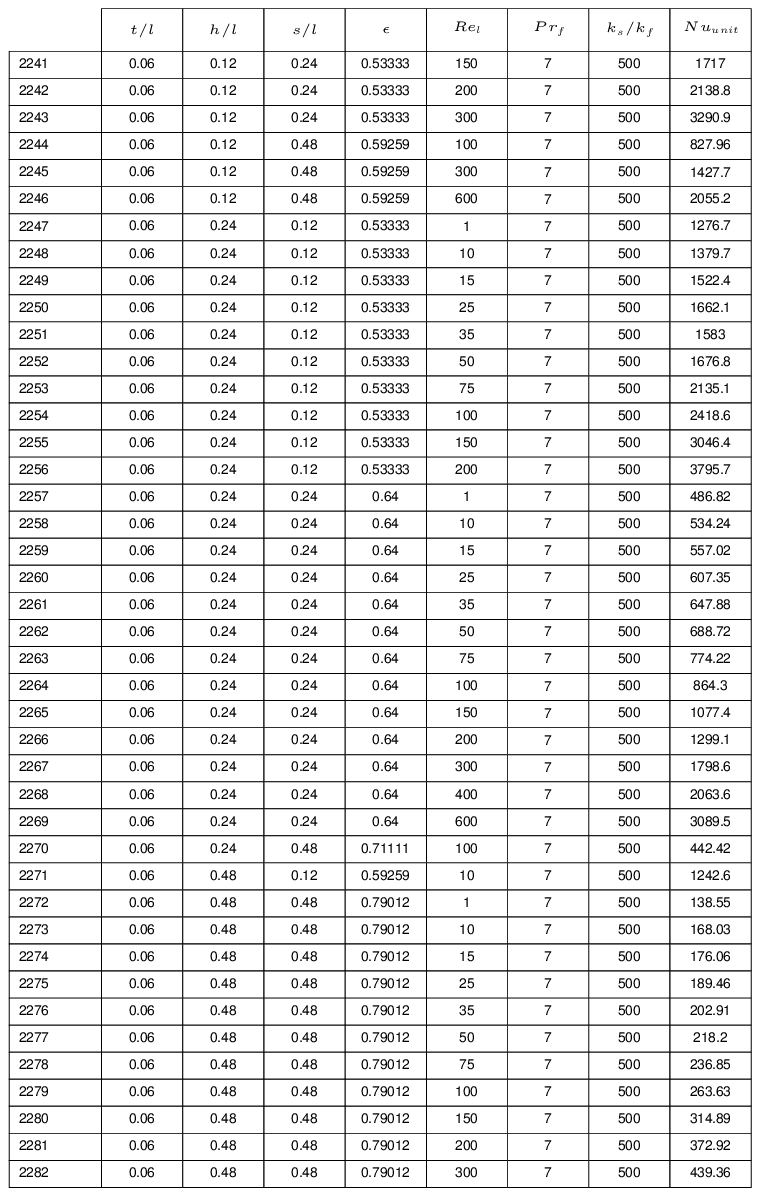}
\end{minipage}
\hfill
\begin{minipage}{.5\textwidth}
\raggedright
\end{minipage}

\newpage
\clearpage

\section{\label{sec:alignedflowdata_extra}Additional periodically developed Nusselt number data for the study on the influence $Pr_f$ and $k_{s}/k_{f}$}

\hspace{-25mm}
\begin{minipage}{.5\textwidth}
\raggedleft
\includegraphics[scale = 1.00]{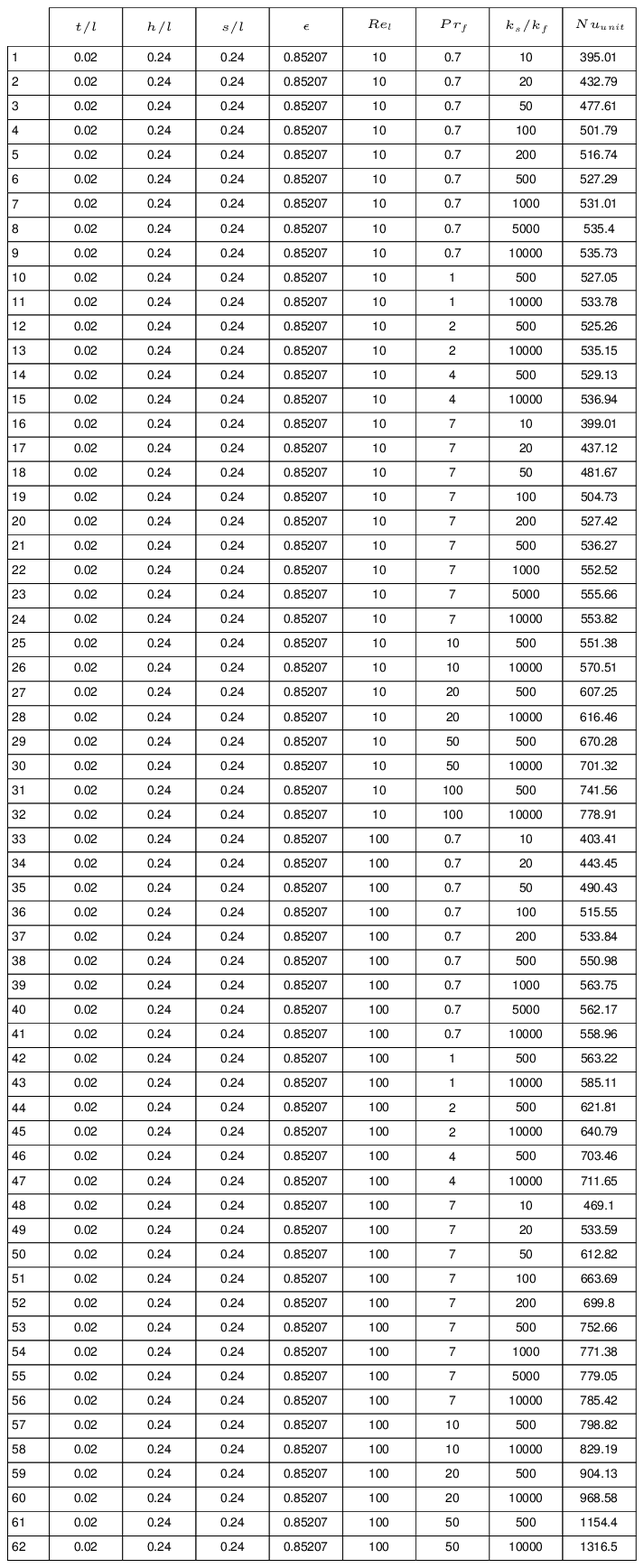}
\end{minipage}
\hfill
\begin{minipage}{.5\textwidth}
\raggedright
\end{minipage}


\nocite{*}
\bibliography{aipsamp}

\end{document}